\documentclass[pdflatex,sn-mathphys-num]{sn-jnl}

\usepackage{graphicx}%
\usepackage{multirow}%
\usepackage{amsmath,amssymb,amsfonts}%
\usepackage{amsthm}%
\usepackage{mathrsfs}%
\usepackage[title]{appendix}%
\usepackage[table,xcdraw]{xcolor}
\usepackage{textcomp}%
\usepackage{manyfoot}%
\usepackage{booktabs}%
\usepackage{algorithm}%
\usepackage{algorithmicx}%
\usepackage{algpseudocode}%
\usepackage{listings}%
\usepackage{subcaption}
\usepackage{siunitx}
\usepackage{graphicx}

\usepackage[version=4,arrows=pgf-filled]{mhchem}
\usepackage{mathtools}
\usepackage{nowidow}
\usepackage[most]{tcolorbox}
\usepackage{tikz}
\newcommand{\IDVG}{{\(I_\mathrm{d}\)-\(V_\mathrm{g}\) }}
\newcommand{\fig}[2]{Fig.~\ref{#1}\textcolor{blue}{#2}}

\theoremstyle{thmstyleone}%
\theoremstyle{thmstyletwo}%

\theoremstyle{thmstylethree}%

\begin{document}

\title[Article Title]{\centering Hysteresis Measurements as a Diagnostic Tool: \\ 
A Systematic Approach for Stability Benchmarking and Performance Projection of 2D-Materials-Based MOSFETs}


\author[1]{\fnm{Alexander} \sur{Karl}}\email{karl@iue.tuwien.ac.at}

\author[1]{\fnm{Dominic} \sur{Waldhoer}}

\author[1]{\fnm{Theresia} \sur{Knobloch}}

\author[1]{\fnm{Axel} \sur{Verdianu}}

\author[1]{\fnm{Jo\"el} \sur{Kurzweil}}

\author[1]{\fnm{Mina} \sur{Bahrami}}

\author[1]{\fnm{Mohammad Rasool} \sur{Davoudi}}

\author[1]{\fnm{Pedram} \sur{Khakbaz}}

\author[1]{\fnm{Bernhard} \sur{Stampfer}}

\author[1]{\fnm{Seyed Mehdi} \sur{Sattari-Esfahlan}}

\author[2]{\fnm{
Yury} \sur{Illarionov}}

\author[3]{\fnm{Aftab} \sur{Nazir}}

\author[3]{\fnm{Changze} \sur{Liu}}

\author[4]{\fnm{Saptarshi} \sur{Das}}

\author[5, 6]{\fnm{Xiao} \sur{Renshaw Wang}}

\author[7]{\fnm{Junchuan} \sur{Tang}}

\author[7]{\fnm{Yichi} \sur{Zhang}}

\author[7]{\fnm{Congwei} \sur{Tan}}

\author[7]{\fnm{Ye} \sur{Li}}

\author[7]{\fnm{Hailin} \sur{Peng}}


\author[1]{\fnm{Michael} \sur{Waltl}}

\author*[1]{\fnm{Tibor} \sur{Grasser}}\email{grasser@iue.tuwien.ac.at}

\affil[1]{\orgdiv{Institute for Microelectronics}, \orgname{Technical University Vienna}, \orgaddress{\country{Austria}}}

\affil[2]{\orgdiv{Department of Materials Science and Engineering}, \orgname{SUSTech}, \orgaddress{\country{China}}}

\affil[3]{\orgname{Huawei Technologies Research and Development Belguim N.V.}, \orgaddress{\country{Belgium}}}

\affil[4]{\orgdiv{Department of Materials Science and Engineering}, \orgname{Penn State University}, \orgaddress{\country{USA}}}

\affil[5]{\orgdiv{Division of Physics and Applied Physics}, \orgname{NTU}, \orgaddress{\country{Singapore}}}
\affil[6]{\orgdiv{School of Electrical and Electronic Engineering}, \orgname{NTU}, \orgaddress{\country{Singapore}}}

\affil[7]{\orgdiv{College of Chemistry and Molecular Engineering}, \orgname{Peking University}, \orgaddress{\country{China}}}


\abstract{


Judging by its omnipresence in the literature, the hysteresis observed in the transfer characteristics of emerging transistors based on 2D-materials is widely accepted as an important metric related to the device quality. The hysteresis is often reported with attributes like ``negligible'' or ``small'' without giving any specifics as to how this was determined and against what reference the measured values were compared to.
Quite surprisingly, there appears to be only a fragmentary understanding of the mechanisms actually contributing to hysteresis and the sensitivity of the actual measurement on various experimental parameters. 
We attempt to close this gap by first providing a comprehensive theoretical analysis of the dominant mechanisms contributing to hysteresis: charge trapping
by defects from the channel or the gate, the drift of mobile charges, and eventually ferroelectricity. We continue by suggesting methods to experimentally distinguishing between these phenomena.
Based on these discussions it becomes clear that previously reported hysteresis values have little meaning as they have been non-systematically recorded under arbitrary conditions. In order to resolve this predicament, we propose a standardized hysteresis measurement scheme to establish the hysteresis as a comparable metric for the assessment of device stability. Our standardized scheme ensures that hysteresis data can be effectively compared
across different technologies and, most importantly, provide a means to extrapolate data obtained on thicker prototypes to subnanometer equivalent oxide thicknesses. This facilitates the systematic benchmarking of insulator/channel combinations in terms of stability, which thereby enables the screening of material systems for more stable and reliable 2D-material-based MOSFETs.

}




\maketitle
\newpage

\section{Introduction}\label{sec1}

Metal-Oxide-Semiconductor Field-Effect Transistors~(MOSFETs) play the main role in shaping the performance and functionality of integrated circuits. The integration of two-dimensional materials (2D-materials) like transition metal dichalcogenides has emerged as a promising approach to push the performance boundaries of MOSFETs, as their atomically thin nature offers a considerable advantage for device scaling. However, currently available 2D-material based MOSFETs (2D-MOSFETs) typically exhibit poor stability. In this paper, we address one of the main stability issues of 2D-MOSFETs, the \textbf{hysteresis in the transfer characteristics} \cite{Late2012, Jin2024, Lee2018, Sheng2024, Kato2023, Si2017, Roh2016, Datye2018}. While hysteresis is widely recognized as a critical metric, many works advertise their achievements using vague terms such as ``negligible hysteresis'' or even ``hysteresis-free'', without providing a clear definition. We will demonstrate here that an accurate interpretation of hysteresis measurements is anything but trivial, since the observed hysteresis can vary dramatically depending on how it is measured. \\
First, we provide a comprehensive theoretical analysis of the dominant mechanisms that potentially contribute to hysteresis. We examine charge trapping due to defects, the drift and diffusion of mobile charges, as well as ferroelectricity, with a focus on how these mechanisms can be experimentally differentiated. This analysis aims to identify the sources of instabilities in emerging 2D-MOSFETs, facilitating the development of effective countermeasures.
Based on the observation that the measured hysteresis is very sensitive to the sweep frequency, sweep voltage range, temperature, vacuum level, as well as insulator thickness \cite{Knobloch2018, DiBartolomeo_2018, Park2017, AlMamun2024}, we highlight the need to standardize hysteresis measurements to establish the hysteresis as a comparable metric for device stability. This is critical because current literature often reports hysteresis values measured under arbitrary conditions, making the reported values unsuitable for comparison. To address this issue, we propose a unified approach to hysteresis measurements, enabling meaningful comparisons across different technologies and allowing for the extrapolation of data obtained from thicker prototypes to sub-nanometer equivalent oxide thicknesses (EOTs). \\

\section{Hysteresis Measurements - The Necessity for Standardization and Normalization}\label{sec2}

\begin{figure}[hbt!]
\begin{subfigure}[b]{.333\linewidth}
    \begin{tikzpicture}
    \node[inner sep=0pt] at (0,0) {
      \begin{minipage}{1.0\textwidth}
        \includegraphics[width=\textwidth]{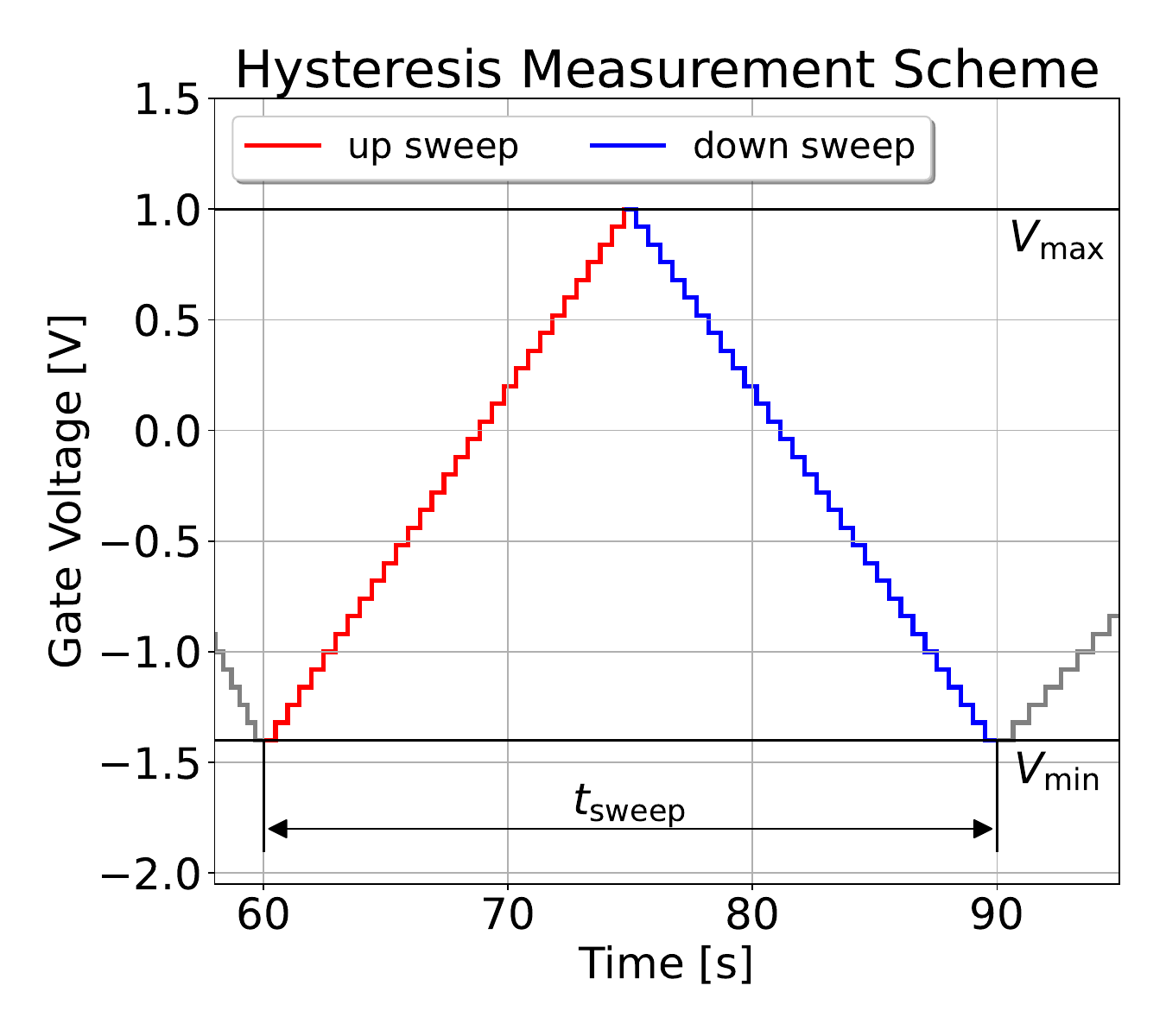} 
    \end{minipage}};
    \node[draw=none] at (-2.0,2.4) {\textbf{(a)}};
    \end{tikzpicture}
\end{subfigure}
\begin{subfigure}[b]{.333\linewidth}
    \begin{tikzpicture}
    \node[inner sep=0pt] at (0,0) {
      \begin{minipage}{1.0\textwidth}
        \includegraphics[width=\textwidth]
        {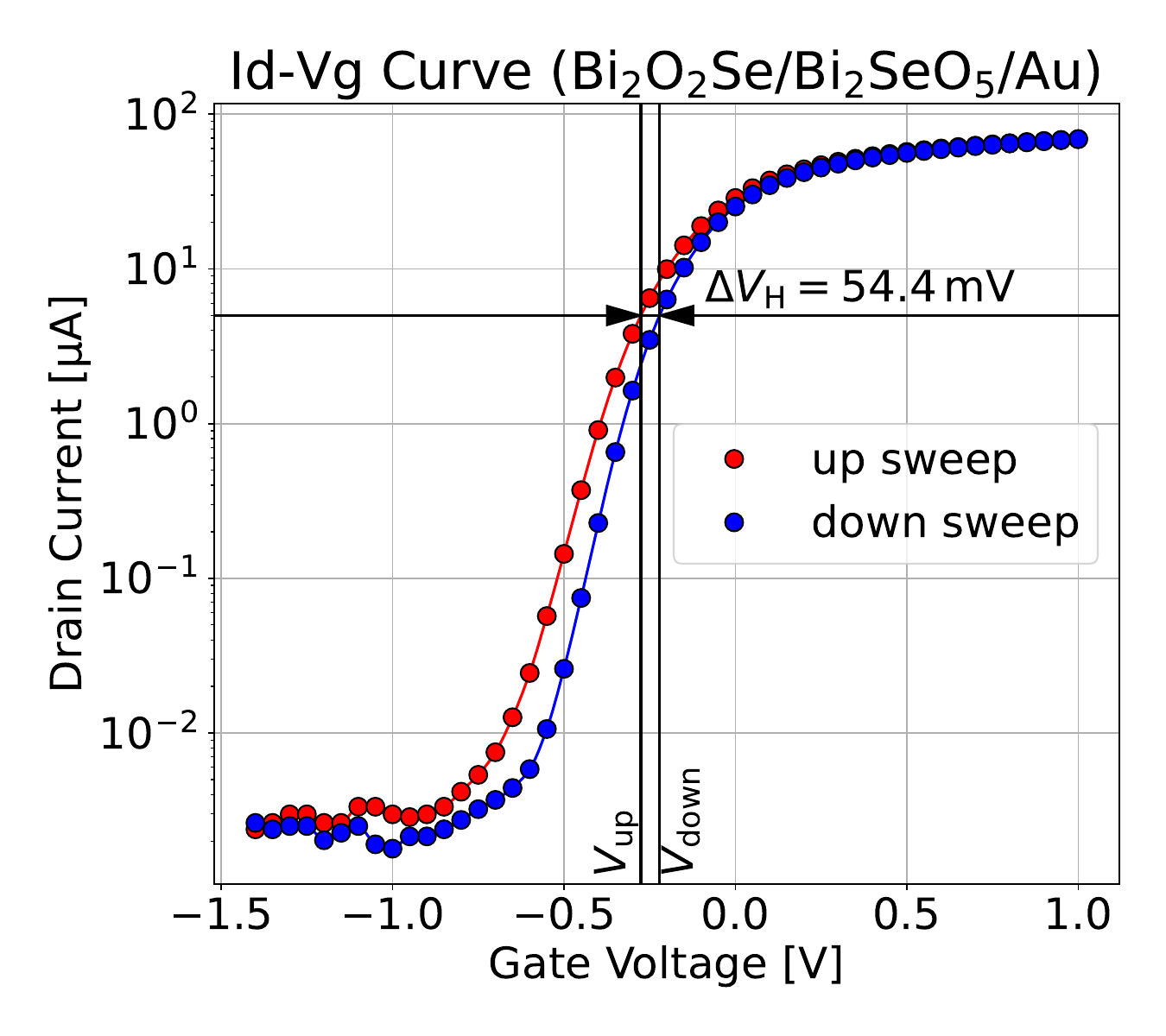}
    \end{minipage}};
    \node[draw=none] at (-2.0,2.4) {\textbf{(b)}};
    \end{tikzpicture}
\end{subfigure}
\begin{subfigure}[b]{.333\linewidth}
    \begin{tikzpicture}
    \node[inner sep=0pt] at (0,0) {
      \begin{minipage}{1.0\textwidth}
        \includegraphics[width=\textwidth]{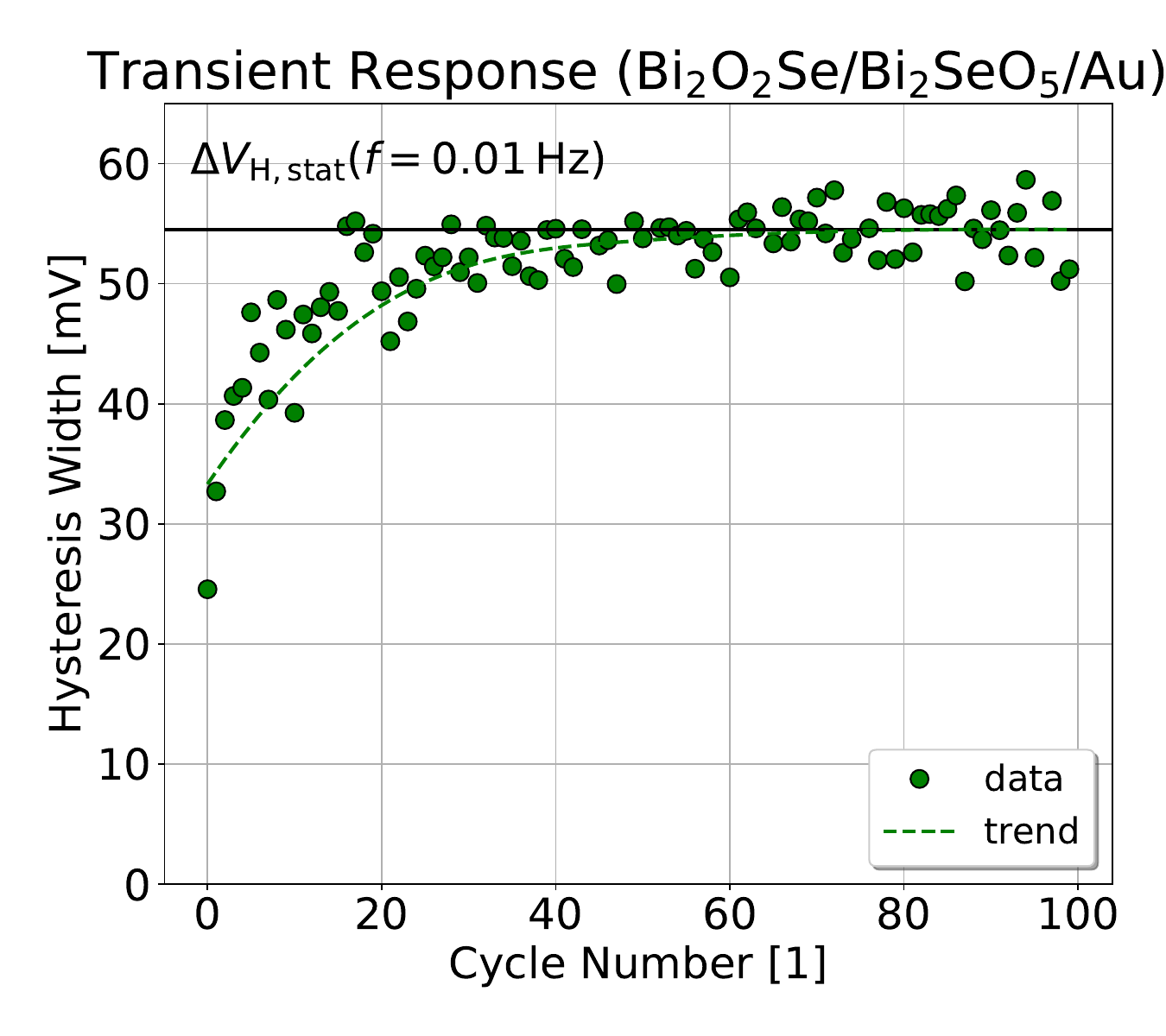}  
    \end{minipage}};
    \node[draw=none] at (-2.0,2.4) {\textbf{(c)}};
    \end{tikzpicture}
\end{subfigure}
\begin{center}
\begin{subfigure}[b]{.333\linewidth}
    \begin{tikzpicture}
    \node[inner sep=0pt] at (0,0) {
      \begin{minipage}{1.0\textwidth}
        \includegraphics[width=\textwidth]{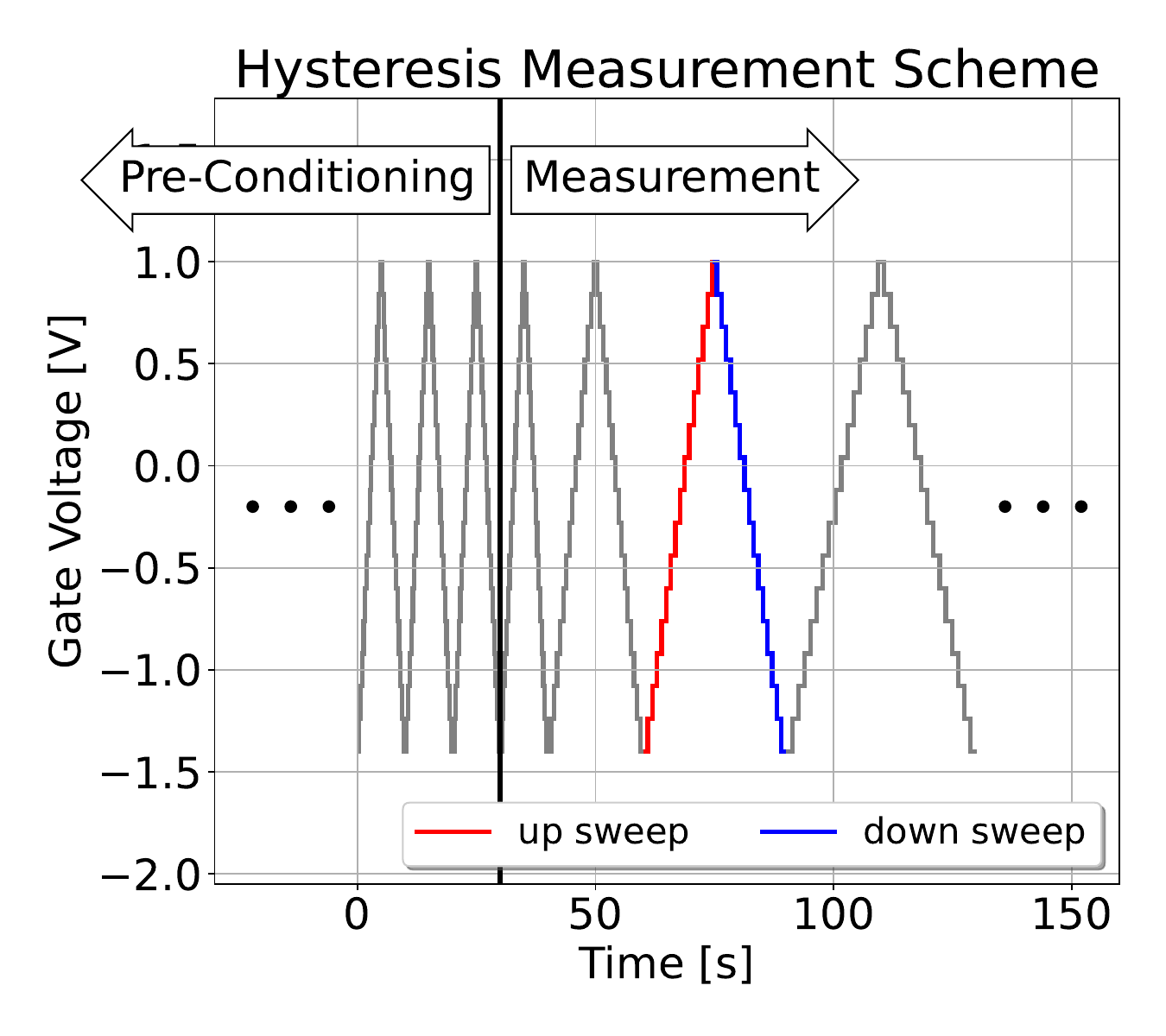}
    \end{minipage}};
    \node[draw=none] at (-2.0,2.4) {\textbf{(d)}};
    \end{tikzpicture}
\end{subfigure}
\begin{subfigure}[b]{.333\linewidth}
    \begin{tikzpicture}
    \node[inner sep=0pt] at (0,0) {
      \begin{minipage}{1.0\textwidth}
        \includegraphics[width=\textwidth]{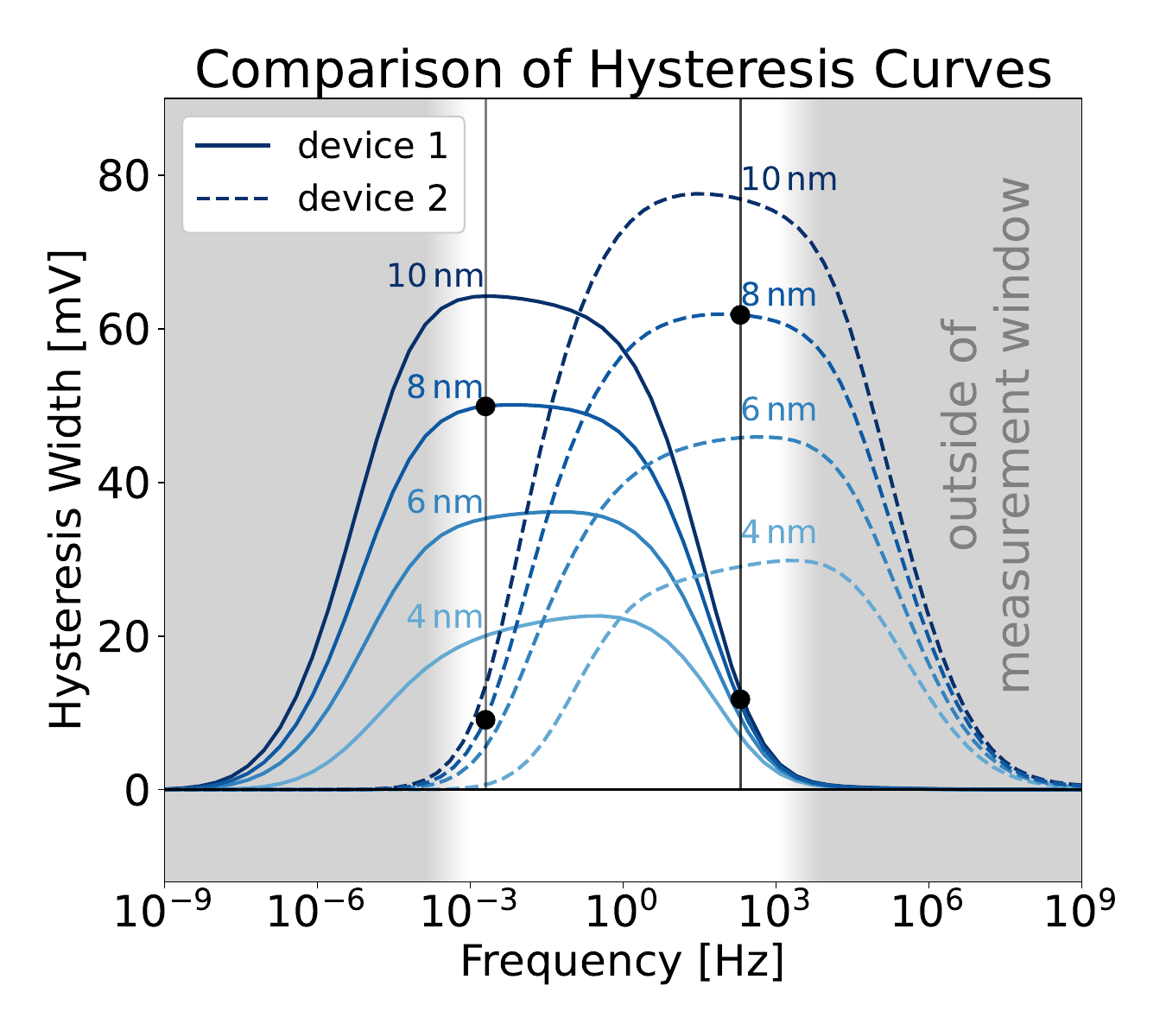}  \end{minipage}};
    \node[draw=none] at (-2.0,2.4) {\textbf{(e)}};
    \end{tikzpicture}
\end{subfigure}
\end{center}
\caption{\textbf{(a)} Single gate sweep defined the by minimum voltage \( V_\mathrm{min} \), maximum voltage \( V_\mathrm{max} \), and frequency \( f = 1/t_\mathrm{sweep} \). \textbf{(b)} Single \IDVG curve recorded on a device with a \ce{Bi2O2Se}/\ce{Bi2SeO5}/\ce{Au} gate stack, showing a clearly visible CW hysteresis of \(\Delta V_\mathrm{H} = \qty{54.4}{\milli\volt}\), where circles represent measurement data whereas lines are only a guide to the eye. Note that the measurement was performed in a fixed current range to ensure precise timing, which however leads to a limited current resolution and thus a supposedly lower on-off ratio. \textbf{(c)} Transient time evolution of the hysteresis measured over 100 sweep cycles on the same device as in (b). \textbf{(d)} Hysteresis measurement scheme, which includes a preconditioning phase, allowing the device to stabilize, followed by a measurement phase where the frequency is gradually reduced between successive sweeps. \textbf{(e)} Simulated hysteresis curves for two \ce{MoS2}/\ce{HfO2}/\ce{Au} devices with various insulator thicknesses assuming different Gaussian defect distributions due to a different fabrication process (device 1: \(E_\mathrm{T} = (-4.4 \pm 0.2) \qty{}{\electronvolt}\), \(E_\mathrm{R} = (2.5 \pm 0.2) \qty{}{\electronvolt}\), \(R = 1\); device 2: \(E_\mathrm{T} = (-4.6 \pm 0.2) \qty{}{\electronvolt}\), \(E_\mathrm{R} = (1.5 \pm 0.2) \qty{}{\electronvolt}\), \(R = 1\)), for modeling details see Sec.~\ref{sec3}).
\label{fig:sec1:sec1:1}}
\end{figure}

\fig{fig:sec1:sec1:1}{a} and \fig{fig:sec1:sec1:1}{b} present a typical hysteresis measurement performed on a 2D-MOSFET with a \ce{Bi2O2Se}/\ce{Bi2SeO5}/\ce{Au} gate stack, as reported in~\cite{Zhang2022}. During the measurement, a triangular staircase signal is applied to the gate, defined by its minimum voltage \( V_\mathrm{min} \), maximum voltage \( V_\mathrm{max} \), and frequency \( f = 1/t_\mathrm{sweep} \). Moreover, a small drain voltage (\( q V_\mathrm{d} \lesssim k_\mathrm{B} T \)) is applied during the sweep to generate a drain current while keeping the electric field along the channel direction small. While not strictly necessary, this restriction enables us to approximate quantities along the channel—such as the quasi Fermi level \(E_\mathrm{F}^\mathrm{ch}\)—as position-independent, thereby simplifying the analysis. The corresponding drain current is recorded throughout the sweep and then plotted against the gate voltage, as shown in \fig{fig:sec1:sec1:1}{b}. Due to non-ideal effects, the curves for the up- and down-sweep differ, which can be quantified by the hysteresis width \( \Delta V_\mathrm{H} \coloneqq V_\mathrm{down}  -  V_\mathrm{up} \), i.e. the shift of the threshold voltage evaluated at a suitable current criterion. According to this definition, curves that are traversed clockwise (CW) lead to a positive sign and those that are traversed counterclockwise (CCW) lead to a negative sign of \( \Delta V _\mathrm{H}\). \\

Hysteresis in the transfer characteristics is a complex, time-dependent phenomenon, often requiring multiple sweep cycles for the device to reach a stable cyclo-stationary state at a given frequency. \fig{fig:sec1:sec1:1}{c} illustrates the evolution of hysteresis over 100 cycles at \qty{0.01}{Hz}. Experimentally, the cyclo-stationary hysteresis value is of interest as it is independent of the device's prior history, reflecting the stability at the given frequency. Proper device characterization requires measuring the cyclo-stationary hysteresis across a broad frequency range. In practice, the accessible frequency range is constrained at the upper end by the limitations of the experimental setup and at the lower end by the time one is willing to dedicate to the measurement. A realistic measurement window that can be covered with commercially available source measure units is given by $f\in [\qty{1}{\milli\hertz}\dots\qty{1}{\kilo\hertz}]$. A suitable measurement procedure is sketched in \fig{fig:sec1:sec1:1}{d}, starting with a preconditioning phase allowing the device to stabilize at the maximum frequency \(f_\mathrm{max} \). During the subsequent measurement phase, the frequency is gradually decreased between successive sweeps until the minimum frequency \(f_\mathrm{min} \) is reached. The frequency change between successive sweeps is chosen to be as small as possible to ensure that the measured hysteresis is a good approximation to the cyclo-stationary hysteresis. \\

The cyclo-stationary hysteresis \(\Delta V_\mathrm{H} \) is best presented as a function of the sweep frequency $f$, as shown in \fig{fig:sec1:sec1:1}{e}, which displays simulated hysteresis curves for hypothetical devices with a similar \ce{MoS2}/\ce{HfO2}/\ce{Au} gate stack but different defect distributions in the insulator. This figure highlights several challenges in defining a metric for device stability. The first issue becomes apparent when we compare the hysteresis values of the devices at the highlighted frequencies for an  insulator thickness of \(d_\mathrm{ins} = \qty{8}{\nano \meter}\). Although device~2 exhibits a significantly smaller hysteresis than device~1 at \qty{2}{\milli\hertz}, the trend reverses at \qty{200}{\hertz}. This demonstrates that single-frequency measurements are inadequate for assessing device stability. Therefore, \textbf{the common practice of reporting hysteresis at a single frequency, with adjectives such as ``negligible'', is misleading}. To overcome this problem, one might be tempted to specify the maximum hysteresis \( \Delta V_\mathrm{H}^\mathrm{max} \) within the measurement window to quantify the worst case. However, \fig{fig:sec1:sec1:1}{e} shows that the magnitude of the hysteresis peaks increases with the thickness of the insulator. Consequently, the naive use of this metric would lead to devices with thinner dielectrics being systematically classified as more stable than those with thicker dielectrics. This highlights the necessity of normalizing \( \Delta V_\mathrm{H}^\mathrm{max} \) to enable meaningful comparisons of devices with varying insulator thicknesses. Therefore, after exploring the underlying physical mechanisms of hysteresis in Sec.~\ref{sec3}, we suggest to normalize \( \Delta V_\mathrm{H}^\mathrm{max} \) by EOT in Sec.~\ref{sec4}.

\section{Physical Mechanisms Contributing to Hysteresis }\label{sec3}

Here, the physical mechanisms of hysteresis are explained using the example of an n-type 2D-MOSFET (Note that for p-type devices, the hysteresis sign / direction is reversed for all mechanisms described in this work due to the opposite polarity of the applied voltages). We assume that the 2D-material is encapsulated, minimizing the influence of environmental factors such as adsorbates on the layer. To accurately describe the device behavior, we extend the model proposed by Marin~et~al.~\cite{8288825} to account for contact resistances and residual charges in the gate insulator (see SI Sec.~\ref{secA1}). In the following, \(d_\mathrm{ins}\) denotes the thickness of the gate insulator and \(C_\mathrm{ins} = \varepsilon_\mathrm{ins} / d_\mathrm{ins} \) its capacitance per area. In the proposed model, the effective threshold voltage is given by:
\begin{equation}\label{sec3:eq2}
\begin{aligned}
   V_\mathrm{th}' = V_\mathrm{th} - q(N_\mathrm{d}^+ - N_\mathrm{a}^-) / C_\mathrm{ins} + \Delta V_\mathrm{ins},
\end{aligned}
\end{equation}
and consists of three contributions: the threshold voltage \(V_\mathrm{th} \) of the ideal device, the impact of ionized channel defects \( q(N_\mathrm{d}^+ - N_\mathrm{a}^-) / C_\mathrm{ins} \) and the voltage shift \(  \Delta V_\mathrm{ins} \) caused by the residual charges in the gate insulator. This voltage shift \(  \Delta V_\mathrm{ins} \) is determined by the total charge \(Q_\mathrm{ins}\) and the charge centroid \(x_\mathrm{ins}\) of the residual charges:
\begin{equation}\label{sec3:eq3}
     \Delta V_\mathrm{ins} = - \frac{Q_\mathrm{ins}}{C_\mathrm{ins}}  \left(1 - \frac{x_\mathrm{ins}}{d_\mathrm{ins}}\right), \qquad 
      Q_\mathrm{ins} =  \int_{0}^{d_\mathrm{ins}} \rho_\mathrm{ins}(x)  \, \mathrm{d}x, \qquad      x_\mathrm{ins} = \frac{1}{Q_\mathrm{ins}} \int_{0}^{d_\mathrm{ins}} \rho_\mathrm{ins}(x) x  \,\mathrm{d}x. 
\end{equation}
Hysteresis in the transfer characteristics arises from a change in the effective threshold voltage between successive up- and down-sweeps (i.e., \(V_\mathrm{H} =  V_\mathrm{th}'(t_\mathrm{down})  -  V_\mathrm{th}'(t_\mathrm{up}) \)), caused by variations in the capacitances and charges of the gate stack \cite{Egginger2009}. In this study, we specifically focus on the properties of the gate insulator and study three time-dependent
processes whose contributions to reliability problems have been repeatedly suggested in the literature: charge trapping by defects in the insulator \cite{Ravichandran2023}, the drift of mobile charges within the insulator \cite{Shirakawa_2016}, and capacitance variations due to ferroelectricity \cite{Lin2016}. As emphasized in SI~Sec.~\ref{secA2} defects within the 2D-material or at the interface between the 2D-material and the insulator are typically too fast to contribute to the hysteresis within the measurement window and are thus omitted in this work.

\subsection{Hysteresis due to Charge Trapping}\label{sec3:sec1}
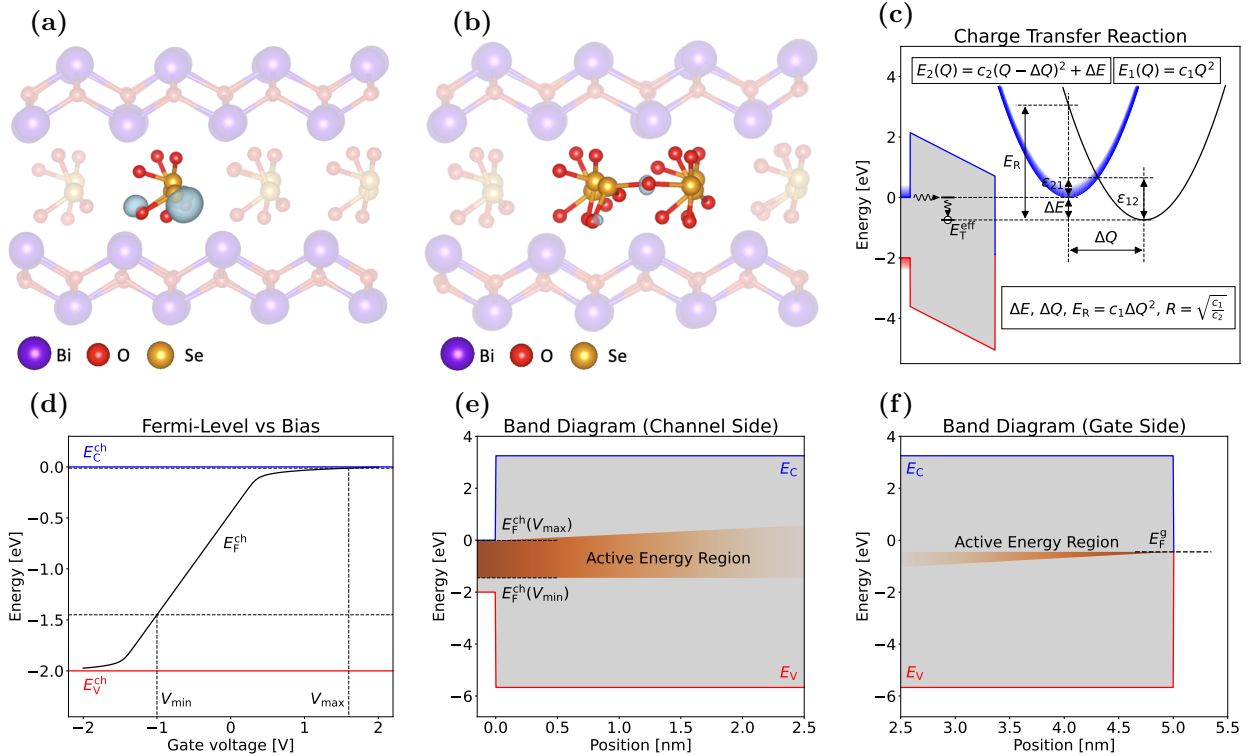
\begin{figure}[hbt!]
\begin{subfigure}[b]{.333\linewidth}
    \begin{tikzpicture}
    \node[inner sep=0pt] at (0,0) {
      \begin{minipage}{1.0\textwidth}
        \includegraphics[width=\textwidth]{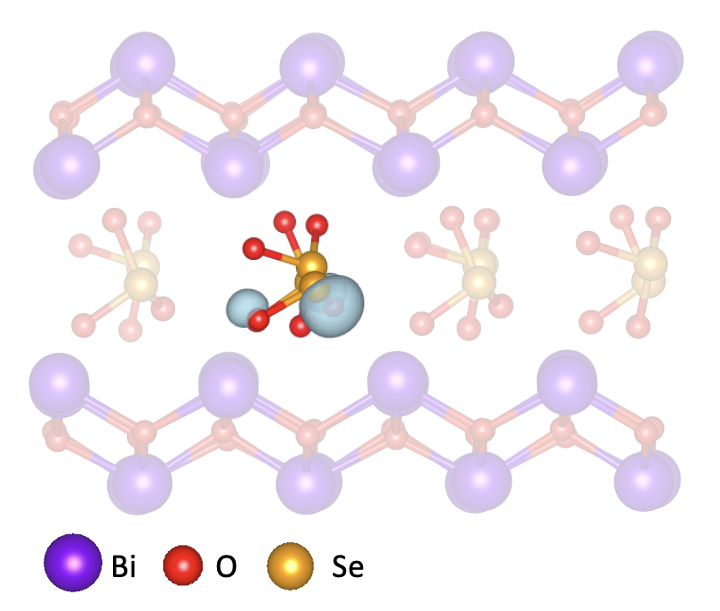} 
    \end{minipage}};
    \node[draw=none] at (-2.0,2.4) {\textbf{(a)}};
    \end{tikzpicture}
\end{subfigure}
\begin{subfigure}[b]{.333\linewidth}
    \begin{tikzpicture}
    \node[inner sep=0pt] at (0,0) {
      \begin{minipage}{1.0\textwidth}
        \includegraphics[width=\textwidth]{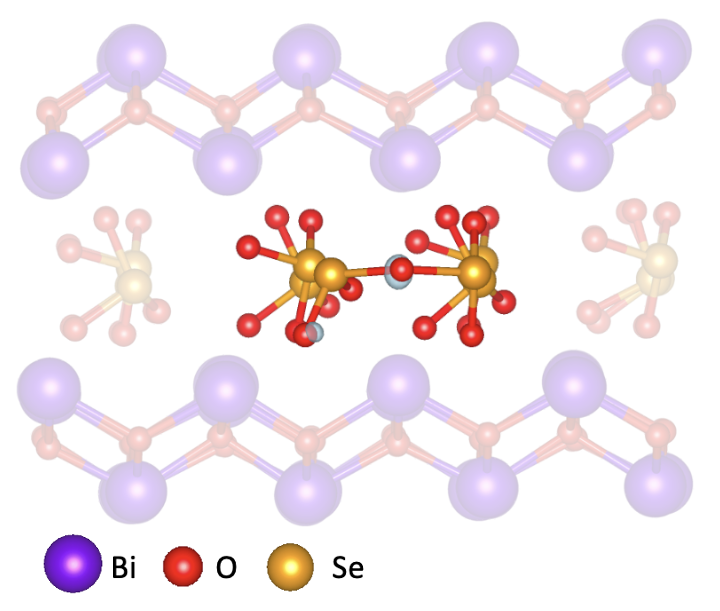} 
    \end{minipage}};
    \node[draw=none] at (-2.0,2.4) {\textbf{(b)}};
    \end{tikzpicture}
\end{subfigure}
\begin{subfigure}[b]{.333\linewidth}
    \begin{tikzpicture}
    \node[inner sep=0pt] at (0,0) {
      \begin{minipage}{1.0\textwidth}
        \includegraphics[width=\textwidth]{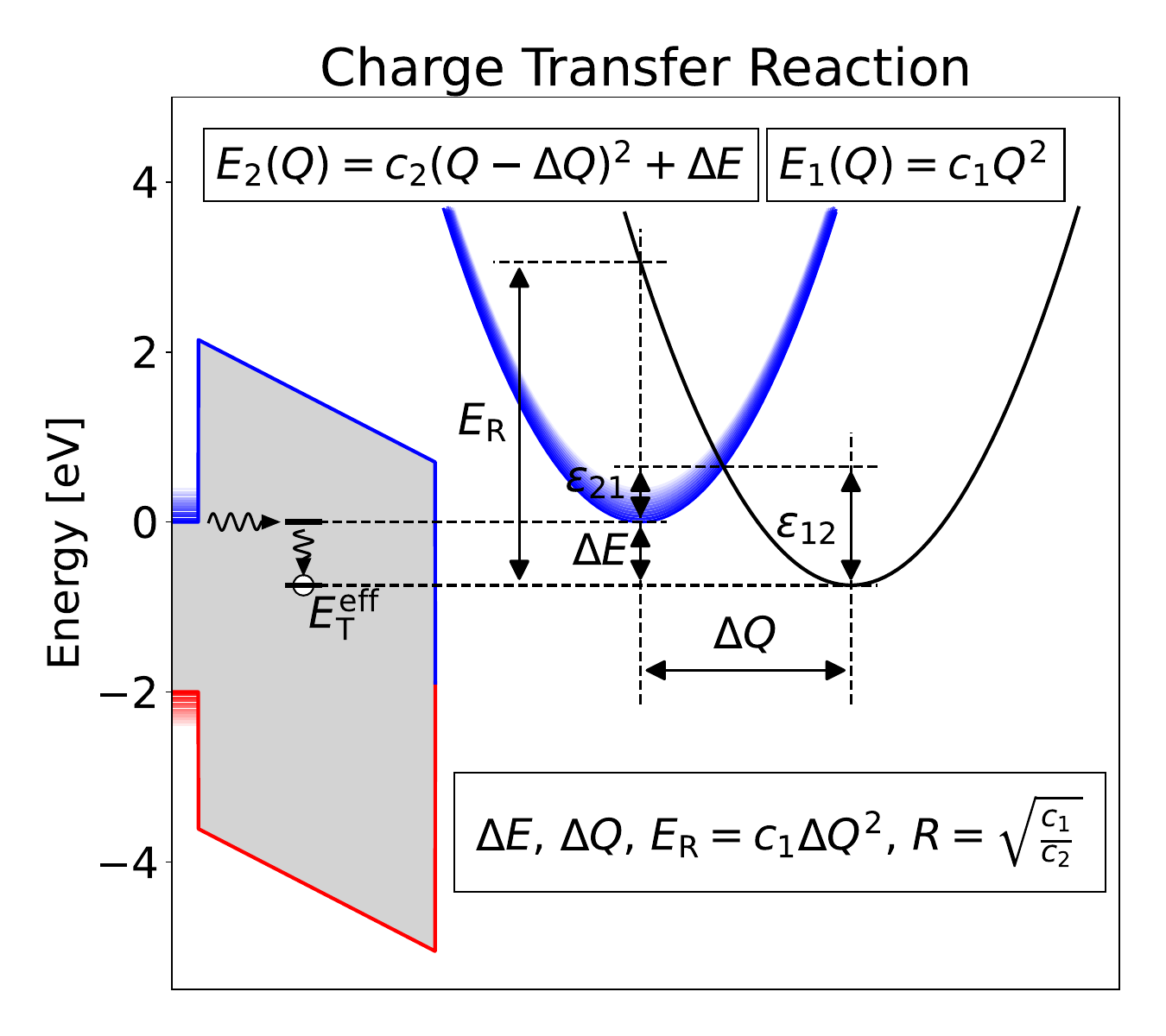} 
    \end{minipage}};
    \node[draw=none] at (-2.0,2.4) {\textbf{(c)}};
    \end{tikzpicture}
\end{subfigure}
\begin{subfigure}[b]{.333\linewidth}
    \begin{tikzpicture}
    \node[inner sep=0pt] at (0,0) {
      \begin{minipage}{1.0\textwidth}
        \includegraphics[width=\textwidth]{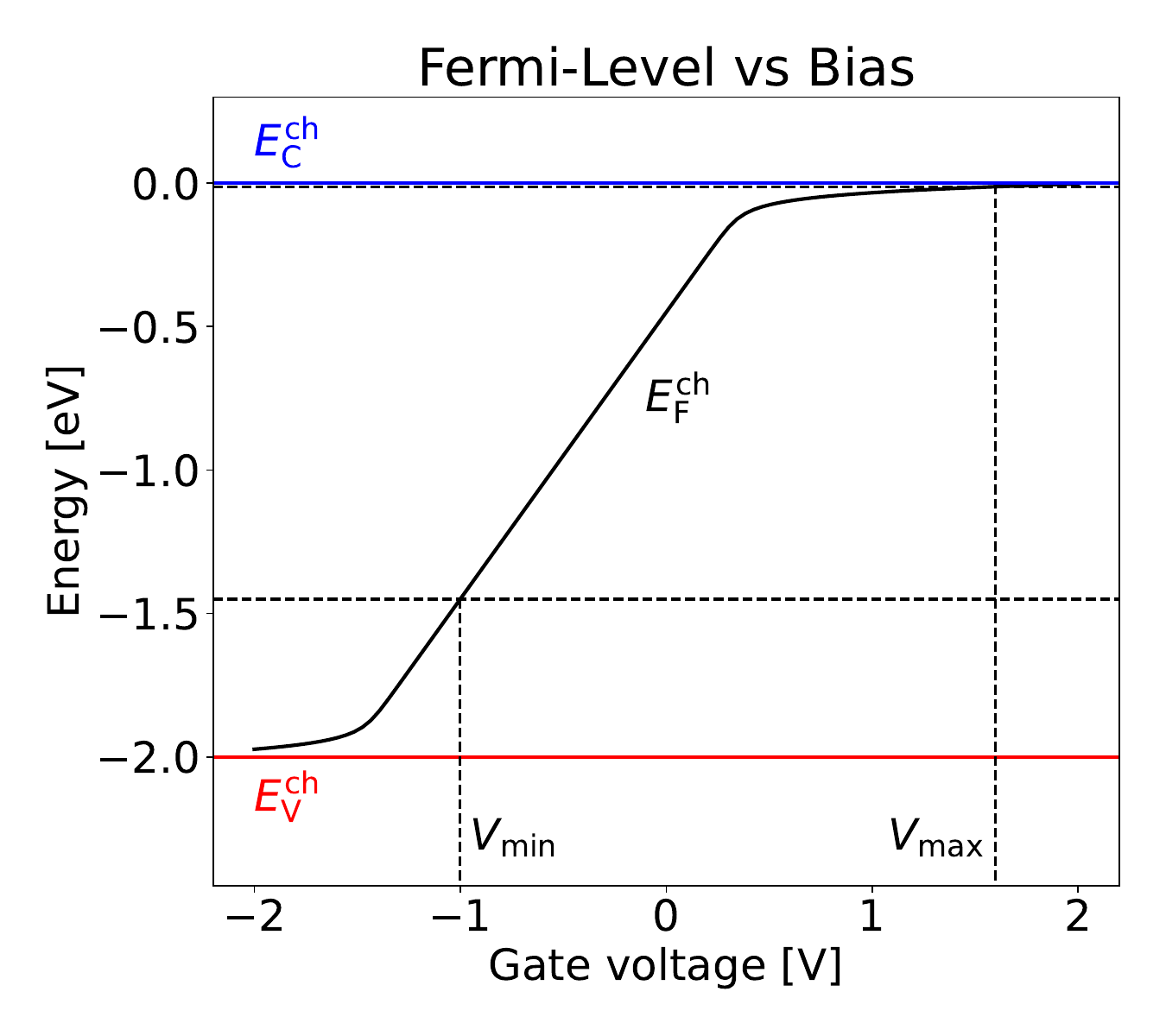} 
    \end{minipage}};
    \node[draw=none] at (-2.0,2.4) {\textbf{(d)}};
    \end{tikzpicture}
\end{subfigure}
\begin{subfigure}[b]{.333\linewidth}
    \begin{tikzpicture}
    \node[inner sep=0pt] at (0,0) {
      \begin{minipage}{1.0\textwidth}
        \includegraphics[width=\textwidth]{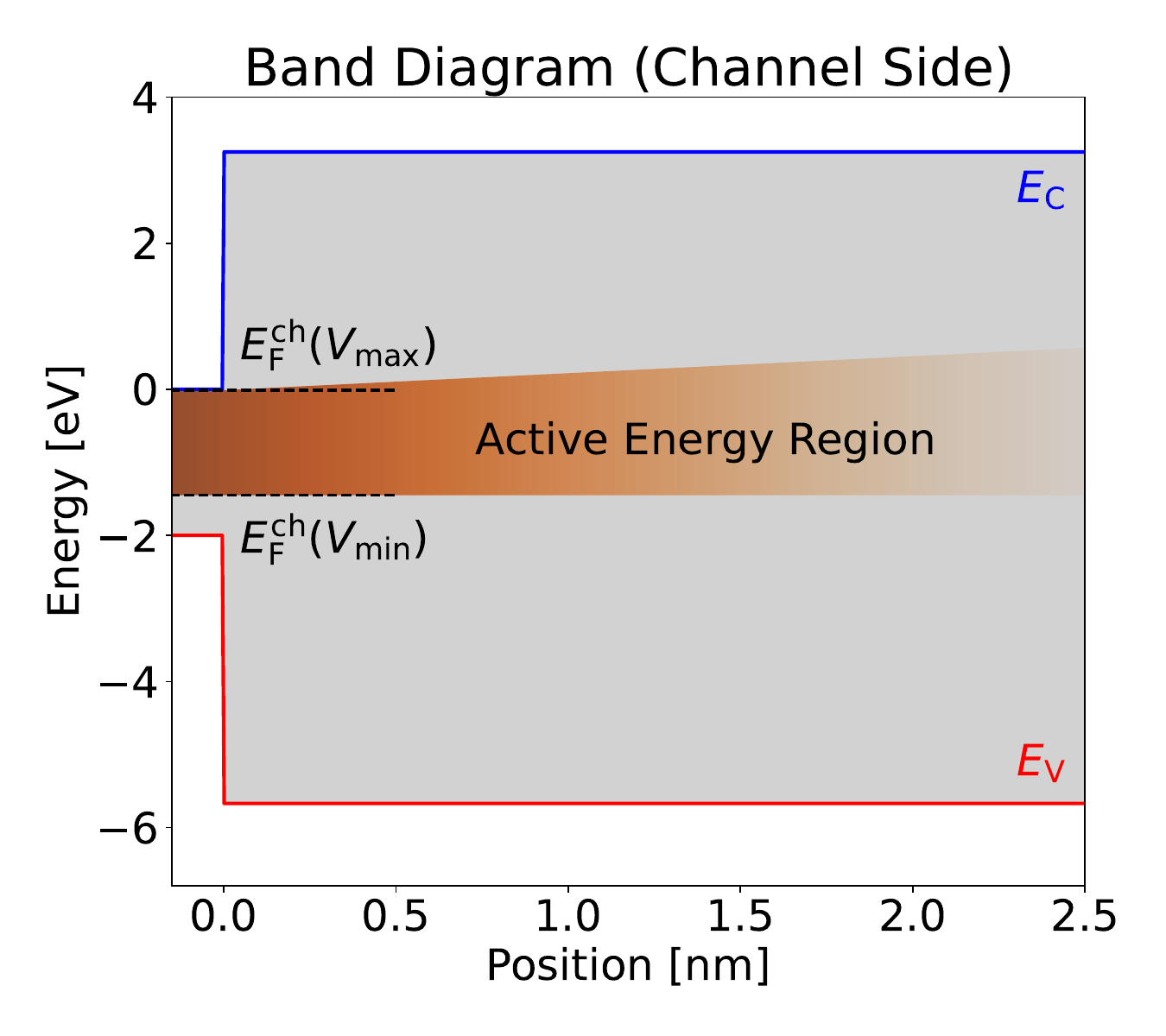}   \end{minipage}};
    \node[draw=none] at (-2.0,2.4) {\textbf{(e)}};
    \end{tikzpicture}
\end{subfigure}
\begin{subfigure}[b]{.333\linewidth}
    \begin{tikzpicture}
    \node[inner sep=0pt] at (0,0) {
      \begin{minipage}{1.0\textwidth}
        \includegraphics[width=\textwidth]{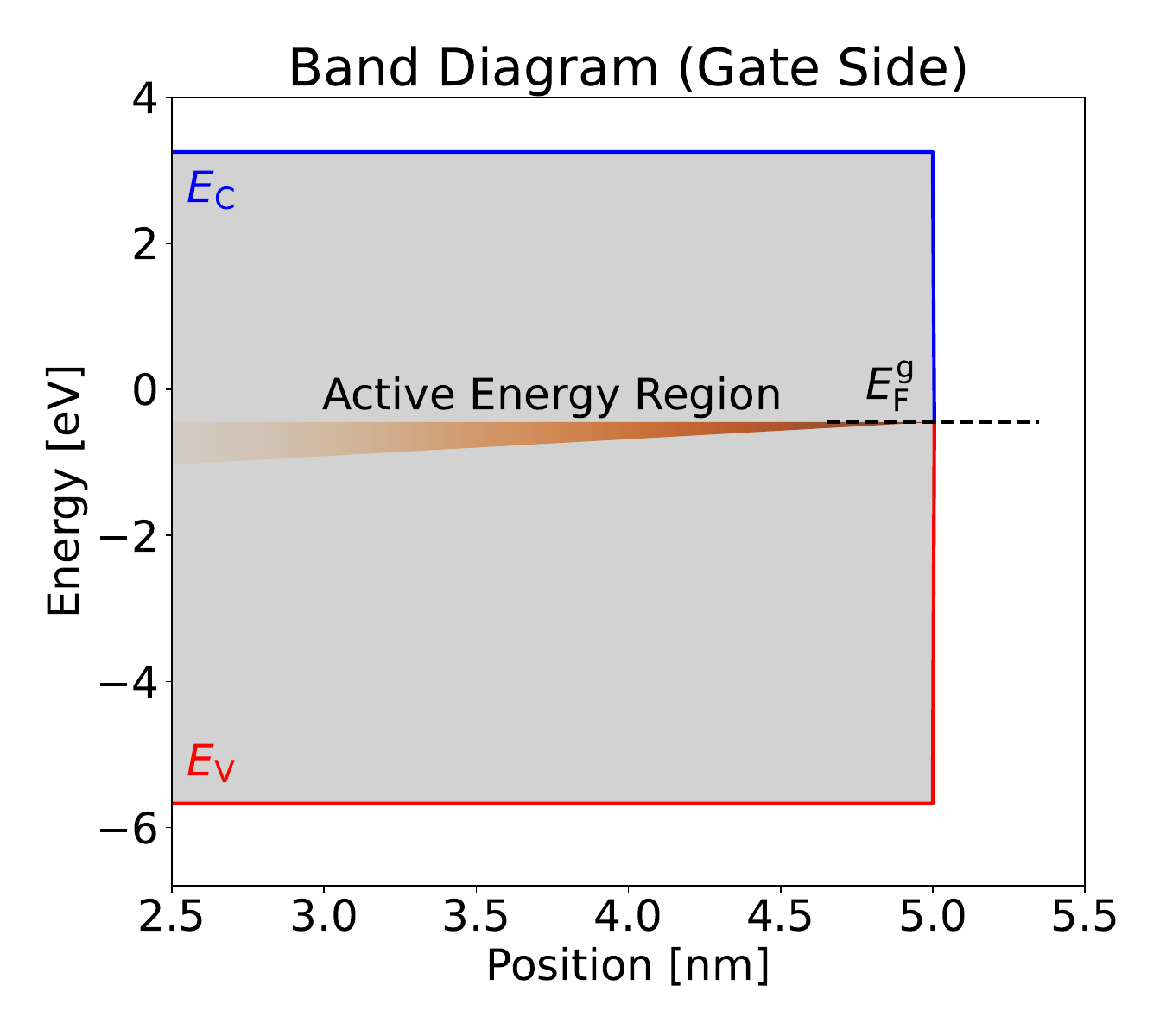}
    \end{minipage}};
    \node[draw=none] at (-2.0,2.4) {\textbf{(f)}};
    \end{tikzpicture}
\end{subfigure}
\caption{\textbf{(a)} Single and \textbf{(b)} double  positively charged oxygen vacancy in \ce{Bi2SeO5}. \textbf{(c)} Configuration coordinate diagram visualizing electron capture by a defect in the gate insulator. In the NMP framework, the system's energy in the two charge states is typically approximated by two harmonic potentials, parameterized by the energy difference \( \Delta E \), the configuration coordinate shift \( \Delta Q \), the relaxation energy \( E_\mathrm{R} \), and the curvature ratio \( R \) (see SI~Sec.~\ref{secA2}). \textbf{(d)} Fermi-level in the channel as a function of the applied gate bias. \textbf{(e)} Active energy region of the band diagram covered by the channel's Fermi level during the switching process. \textbf{(f)} Active energy region of the band diagram covered by the gate's Fermi level during the switching process.}
\label{fig:sec3:sec1:1}
\end{figure}

Charge trapping results in a time-dependent charge density \(\rho_\mathrm{ins}(x, t)\) within the insulator, which produces an observable hysteresis (Eq.~\ref{sec3:eq2} and Eq.~\ref{sec3:eq3}). Notably, only charge differences contribute to hysteresis, while the effect of time-independent charge distributions (fixed charges) cancels out. The capture of an electron by an insulator defect conceptually occurs in two steps: First, the electron tunnels to the defect, changing its charge state and disturbing the equilibrium of the original atomic configuration. Consequently, in the second step, the atoms relax towards their new equilibrium positions. \fig{fig:sec3:sec1:1}{a} and \fig{fig:sec3:sec1:1}{b} illustrate an example of this process with an oxygen vacancy in \ce{Bi2SeO5} deforming due to the capture of an electron. \\

The configuration coordinate diagram shown in \fig{fig:sec3:sec1:1}{c} illustrates the capture of an electron from the conduction band by an insulator defect. The dynamics of this process are governed by the capture rate \(k_{21}^\mathrm{CB}\) and the emission rate \(k_{12}^\mathrm{CB}\), both of which can be calculated by non-radiative multiphonon (NMP) theory \cite{PhysRevB.90.075202, TURIANSKY2021108056, GOES2018286, GRASSER201239}, as summarized in SI~Sec.~\ref{secA2}. In general, every defect interacts not only with the conduction band but also with the valence band and gate. Thus the total rates for capture and emission are given by \(k_{21} = k_{21}^\mathrm{CB} + k_{21}^\mathrm{VB} + k_{21}^\mathrm{G} \) and \(k_{12} = k_{12}^\mathrm{CB} + k_{12}^\mathrm{VB} + k_{12}^\mathrm{G} \) respectively.  The mean time that elapses until the charge is captured or emitted is given by \(\tau_\mathrm{c} = 1 / k_{21}\) or \(\tau_\mathrm{e} = 1 / k_{12}\) respectively. Each rate is proportional to a tunneling probability, which decreases exponentially with distance from the corresponding charge reservoir. Consequently, the rates \( k_{12}^\mathrm{CB} \), \( k_{21}^\mathrm{CB} \), \( k_{12}^\mathrm{VB} \), \( k_{21}^\mathrm{VB} \) decrease exponentially with the distance \( x_\mathrm{T} \) from the channel, while the rates \( k_{12}^\mathrm{G}\), \( k_{21}^\mathrm{G}\) decrease exponentially with the distance \(  d_\mathrm{ins} - x_\mathrm{T} \) from the gate. Therefore, in a sufficiently thick insulator, the gate's influence on channel-sided defects and the channel's influence on gate-sided defects can be neglected. \\

\subsubsection{Equilibrium Occupation \& Active Energy Region (AER)}

For defects near the channel in a sufficiently thick insulator the equilibrium occupation \(f_\mathrm{1}(t \to \infty)\) is simply given by Fermi-Dirac statistics:

\begin{equation}\label{eq:sec3:sec1:9}
\begin{aligned}
f_\mathrm{1}(t \to \infty) \Big|_\mathrm{Channel} = \frac{1}{1 + \exp{ \left( \frac{E_\mathrm{T}^\mathrm{eff} - E_\mathrm{F}^\mathrm{ch}}{k_\mathrm{B} T} \right) }},
\end{aligned}
\end{equation}
where the channel's Fermi-level \( E_\mathrm{F}^\mathrm{ch}\) plays the role of the chemical potential. As a consequence, a change in the defect's charge state is initiated when the effective defect level \(E_\mathrm{T}^\mathrm{eff}\) crosses the Fermi-level \(E_\mathrm{F}^\mathrm{ch}\). \fig{fig:sec3:sec1:1}{d}  displays the channel's Fermi level as a function of the applied gate voltage. When the transistor is turned on by increasing the voltage from \(V_\mathrm{min} \) to \(V_\mathrm{max} \), the Fermi level is shifted throughout the bandgap of the semiconductor and covers a certain region of the band diagram, referred to as the channel-sided active energy region (AER), highlighted in orange in \fig{fig:sec3:sec1:1}{e}. The channel-sided AER marks the region near the channel in which defects can change their charge state during the switching process  and thus contribute to hysteresis.\\

Conversely, for defects near the gate in a sufficiently thick insulator the equilibrium occupation \(f_\mathrm{1}(t \to \infty)\) also corresponds to the Fermi-Dirac distribution,
where the gate's Fermi level \( E_\mathrm{F}^\mathrm{g}\) acts as the relevant chemical potential. Similarly, when the transistor is turned on, the Fermi level \( E_\mathrm{F}^\mathrm{g}\) covers a specific region of the band diagram referred to as the gate-sided active energy region (AER), highlighted in orange in \fig{fig:sec3:sec1:1}{f}. However, since the Fermi level \(E_\mathrm{F}^\mathrm{g} \) is pinned relative to the insulator (assuming a metal gate with no depletion layer), the gate-sided AER is much smaller than the channel-sided AER. Nevertheless, the gate-sided AER has a similar relevance for the hysteresis, as it marks the region near the gate in which defects can change their charge state  during the switching process and thus contribute to hysteresis.  \\

\subsubsection{Gate versus Channel Interaction}

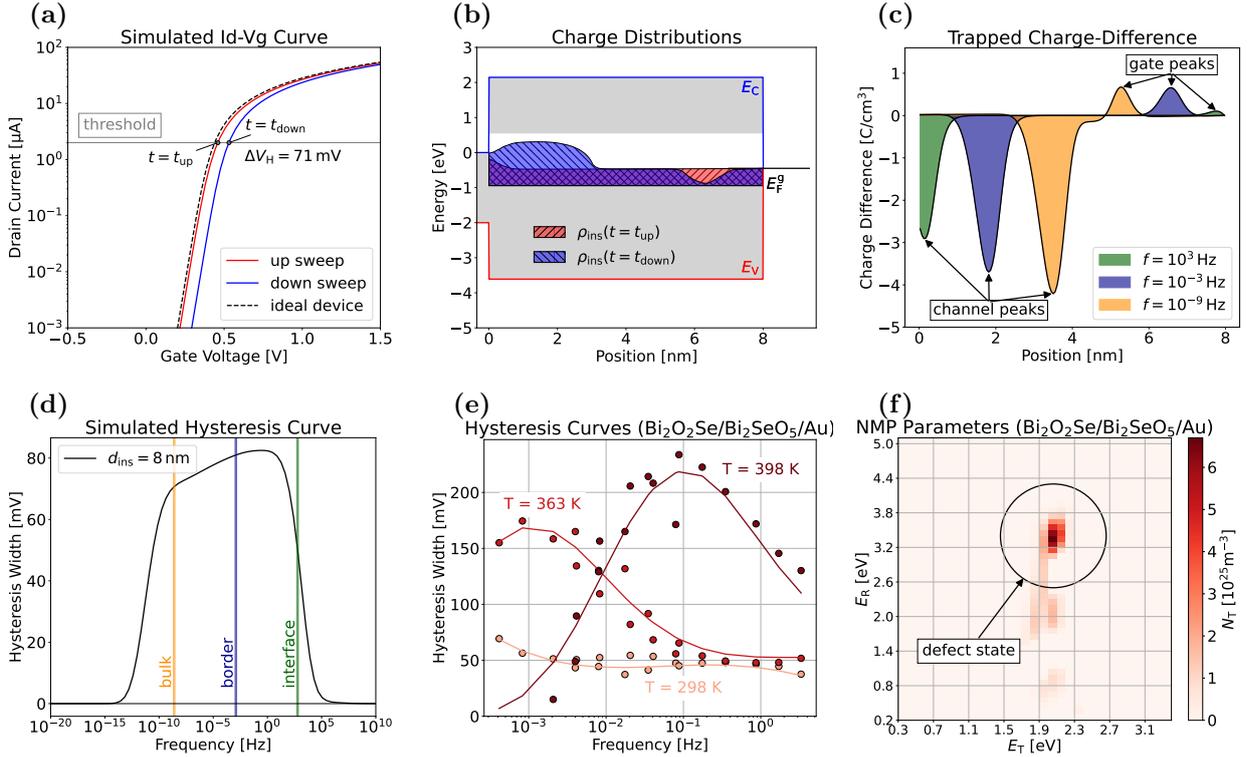
\begin{figure}[hbt!]
\begin{subfigure}[b]{.333\linewidth}
    \begin{tikzpicture}
    \node[inner sep=0pt] at (0,0) {
      \begin{minipage}{1.0\textwidth}
        \includegraphics[width=\textwidth]{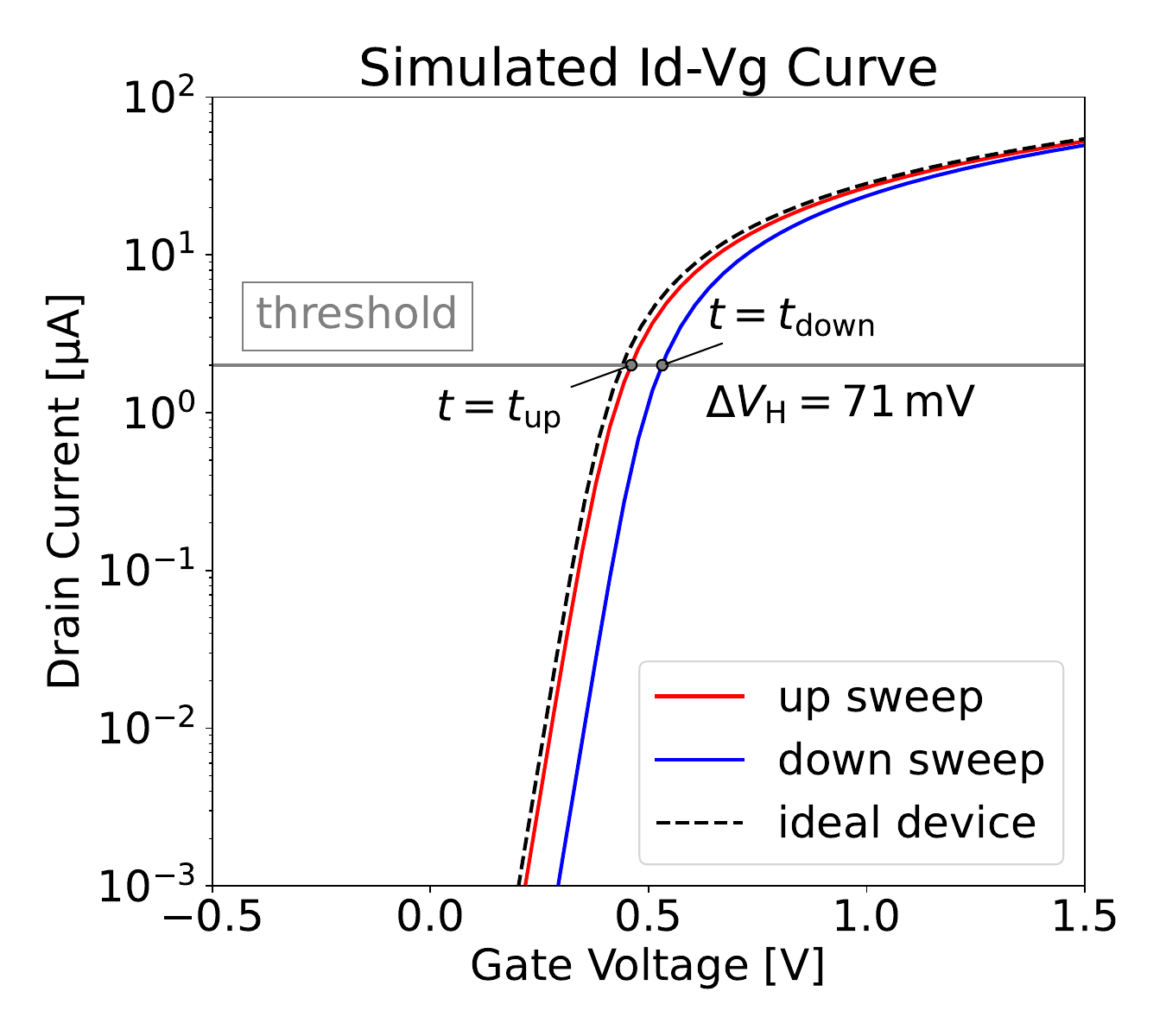}
    \end{minipage}};
    \node[draw=none] at (-2.0,2.4) {\textbf{(a)}};
    \end{tikzpicture}
\end{subfigure}
\begin{subfigure}[b]{.333\linewidth}
    \begin{tikzpicture}
    \node[inner sep=0pt] at (0,0) {
      \begin{minipage}{1.0\textwidth}
        \includegraphics[width=\textwidth]{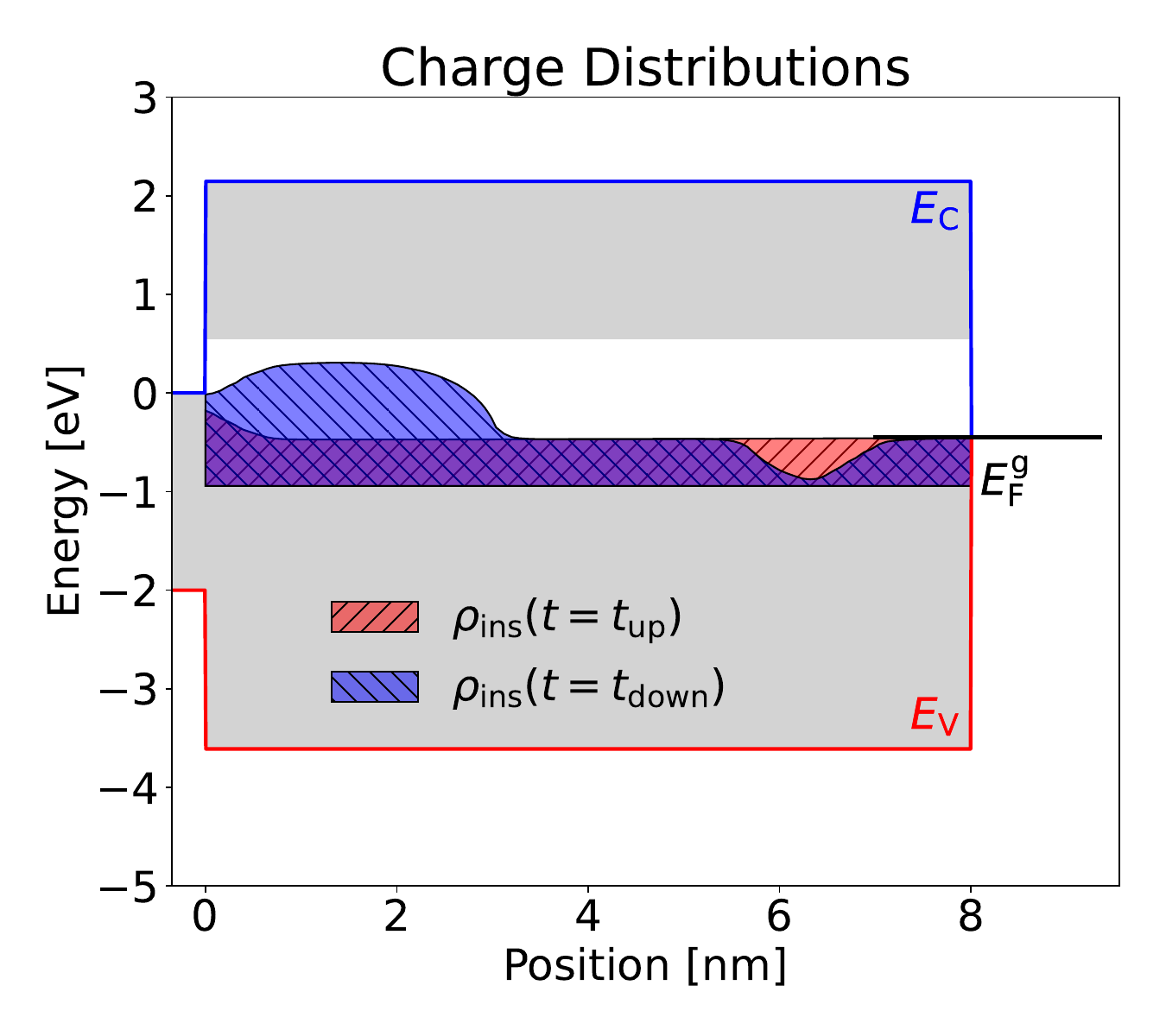}
    \end{minipage}};
    \node[draw=none] at (-2.0,2.4) {\textbf{(b)}};
    \end{tikzpicture}
\end{subfigure}
\begin{subfigure}[b]{.333\linewidth}
    \begin{tikzpicture}
    \node[inner sep=0pt] at (0,0) {
      \begin{minipage}{1.0\textwidth}
        \includegraphics[width=\textwidth]{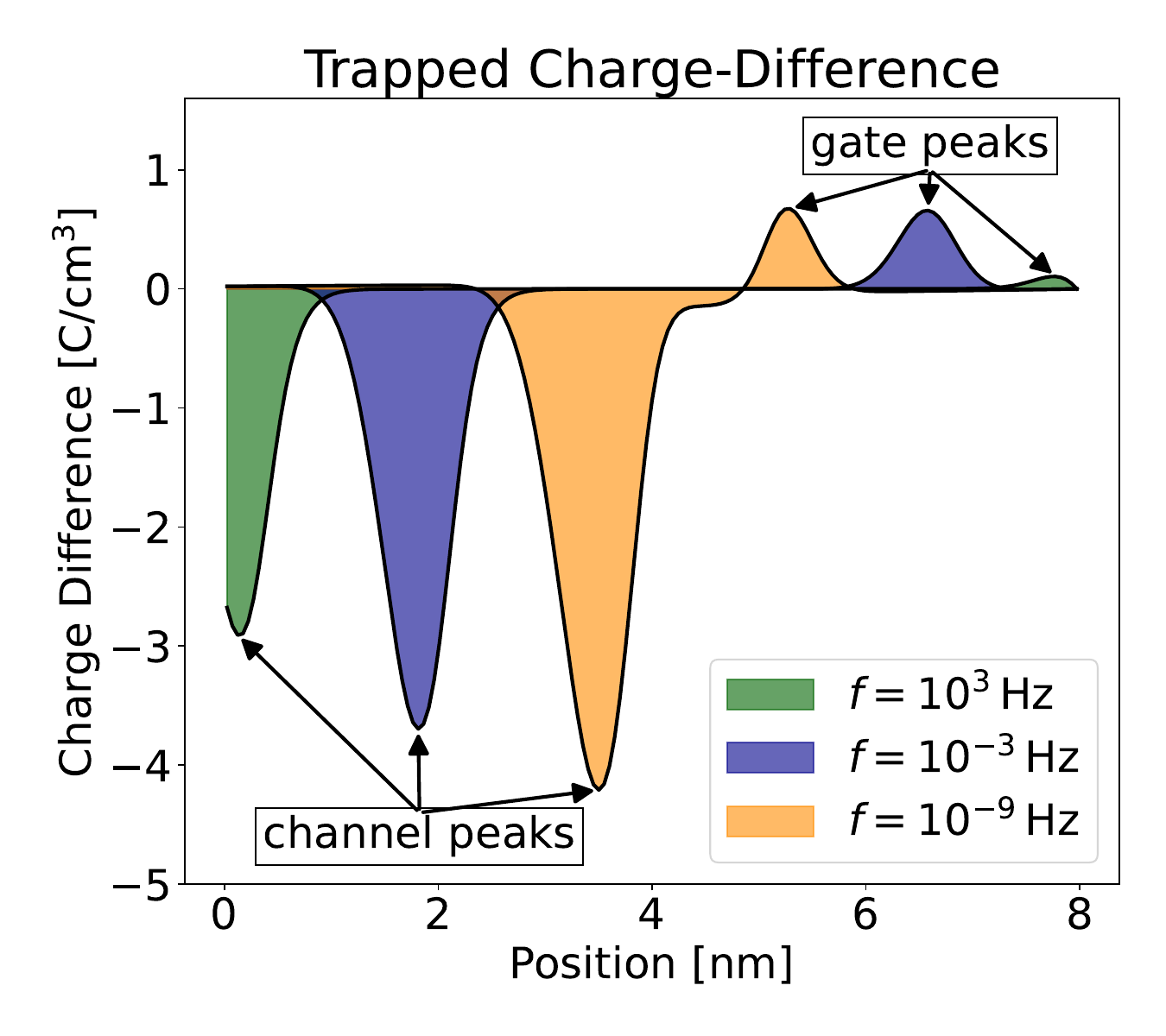}
    \end{minipage}};
    \node[draw=none] at (-2.0,2.4) {\textbf{(c)}};
    \end{tikzpicture}
\end{subfigure}
\begin{subfigure}[b]{.333\linewidth}
    \begin{tikzpicture}
    \node[inner sep=0pt] at (0,0) {
      \begin{minipage}{1.0\textwidth}
        \includegraphics[width=\textwidth]{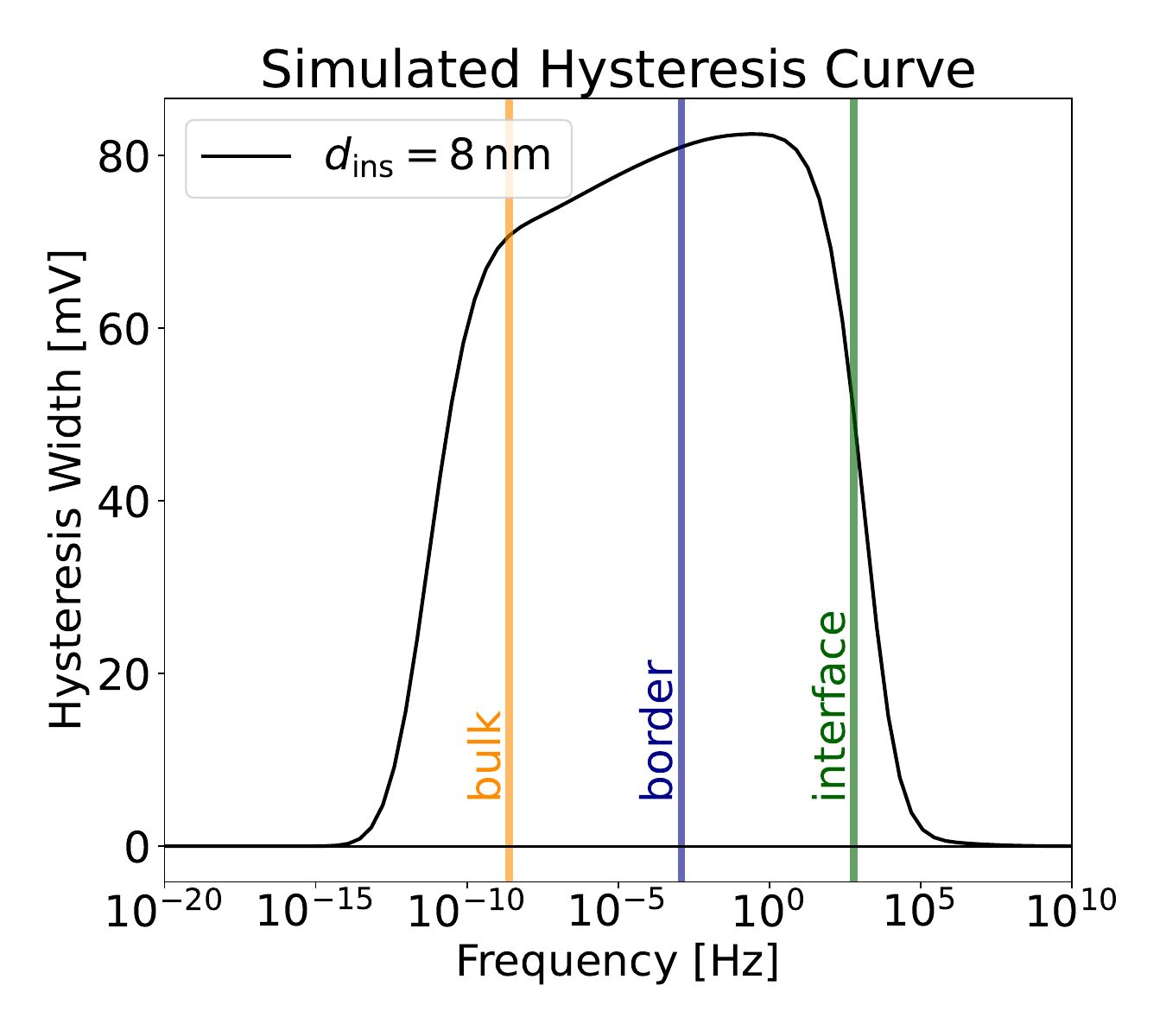}
    \end{minipage}};
    \node[draw=none] at (-2.0,2.4) {\textbf{(d)}};
    \end{tikzpicture}
\end{subfigure}
\begin{subfigure}[b]{.333\linewidth}
    \begin{tikzpicture}
    \node[inner sep=0pt] at (0,0) {
      \begin{minipage}{1.0\textwidth}
        \includegraphics[width=\textwidth]
        {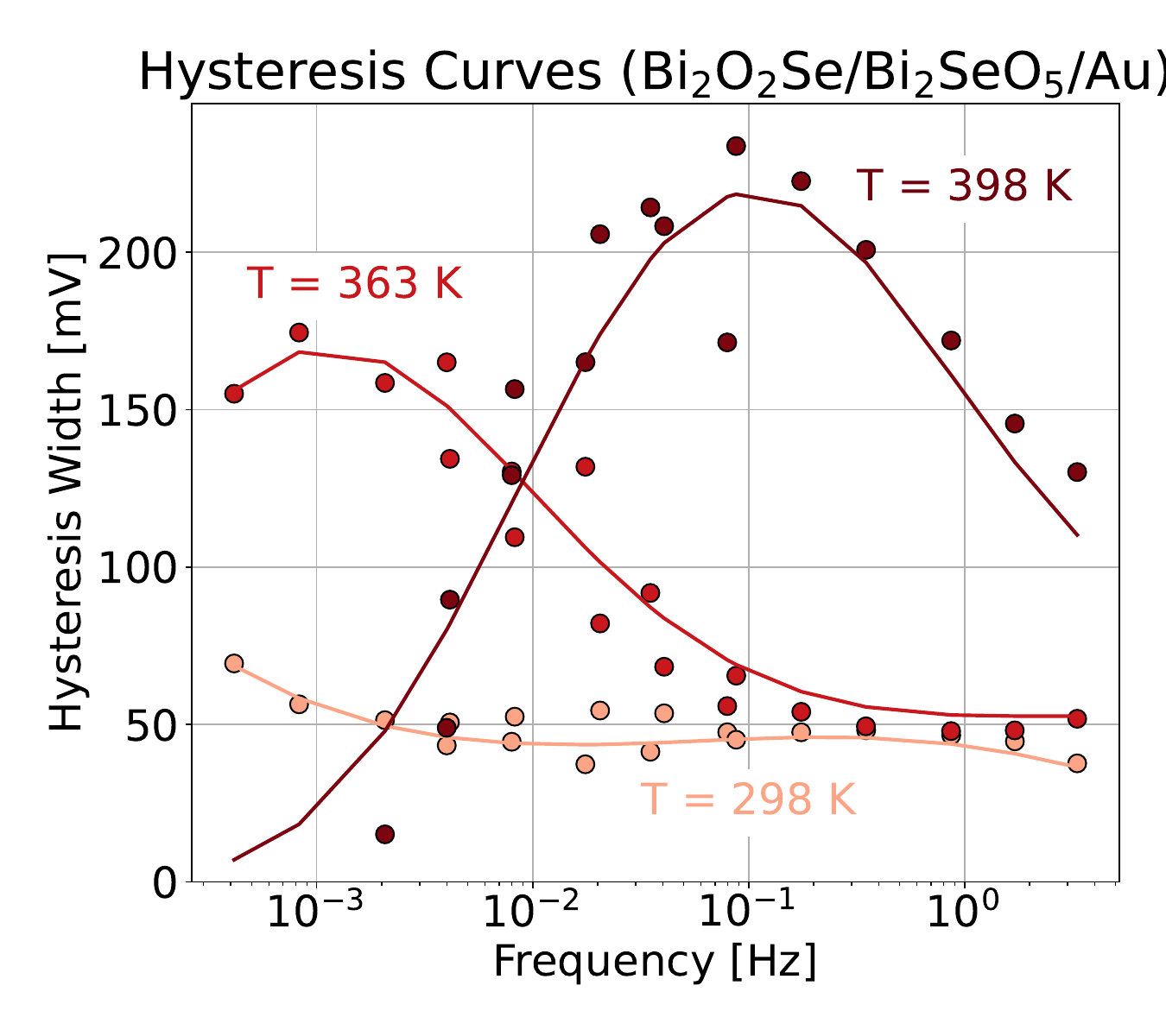}
        \end{minipage}};
    \node[draw=none] at (-2.0,2.4) {\textbf{(e)}};
    \end{tikzpicture}
\end{subfigure}
\begin{subfigure}[b]{.333\linewidth}
    \begin{tikzpicture}
    \node[inner sep=0pt] at (0,0) {
      \begin{minipage}{1.0\textwidth}
        \includegraphics[width=\textwidth]
        {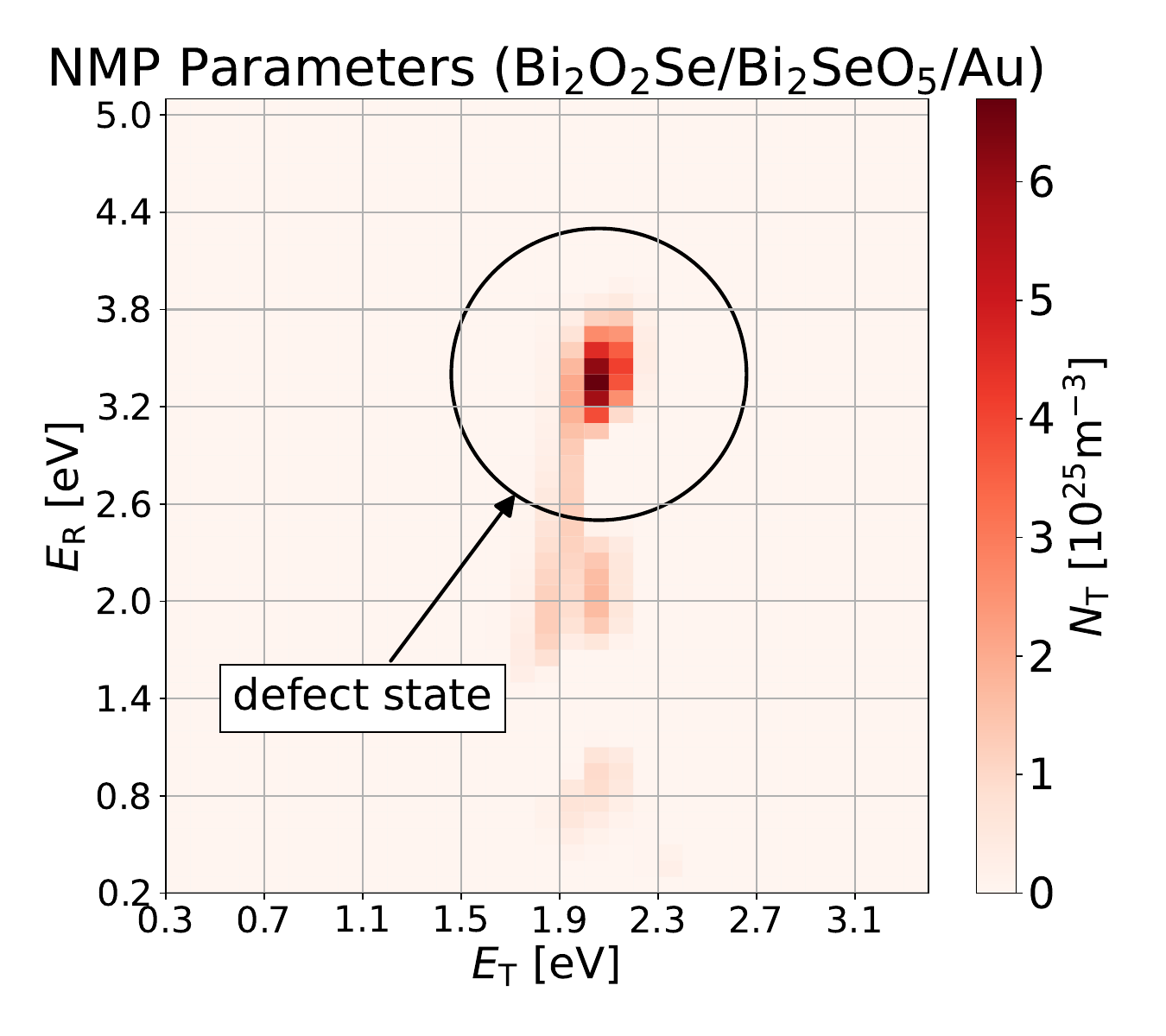}
    \end{minipage}};
    \node[draw=none] at (-2.0,2.4) {\textbf{(f)}};
    \end{tikzpicture}
\end{subfigure}
\caption{\textbf{(a)} Exemplary \IDVG curve for a device with a \ce{MoS2}/\ce{HfO2}/\ce{Au} gate stack, simulated with a Gaussian defect band (\(E_\mathrm{T} = (-4.4 \pm 0.2)\qty{}{\electronvolt}\),  \(E_\mathrm{R} = (2.25 \pm 0.2)\qty{}{\electronvolt}\), \(R = 1\)) placed across the gate insulator. The points in time at which the current criterion is reached are marked by \(t_\mathrm{up}\) and \(t_\mathrm{down}\). \textbf{(b)} Corresponding band diagram of the device showing the charge distributions \( \rho_\mathrm{ins}(t_\mathrm{up}) \) (in red) and \(\rho_\mathrm{ins}(t_\mathrm{down}) \) (in blue) in the insulator. \textbf{(c)}  Corresponding charge differences \(\Delta \rho = \rho_\mathrm{ins}(t_\mathrm{down}) - \rho_\mathrm{ins}(t_\mathrm{up}) \) induced by cyclo-stationary hysteresis measurements at various frequencies. The channel interaction leads to a negative peak near the channel, while the gate interaction leads to a positive peak in the proximity of the gate. Furthermore, the channel and gate peaks shift further into the insulator at lower frequencies, as slower defects with larger time constants (\( \tau_\mathrm{c} \)~and~\( \tau_\mathrm{e} \)) reside deeper in the insulator due to the exponentially decaying tunneling probability. \textbf{(d)}  Corresponding hysteresis curve of the device. The colored vertical lines correspond to the frequencies already shown in \textbf{(c)}. \textbf{(e)}  Measured (circles) and simulated (lines) hysteresis curves for a device with a \ce{Bi2O2Se}/\ce{Bi2SeO5}/\ce{Au}  gate stack \cite{Zhang2022}, showing a good agreement for varying temperatures. \textbf{(f)} Extracted defect distribution for the device with the \ce{Bi2O2Se}/\ce{Bi2SeO5}/\ce{Au} gate stack.} 
\label{fig:sec3:sec1:2}
\end{figure}

To highlight the channel and gate interaction, \fig{fig:sec3:sec1:2}{a} shows a hysteresis simulation for a hypothetical device with a \ce{MoS2}/\ce{HfO2}/\ce{Au} gate stack. For the simulation a very broad Gaussian acceptor-like defect band was placed across the insulator. \fig{fig:sec3:sec1:2}{b} visualizes the corresponding band diagram including the charge distribution \(\rho_\mathrm{ins}(t_\mathrm{up}) \) during the up-sweep in red and the charge distribution \(\rho_\mathrm{ins}(t_\mathrm{down}) \) during the down-sweep in blue. The common area displayed in violet represents the majority of the defects whose charge state remains unaffected by the cyclo-stationary sweep, and thus does not contribute to the hysteresis. When the device is turned on, the channel's Fermi level is raised w.r.t. the insulator, and all defects contained in the channel-sided AER with \( 1/f_\mathrm{sweep} \gtrsim \tau_\mathrm{c}\) will capture electrons from the channel. At the same time, the gate's Fermi level is lowered w.r.t. the insulator and all defects contained in the gate-sided AER with \( 1/f_\mathrm{sweep} \gtrsim \tau_\mathrm{e}\) will lose electrons to the gate.  As a result, the channel interaction leads to a buildup of a negative charge difference near the channel, while the gate interaction leads to a positive charge difference in the proximity of the gate. \\

This behavior is highlighted in \fig{fig:sec3:sec1:2}{c}, where the total charge difference \( \rho_\mathrm{ins}(t_\mathrm{down}) - \rho_\mathrm{ins}(t_\mathrm{up}) \), induced by cyclo-stationary hysteresis measurements, is plotted as a function of position for varying frequencies. At a given frequency, a negative peak due to the channel interaction and a positive peak due to the gate interaction is formed. Given that the threshold voltage shift has the opposite sign to the trapped charge (Eq.~\ref{sec3:eq3}), we conclude that the \textbf{channel interaction contributes to CW hysteresis}, while the \textbf{gate interaction contributes to CCW hysteresis}. The overall hysteresis of the device shown in  \fig{fig:sec3:sec1:2}{d} is positive at all frequencies, as the channel interaction dominates over the gate interaction. The right edge of the plateau-shaped hysteresis curve, highlighted in green, starts when the fastest defects near the insulator interface are probed. In contrast, the left edge, highlighted in yellow, is reached when the slowest defects, located in the bulk of the insulator dominate. A broad homogeneous spatial defect distribution in the insulator therefore leads to a broad plateau-shaped hysteresis peak, since the time constants are broadly distributed. \\

\fig{fig:sec3:sec1:2}{e} compares the measured and simulated hysteresis data of a device with a \ce{Bi2O2Se}/\ce{Bi2SeO5}/Au stack \cite{Zhang2022}. In the simulation, the CW hysteresis was attributed to defects located within the first \qty{2}{\nano\meter} of the insulator near the channel interface. Both the simulation and experimental results consistently show that the hysteresis peak shifts to higher frequencies as the temperature increases, which can be explained by the exponential temperature dependence of the rates (SI~Sec.~\ref{secA2}). Therefore, specifying the temperature at which a hysteresis curve was measured is essential. In addition, \fig{fig:sec3:sec1:2}{f} presents the corresponding defect parameter distribution extracted with the ESiD algorithm~\cite{9472877}. The strongly localized distribution with a pronounced maximum at (\(E_\mathrm{T}\), \(E_\mathrm{R}\)) = (\qty{2.0}{\electronvolt}, \qty{3.3}{\electronvolt}) suggests that a single defect species dominates the hysteresis in this device.

\subsection{Hysteresis due to Mobile Insulator Charges}\label{sec3:sec2}

\begin{figure}[hbt!]
\begin{subfigure}[b]{.333\linewidth}
    \begin{tikzpicture}
    \node[inner sep=0pt] at (0,0) {
      \begin{minipage}{1.0\textwidth}
        \includegraphics[width=\textwidth]{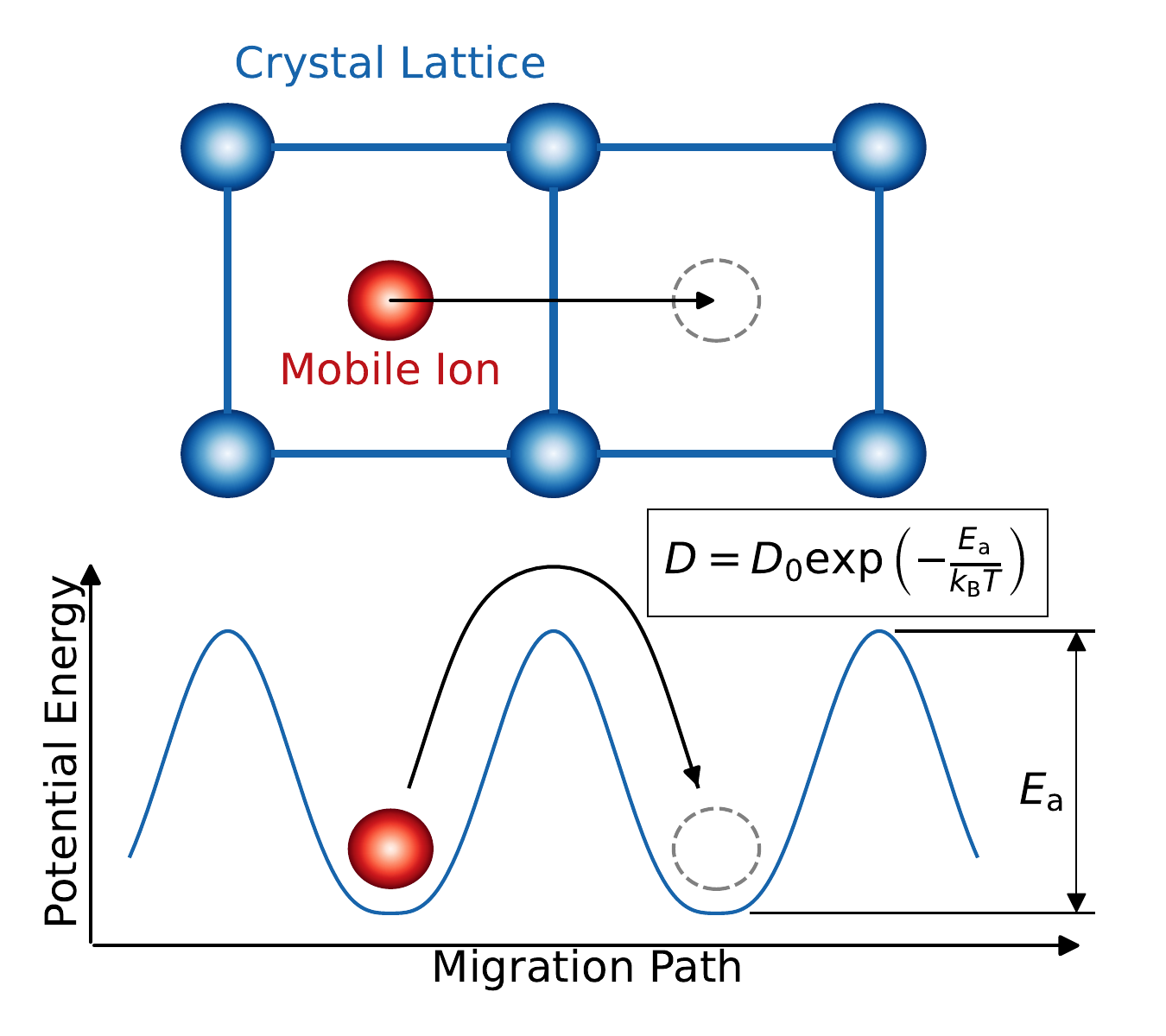}  
    \end{minipage}};
    \node[draw=none] at (-2.0,2.4) {\textbf{(a)}};
    \end{tikzpicture}
\end{subfigure}
\begin{subfigure}[b]{.333\linewidth}
    \begin{tikzpicture}
    \node[inner sep=0pt] at (0,0) {
      \begin{minipage}{1.0\textwidth}
        \includegraphics[width=\textwidth]{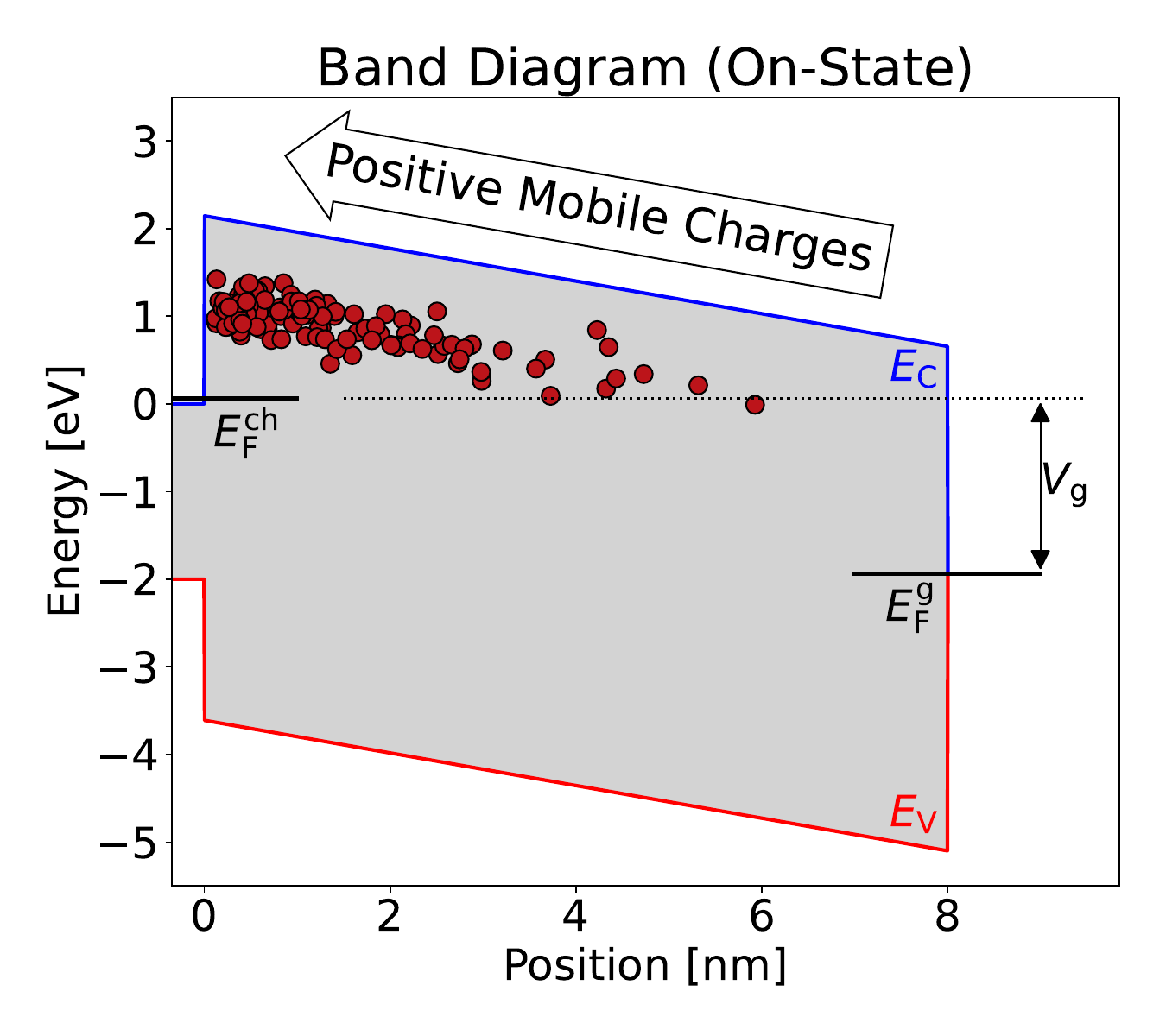} 
    \end{minipage}};
    \node[draw=none] at (-2.0,2.4) {\textbf{(b)}};
    \end{tikzpicture}
\end{subfigure}
\begin{subfigure}[b]{.333\linewidth}
    \begin{tikzpicture}
    \node[inner sep=0pt] at (0,0) {
      \begin{minipage}{1.0\textwidth}
        \includegraphics[width=\textwidth]{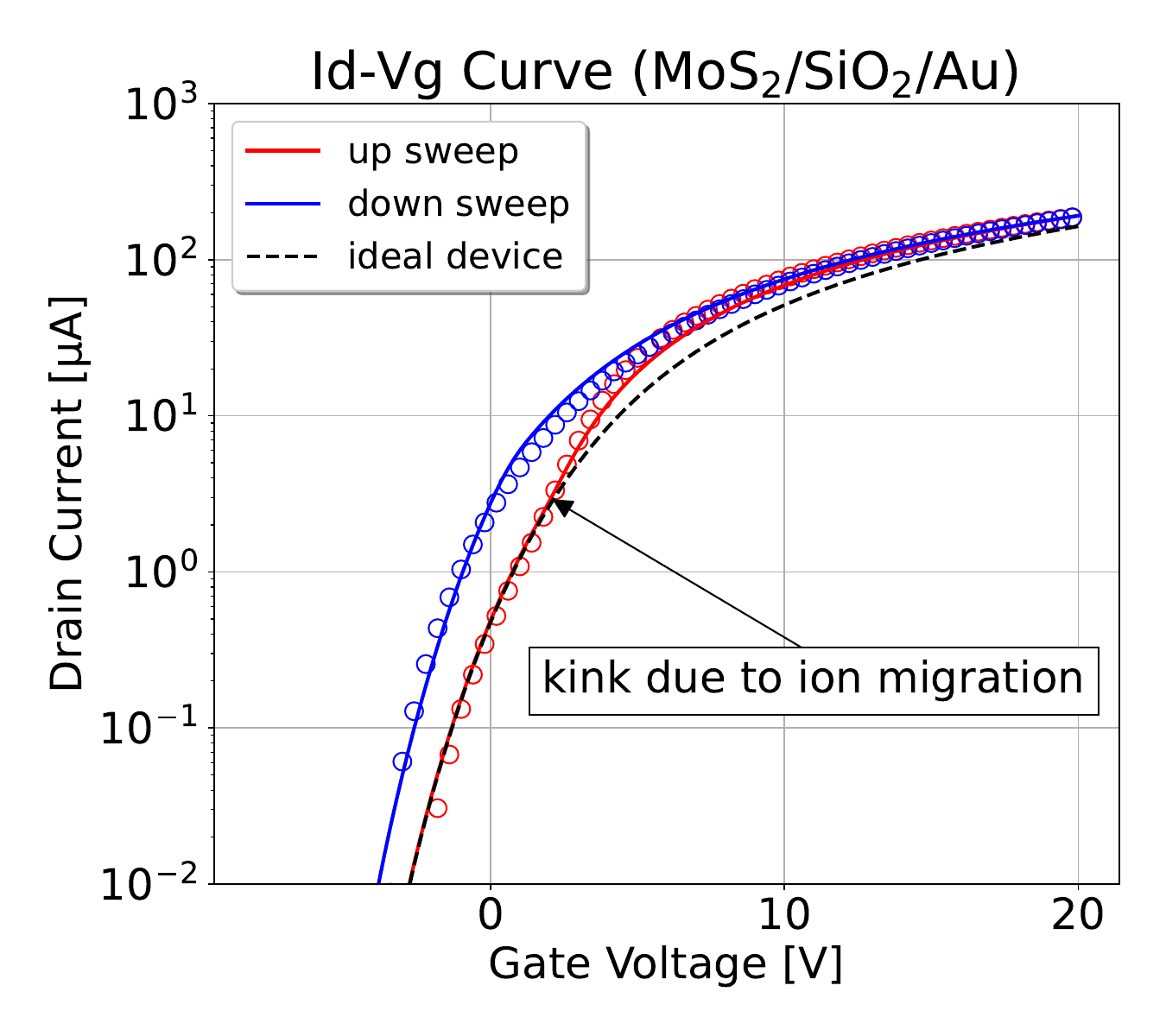}
    \end{minipage}};
    \node[draw=none] at (-2.0,2.4) {\textbf{(c)}};
    \end{tikzpicture}
\end{subfigure}
\begin{subfigure}[b]{.333\linewidth}
    \begin{tikzpicture}
    \node[inner sep=0pt] at (0,0) {
      \begin{minipage}{1.0\textwidth}
        \includegraphics[width=\textwidth]{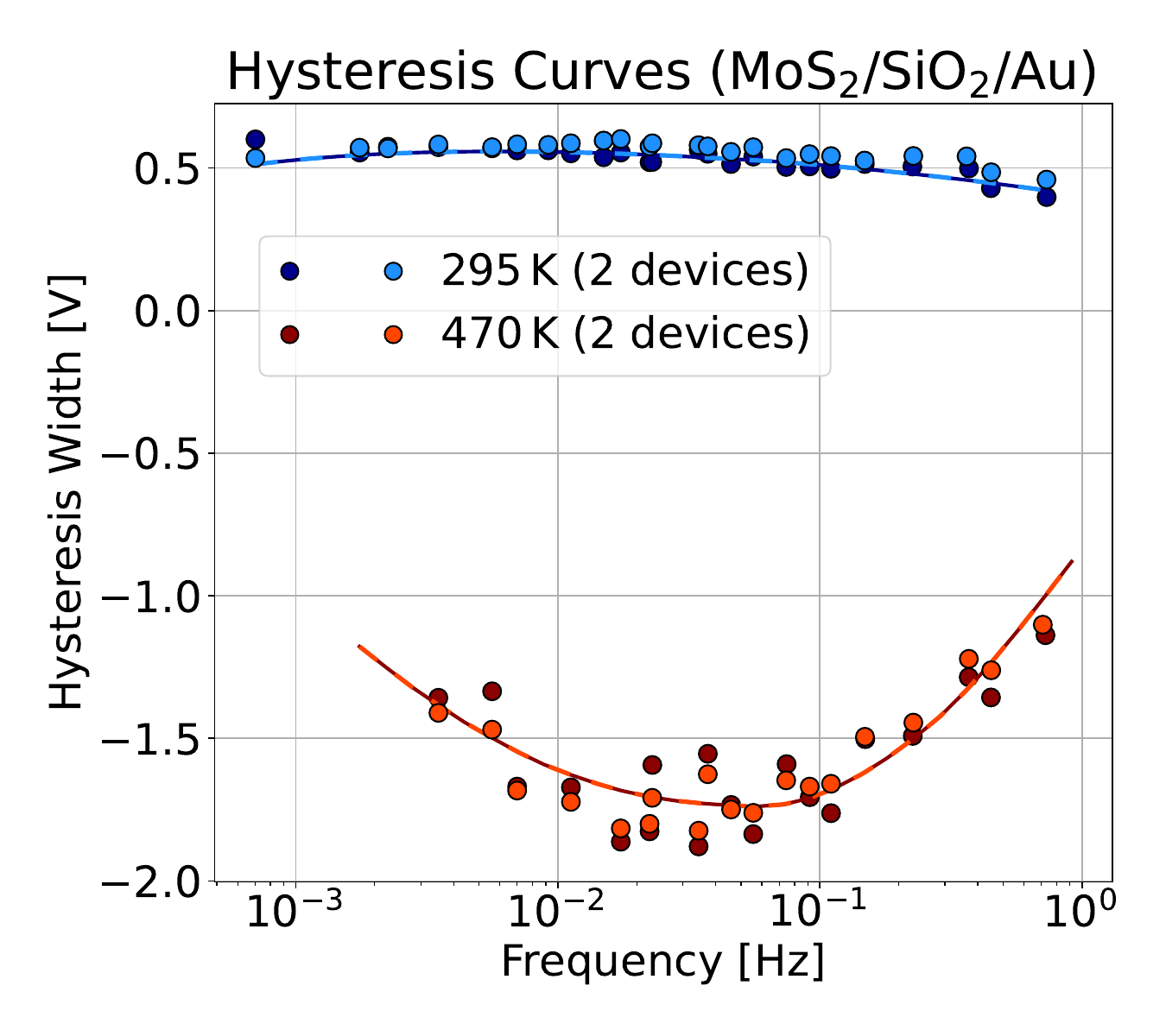}
    \end{minipage}};
    \node[draw=none] at (-2.0,2.4) {\textbf{(d)}};
    \end{tikzpicture}
\end{subfigure}
\begin{subfigure}[b]{.333\linewidth}
    \begin{tikzpicture}
    \node[inner sep=0pt] at (0,0) {
      \begin{minipage}{1.0\textwidth}
        \includegraphics[width=\textwidth]{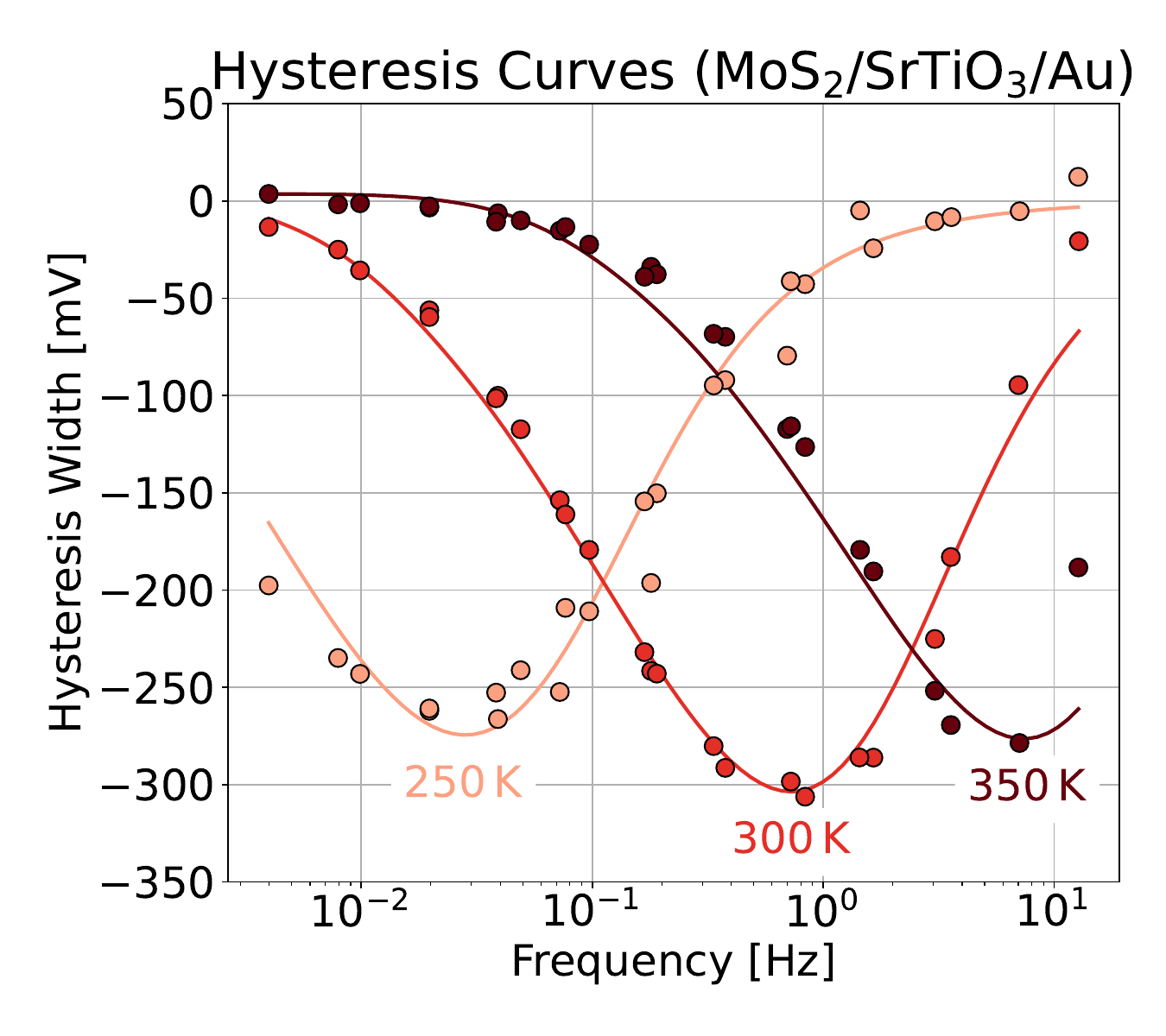}
        \end{minipage}};
    \node[draw=none] at (-2.0,2.4) {\textbf{(e)}};
    \end{tikzpicture}
\end{subfigure}
\begin{subfigure}[b]{.333\linewidth}
    \begin{tikzpicture}
    \node[inner sep=0pt] at (0,0) {
      \begin{minipage}{1.0\textwidth}
        \includegraphics[width=\textwidth]{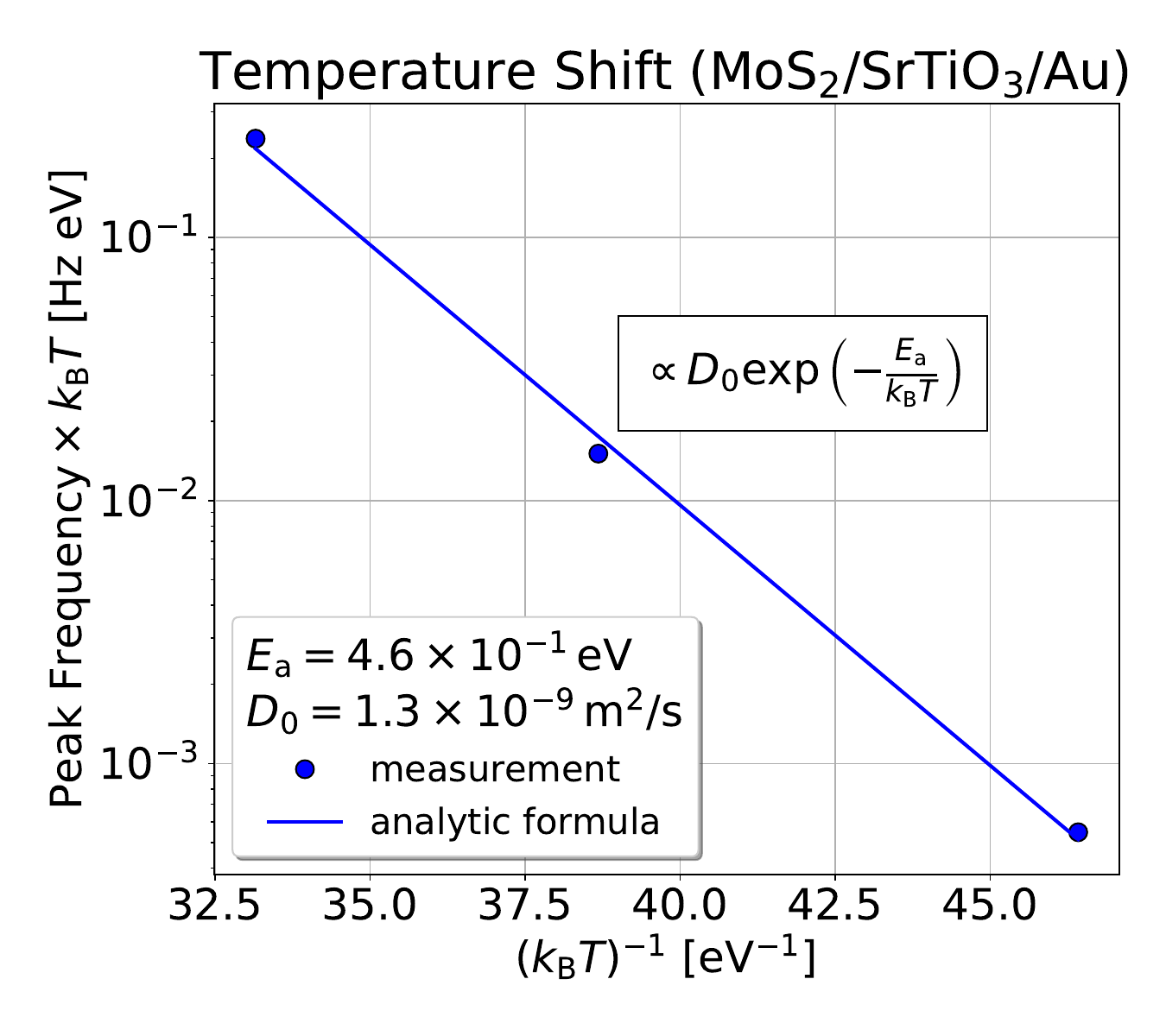}
    \end{minipage}};
    \node[draw=none] at (-2.0,2.4) {\textbf{(f)}};
    \end{tikzpicture}
\end{subfigure}
\caption{(\textbf{a}) Microscopic representation of the movement of mobile charges in a solid. Microscopically, the mobile charges perform jumps between stable lattice sides, requiring them to overcome a specific migration barrier \(E_\mathrm{a}\). (\textbf{b}) Band diagram visualizing the drift of positive mobile charges towards the channel. (\textbf{c}) Comparison of measured (circles) and simulated (lines) \IDVG curves for a device with a \ce{MoS2}/\ce{SiO2}/\ce{Au} gate stack.  Both the simulated and the measured \IDVG curves exhibit a characteristic kink in the up-sweep caused by the onset of the drift of the mobile charges.
(\textbf{d}) Comparison of corresponding measured (circles) and simulated (lines) hysteresis curves of the device with a temperature induced inversion of the hysteresis sign. The CCW hysteresis was attributed to mobile \ce{K+} ions in the \ce{SiO2} insulator, with parameters \( E_\mathrm{a} = \qty{0.98}{\electronvolt} \) and \( D_\mathrm{0} = \qty{1.87E-06}{\m^2s^{-1}} \). The CW hysteresis was attributed to defects in the first \qty{2}{\nano\meter} of the insulator, using parameters  reported in literature \cite{9472877}. (\textbf{e}) Comparison of measured (circles) and simulated (lines) hysteresis curves of a device with a \ce{MoS2}/\ce{SrTiO3}/\ce{Au} gate stack. The CCW hysteresis was attributed to mobile charges with a migration barrier of \mbox{\(E_\mathrm{a} = \qty{0.49}{\electronvolt}\)}. (\textbf{f}) Arrhenius plot of the temperature induced shift of the hysteresis peak. } 
\label{fig:sec3:sec2:1}
\end{figure}
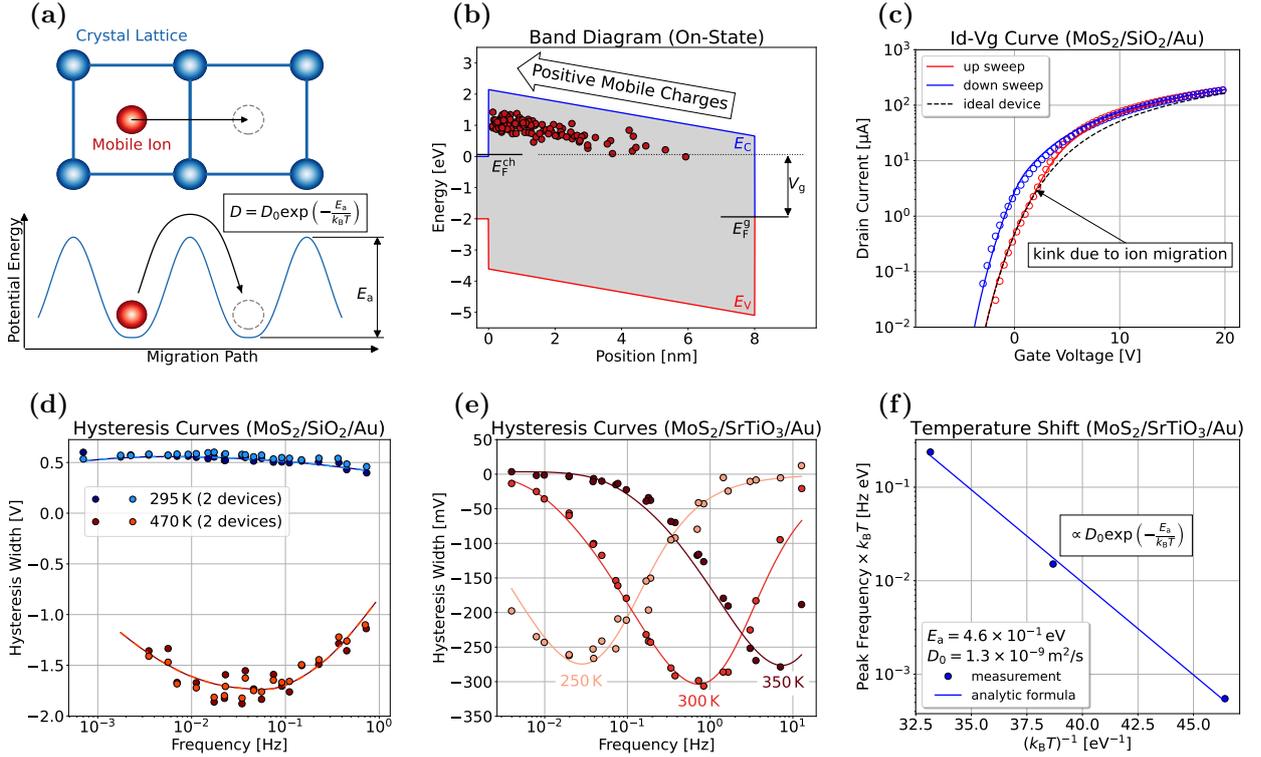

Similar to charge trapping, mobile charges (external contaminants such as \ce{Na+} and \ce{K+} or charged intrinsic defects such as oxygen vacancies \(V_\mathrm{O}^{+}\)) result in a time-dependent charge density \(\rho_\mathrm{ins}(x, t)\) in the insulator. Notably, such contaminants were a serious issue in early silicon-based devices, as their presence adversely affected device reliability and performance \cite{Hillen1979, Greeuw1984}.\\ 

From a microscopic viewpoint, diffusion occurs as the charges hop between stable lattice sites, requiring them to overcome a specific migration barrier \(E_\mathrm{a}\) (\fig{fig:sec3:sec2:1}{a}). Macroscopically, the charge movement in the insulator is assumed to be governed by the drift-diffusion (DD) equation \cite{heitjans_diffusion_2005}, with a spatial independent  diffusion constant \(D = D_0 \exp\left( E_\mathrm{a} / k_\mathrm{B}T \right) \) (SI~Sec.~\ref{secA3}). When the device is turned off, the electric field in the insulator is weak, causing the charges to strive for a mostly uniform distribution throughout the insulator. However, when the device is turned on, the drift term dominates the DD equation, driving positive charges towards the channel and negative charges towards the gate, resulting in a positive charge difference near the channel in both cases (see \fig{fig:sec3:sec2:1}{b}). Thus, according to Eq.~\ref{sec3:eq3} \textbf{ mobile charges within the insulator induce CCW hysteresis, irrespective of their charge state}.\\

\fig{fig:sec3:sec2:1}{c} shows a comparison of measured and simulated \IDVG curves of a device with a \ce{MoS2}/\ce{SiO2}/Au stack, which clearly illustrates the discussed behavior. As soon as the device is switched on and a significant field builds up in the insulator, positive charges are driven to the channel side, which leads to a characteristic kink in the up-sweep of the \IDVG curve. During the down-sweep, the majority of the charges are still close to the channel interface, resulting in a clearly visible CCW hysteresis.
\fig{fig:sec3:sec2:1}{d} presents the corresponding measured and simulated hysteresis curves of the device, with an excellent agreement between measurement  and simulation. In the simulation, the CW hysteresis was attributed to defects in the first \qty{2}{\nano\meter} of the insulator near the channel interface, using electron trap parameters from literature \cite{9472877}. The CCW hysteresis was attributed to mobile \ce{K+} ions in the \ce{SiO2} insulator, with parameters \( E_\mathrm{a} = \qty{0.98}{\electronvolt} \) and \( D_\mathrm{0} = \qty{1.87E-7}{\m^2s^{-1}} \), which align well with experimental values reported in literature \cite{Hillen1979}. The hysteresis of this device changes from CCW to CW when its cooled down, which can be explained by the simulation: The mobility of the ions is suppressed exponentially with decreasing temperature. Consequently, when the temperature decreases from \qty{470}{\kelvin} to \qty{295}{\kelvin}, the ions become immobile and cease to contribute to hysteresis within the measurement window, making charge trapping the dominant mechanism and resulting in a reversal of the hysteresis sign.\\

Furthermore, \fig{fig:sec3:sec2:1}{e} presents another example of hysteresis curves for a device with a \ce{MoS2}/\ce{SrTiO3}/Au gate stack with excellent agreement between simulation and measurement. In the simulation, the CCW hysteresis was attributed to mobile charges with a migration barrier of \mbox{\(E_\mathrm{a} = \qty{0.49}{\electronvolt}\)}, consistent with the predicted migration barrier of charged oxygen vacancies in \ce{SrTiO3} \cite{PhysRevB.83.220301}. At a given temperature, these vacancies generate a distinct CCW hysteresis peak at a characteristic frequency \( f_\mathrm{peak}\). As the temperature rises, \( f_\mathrm{peak}\) shifts to higher frequencies, while the peak height \(\Delta V_\mathrm{H}(f_\mathrm{peak})\) remains nearly constant. The frequency \( f_\mathrm{peak} \) can be approximated by the inverse time required for a single ion to traverse the entire insulator, resulting in
\begin{equation}\label{eq:sec3:sec2:2}
 f_\mathrm{peak} \approx  \frac{D_0 E_\mathrm{eff}}{d_\mathrm{ins}} \frac{\left|q\right|}{k_\mathrm{B} T} \exp\left(- \frac{E_\mathrm{a}}{k_\mathrm{B} T }  \right),
\end{equation}
where \(E_\mathrm{eff} \) represents the time-averaged electric field during the sweep. \fig{fig:sec3:sec2:1}{f} presents the experimentally extracted values of \( f_\mathrm{peak} \times k_\mathrm{B} T\) on an Arrhenius plot, resulting in a straight line. According to  Eq.~\ref{eq:sec3:sec2:2} the slope of this line directly corresponds to the migration barrier of the mobile charges. The value extracted in this manner also matches the predicted migration barrier of charged oxygen vacancies in \ce{SrTiO3} to a good approximation \cite{PhysRevB.83.220301}. This simulation-free approach considerably simplifies the parameter extraction, thereby aiding in the identification of mobile charges within the gate stack.

\subsection{Hysteresis due to Ferroelectricity}\label{sec3:sec3}

\begin{figure}[hbt!]
\begin{subfigure}[b]{.333\linewidth}
    \begin{tikzpicture}
    \node[inner sep=0pt] at (0,0) {
      \begin{minipage}{1.0\textwidth}
        \includegraphics[width=\textwidth]{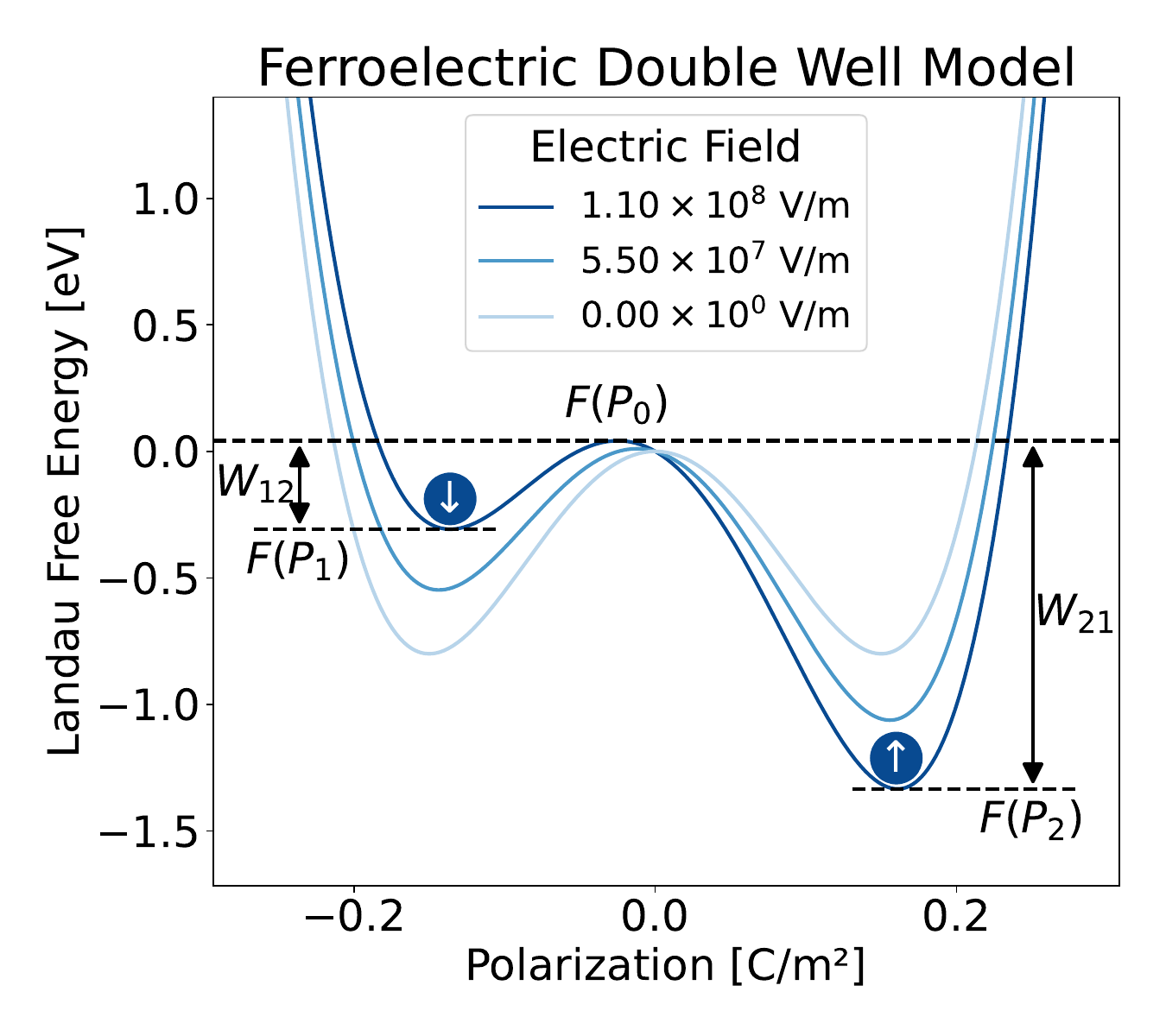} 
    \end{minipage}};
    \node[draw=none] at (-2.0,2.4) {\textbf{(a)}};
    \end{tikzpicture}
\end{subfigure}
\begin{subfigure}[b]{.333\linewidth}
    \begin{tikzpicture}
    \node[inner sep=0pt] at (0,0) {
      \begin{minipage}{1.0\textwidth}
        \includegraphics[width=\textwidth]{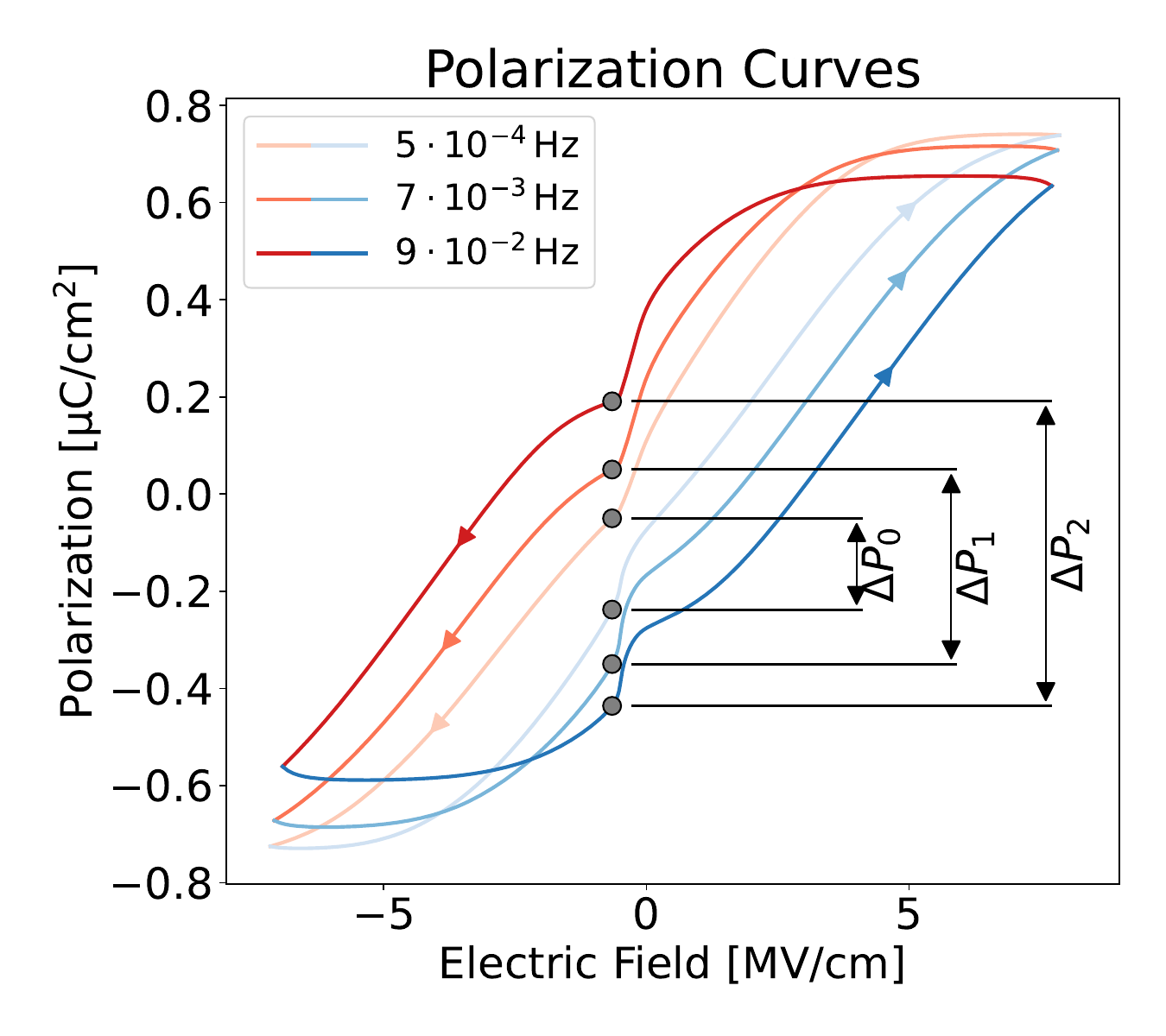}
    \end{minipage}};
    \node[draw=none] at (-2.0,2.4) {\textbf{(b)}};
    \end{tikzpicture}
\end{subfigure}
\begin{subfigure}[b]{.333\linewidth}
    \begin{tikzpicture}
    \node[inner sep=0pt] at (0,0) {
      \begin{minipage}{1.0\textwidth}
        \includegraphics[width=\textwidth]{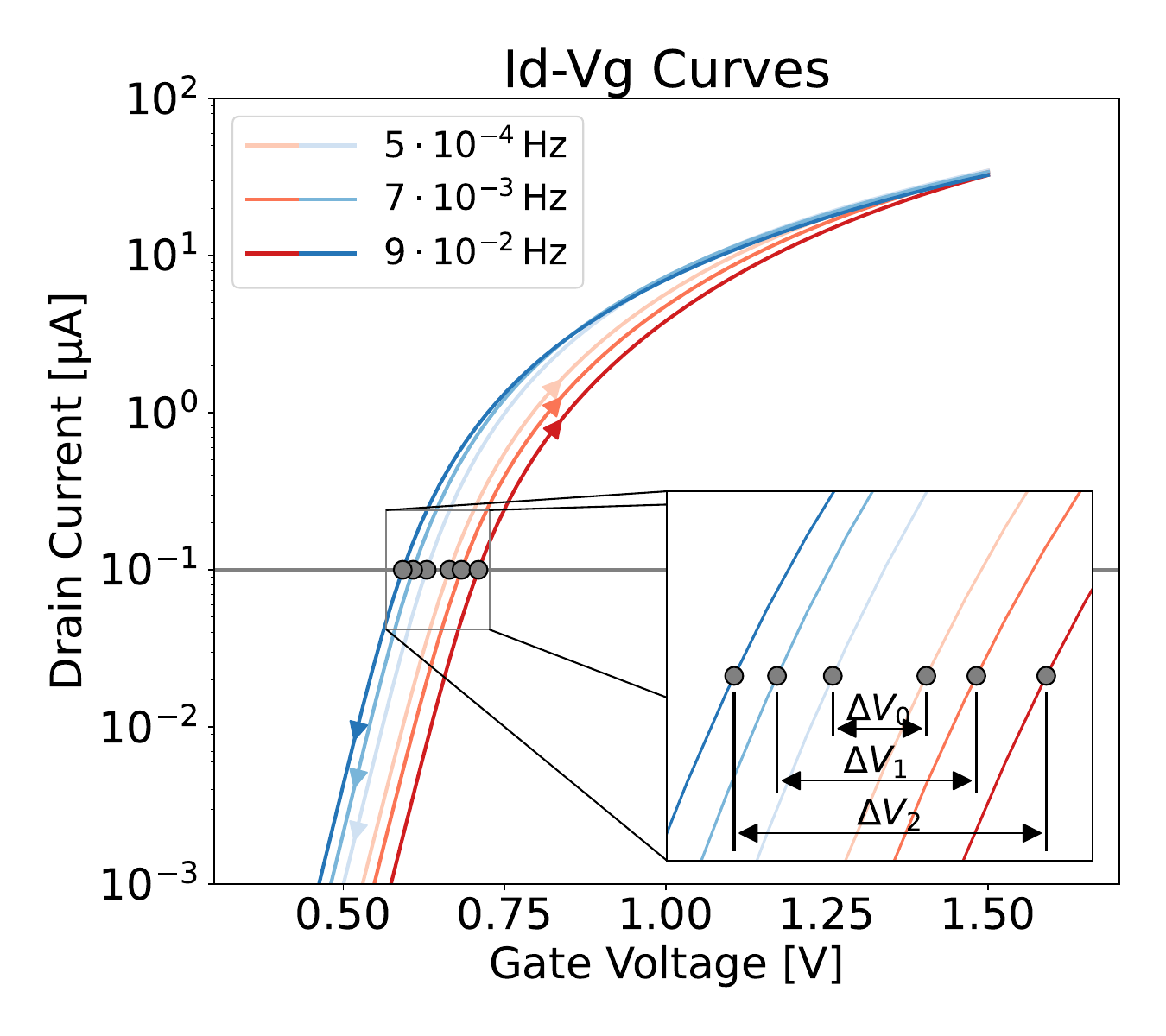}
    \end{minipage}};
    \node[draw=none] at (-2.0,2.4) {\textbf{(c)}};
    \end{tikzpicture}
\end{subfigure}
\begin{subfigure}[b]{.333\linewidth}
    \begin{tikzpicture}
    \node[inner sep=0pt] at (0,0) {
      \begin{minipage}{1.0\textwidth}
        \includegraphics[width=\textwidth]{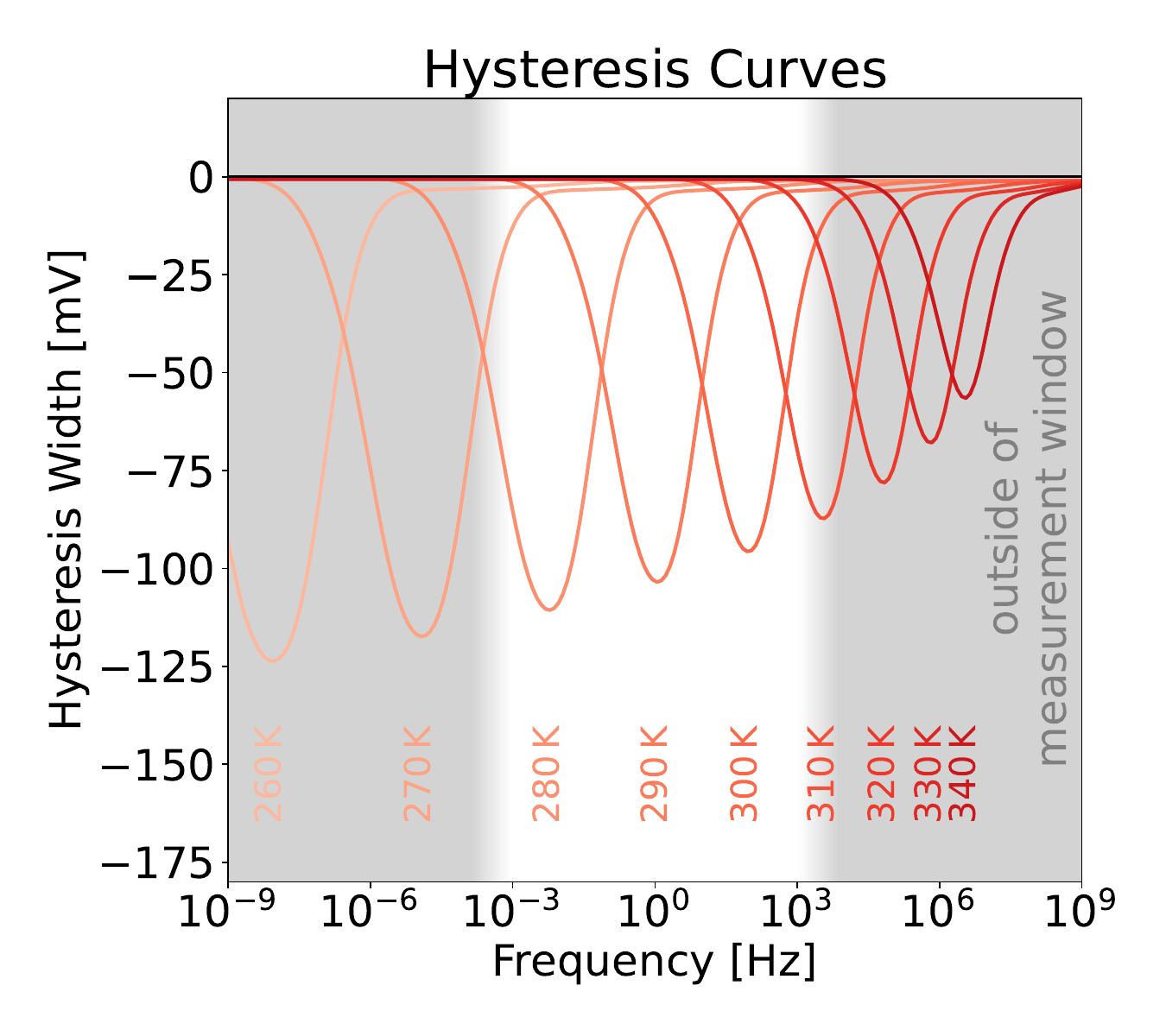}
    \end{minipage}};
    \node[draw=none] at (-2.0,2.4) {\textbf{(d)}};
    \end{tikzpicture}
\end{subfigure}
\begin{subfigure}[b]{.333\linewidth}
    \begin{tikzpicture}
    \node[inner sep=0pt] at (0,0) {
      \begin{minipage}{1.0\textwidth}
        \includegraphics[width=\textwidth]{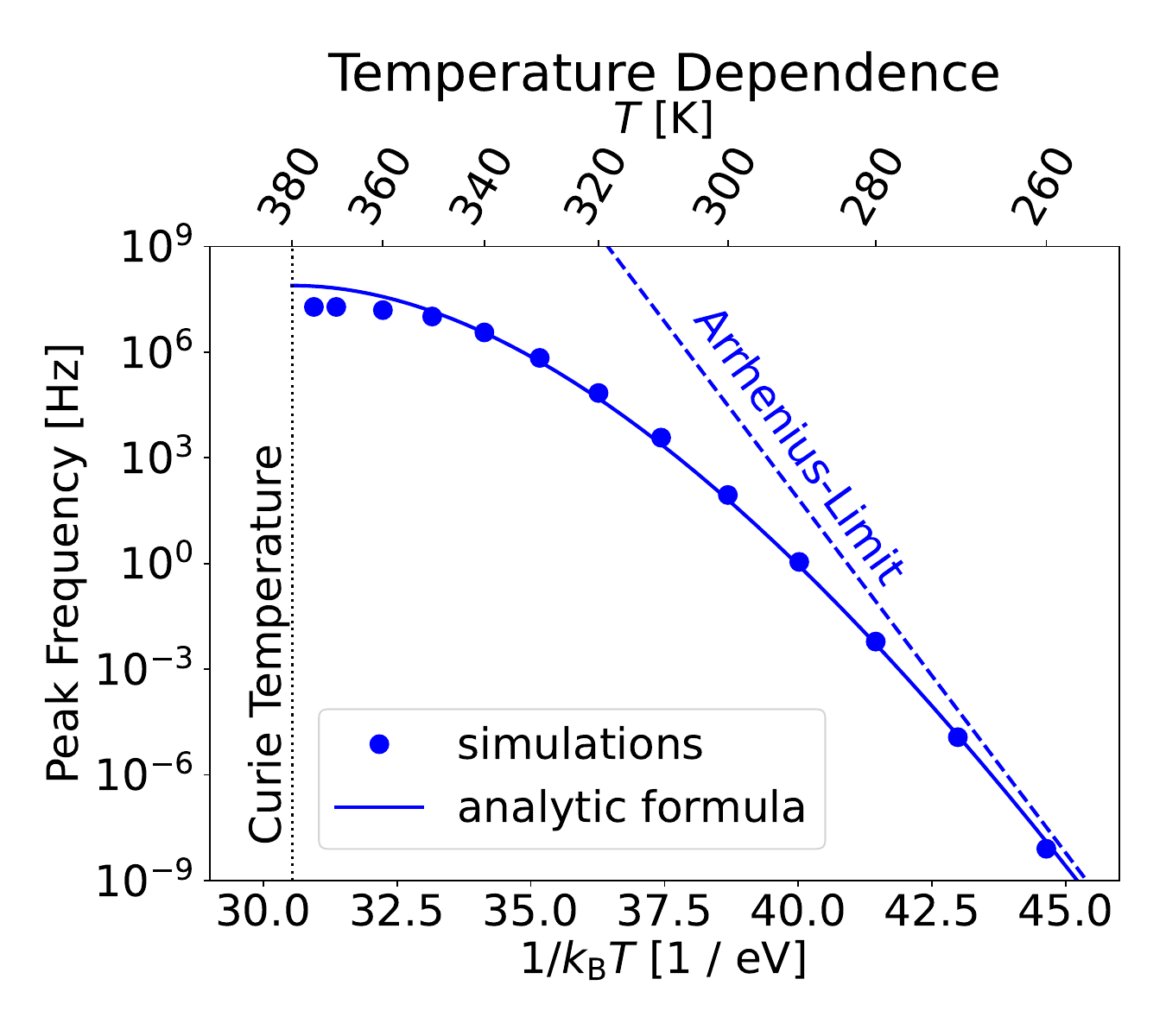}   \end{minipage}};
    \node[draw=none] at (-2.0,2.4) {\textbf{(e)}};
    \end{tikzpicture}
\end{subfigure}
\begin{subfigure}[b]{.333\linewidth}
    \begin{tikzpicture}
    \node[inner sep=0pt] at (0,0) {
      \begin{minipage}{1.0\textwidth}
        \includegraphics[width=\textwidth]{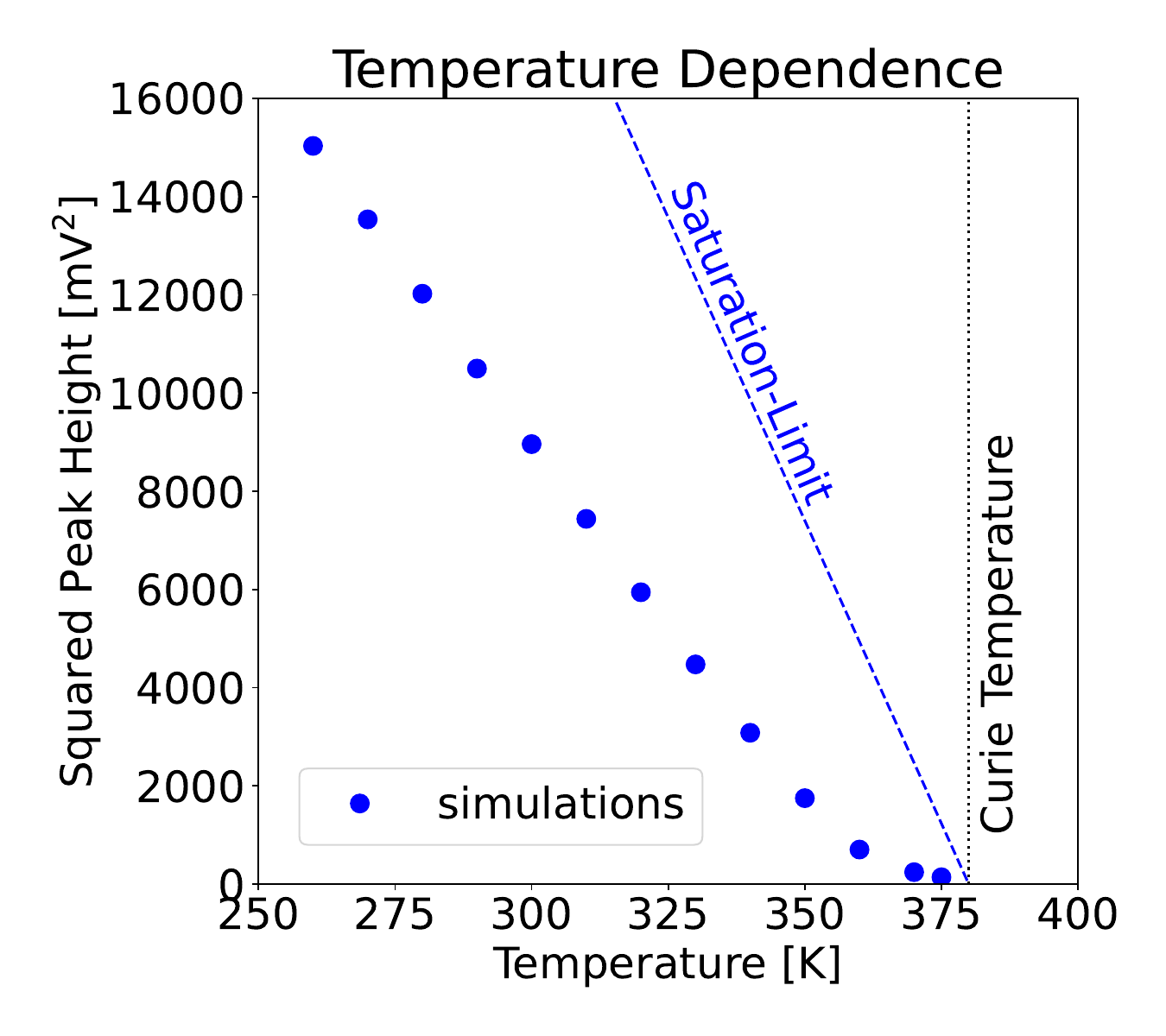}
    \end{minipage}};
    \node[draw=none] at (-2.0,2.4) {\textbf{(f)}};
    \end{tikzpicture}
\end{subfigure}
\caption{\textbf{(a)} Landau free energy landscape for a material in its ferroelectric phase with its two polarization states \(P_1\) and \(P_2\). \textbf{(b)} Polarization curves for a MOSFET with a slightly ferroelectric gate insulator, simulated with the parameters \(P_\mathrm{S} = \qty{0.75}{\micro\coulomb\centi\meter^{-2}} \), \(W_\mathrm{B} = \qty{8E20}{\electronvolt \centi\meter^{-3}} \) and  \(V = \qty{0.5E-20}{\centi\meter^{3}} \) (for further details on the model parameters see SI~Sec.~\ref{secA4}). \textbf{(c)} Corresponding \IDVG curves of the device plotted for various frequencies. The polarization \(\Delta P\) translates into a CCW hysteresis \( \Delta V = \Delta P \, d_\mathrm{ferro} / \varepsilon_\mathrm{ferro} \) in the \IDVG curve. \textbf{(d)} Corresponding hysteresis curves of the device plotted for various temperatures.  \textbf{(e)} Arrhenius plot of the peak frequency \(f_\mathrm{peak}\). The frequency \(f_\mathrm{peak}\) reaches its maximum near the Curie temperature and exhibits Arrhenius-like behavior at lower temperatures, which is consistent with the analytic formula given by Eq.~\ref{eq:sec3:sec3:5}. \textbf{(f)} Squared peak height as a function of temperature. Since the polarization of the ferroelectric layer saturates when the spontaneous polarization is reached, 
\(\Delta V  \lesssim  (d_\mathrm{ferro} /\varepsilon_\mathrm{ferro}) 2 P_\mathrm{S} 
\) represents an upper limit for the peak height.
}
\label{fig:sec3:sec3:1}
\end{figure}

In commercial silicon technology, dielectric materials are used as gate insulators because their properties are ideally independent of the device's history, ensuring stable performance. In contrast, ferroelectric materials (such as perovskites like \ce{SrTiO3} and \ce{BaTiO3}) can retain their polarization and thus exhibit a complex, time-dependent polarization response to  electric fields \cite{KittelCharles2018Itss}. While ferroelectric materials can be deliberately used for neuromorphic \cite{9911305} or memory devices \cite{8993642}, their unintended use in conventional MOSFETs introduces instabilities in the transfer characteristics. This risk is significant as experimental devices often use novel gate insulators that may exhibit ferroelectric phases. For instance, the commonly used material \ce{SrTiO3} can exist in either its paraelectric or ferroelectric phase within a device, depending on factors such as thickness, strain, and temperature \cite{haeni_room-temperature_2004}. \\

\fig{fig:sec3:sec3:1}{a} illustrates the Landau free energy landscape for a material in its ferroelectric phase with its two polarization states \(P_1\) and \(P_2\). When an electric field is applied, one of the states is lowered in energy, while the other is raised, thereby facilitating the transition between the two polarization states. We follow the approach of Vopsaroiu et al. \cite{PhysRevB.82.024109} and describe the transition as a thermally activated process, which has been effective in replicating experimental data for thin films \cite{CHEN2020100919, Zhu2016, Nomura2015} (for further details, see SI~Sec.~\ref{secA4}). \fig{fig:sec3:sec3:1}{b} illustrates the polarization of a ferroelectric gate insulator in a MOSFET as a function of the MOSFET’s surface field. Under the effect of a positive field, the ferroelectric layer polarizes in the positive direction until saturation is reached. Likewise, when a negative field is applied, the ferroelectric layer is polarized in the negative direction, eventually saturating as well. However, the up-sweep (shown in red) and down-sweep (shown in blue) follow different paths, which results in a clear hysteresis loop, characterized by the polarization difference \(\Delta P = P(t_\mathrm{down}) - P(t_\mathrm{up})\). As shown in \fig{fig:sec3:sec3:1}{c}, \textbf{the hysteresis in the polarization of the ferroelectric gate insulators translates to a CCW hysteresis in the transfer characteristic},  given by  \( \Delta V = \Delta P \, d_\mathrm{ferro} / \varepsilon_\mathrm{ferro} \), where \( d_\mathrm{ferro} \) and \( \varepsilon_\mathrm{ferro} \) represent the ferroelectric layer's thickness and permittivity.\\

\fig{fig:sec3:sec3:1}{d} illustrates simulated hysteresis curves for a hypothetical device with a weakly ferroelectric gate insulator. While hysteresis curves of real systems may deviate from those depicted due to the impact of additional terms in the Landau free energy, two general trends can be observed with increasing temperature: First, the hysteresis peak shifts to higher frequencies, which results from the temperature activation of the transition rates. Second, the height of the hysteresis peak decreases due to the reduction of the spontaneous polarization. As already demonstrated for mobile charges (see Sec.~\ref{sec3:sec1}), valuable information can be extracted from the temperature-induced shift of the hysteresis peak. When the applied electric fields are moderate, we can approximate the peak frequency by the rate at zero field and obtain

\begin{equation}\label{eq:sec3:sec3:5}
 f_\mathrm{peak} \approx \nu_0 \exp\left( \frac{1}{4}  \frac{a_0^2 V }{b } \frac{(T_\mathrm{C} - T)^2}{k_\mathrm{B}T} \right).
\end{equation}

\fig{fig:sec3:sec3:1}{e} shows the simulated peak frequencies plotted on an Arrhenius plot, demonstrating that the analytical formula given by Eq.~\ref{eq:sec3:sec3:5} effectively captures the temperature dependence observed in the simulation. The peak frequency reaches its maximum near the Curie temperature and exhibits Arrhenius-like behavior at lower temperatures. In practice, this analytical formula can be fitted to experimental data to determine both the ratio \(a_0^2 V / b \) and the Curie temperature \( T_\mathrm{C} \) (for further details see SI~Sec.~\ref{secA4}). In addition, \fig{fig:sec3:sec3:1}{f} displays the squared peak height as a function of temperature. Unfortunately, there is no straightforward analytical expression for the peak height, as it depends on whether the applied voltage drives the ferroelectric layer into saturation or not.

\subsection{Experimental identification of individual mechanisms}\label{sec3:sec4}
\begin{table}[hbt!]
\caption{Physical mechanisms contributing to hysteresis. \\ (the sign/direction is given for n-type devices and reverses for p-type devices)}
\label{tab:my-table}
\renewcommand{\arraystretch}{1}
\begin{tabular}{lll}
\hline
\rowcolor[HTML]{C0C0C0} 
\multicolumn{3}{c}{\cellcolor[HTML]{C0C0C0}{\color[HTML]{333333} Contributions to hysteresis}}                                                                                                                                                                                                                                                                                                                           \\ \hline
\rowcolor[HTML]{EFEFEF} 
\multicolumn{1}{|l|}{\cellcolor[HTML]{EFEFEF}{\color[HTML]{000000} physical effect}}                         & \multicolumn{1}{l|}{\cellcolor[HTML]{EFEFEF}{\color[HTML]{000000} sign}}     & \multicolumn{1}{l|}{\cellcolor[HTML]{EFEFEF}{\color[HTML]{000000} comment on identification}}                                                                                                                              \\ \hline
\rowcolor[HTML]{FFFFFF} 
\multicolumn{1}{|l|}{\cellcolor[HTML]{FFFFFF}{\color[HTML]{000000} charge trapping by defects near channel}} & \multicolumn{1}{l|}{\cellcolor[HTML]{FFFFFF}{\color[HTML]{000000} positive}} & \multicolumn{1}{l|}{\cellcolor[HTML]{FFFFFF}{\color[HTML]{000000} can be identified by sign}}                                                                                                                              \\ \hline
\rowcolor[HTML]{FFFFFF} 
\multicolumn{1}{|l|}{\cellcolor[HTML]{FFFFFF}{\color[HTML]{000000} charge trapping by defects near gate}}    & \multicolumn{1}{l|}{\cellcolor[HTML]{FFFFFF}{\color[HTML]{000000} negative}} & \multicolumn{1}{l|}{\cellcolor[HTML]{FFFFFF}{\color[HTML]{000000} can often be excluded by required high defect density}}                                                                                                       \\ \hline
\rowcolor[HTML]{FFFFFF} 
\multicolumn{1}{|l|}{\cellcolor[HTML]{FFFFFF}{\color[HTML]{000000} drift/diffusion of mobile charges}}       & \multicolumn{1}{l|}{\cellcolor[HTML]{FFFFFF}{\color[HTML]{000000} negative}} & \multicolumn{1}{l|}{\cellcolor[HTML]{FFFFFF}{\color[HTML]{000000} \begin{tabular}[c]{@{}l@{}}can be identified by characteristic kink in the \IDVG curve, \\ temperature induced shift of peaks follows Arrhenius law,\\ height of hysteresis peak is temperature independent  \end{tabular}}} \\ \hline
\rowcolor[HTML]{FFFFFF} 
\multicolumn{1}{|l|}{\cellcolor[HTML]{FFFFFF}{\color[HTML]{000000} ferroelectric gate insulator}}            & \multicolumn{1}{l|}{\cellcolor[HTML]{FFFFFF}{\color[HTML]{000000} negative}} & \multicolumn{1}{l|}{\cellcolor[HTML]{FFFFFF}{\color[HTML]{000000} \begin{tabular}[c]{@{}l@{}}can be identified by phase transition, \\ temperature induced shift of peak deviates from Arrhenius
law, \\ height of the hysteresis peak decreases with increasing temperature  \end{tabular}}}          \\ \hline
\end{tabular}
\end{table}

While, in real devices, several mechanisms can contribute to hysteresis simultaneously, in many cases a single mechanism dominates the hysteresis process. Identifying this dominant mechanism is essential for improving device stability by implementing appropriate countermeasures. Table \ref{tab:my-table} summarizes the main features of the mechanisms discussed in this work, and provides the basis for the following discussion: 

\subsubsection{Positive (Clockwise) Hysteresis}\label{sec3:sec3}

CW hysteresis is typically associated with charge trapping by defects near the channel, as other mechanisms presented cannot account for the positive sign of the hysteresis. In general, as shown in \fig{fig:sec3:sec1:2}{e}, the observed hysteresis peak shifts to higher frequencies due to the temperature activation of the rates. As only defects within the AER can contribute to hysteresis, DFT calculations can be used to verify whether a particular defect type falls within the AER. Conversely, the NPM parameters extracted from experiments can be compared with DFT predictions to identify the most likely defect type.

\subsubsection{Negative (Counterclockwise) Hysteresis}\label{sec3:sec3}

One contribution to CCW hysteresis may originate from charge trapping due to defects near the gate. However, the contribution of defects to the hysteresis decreases linearly the closer the defects are to the gate (see Eq.~\ref{sec3:eq3}). As a result, in many practical cases, the defect density required to account for the observed CCW hysteresis is unreasonably high, i.e. \(\geq \qty{E21}{\centi\meter^{-3}}\), which rules out charge trapping by gate-sided defects as the primary mechanism (a possible exception where such a high defect density might still occur is if the insulator is damaged during the deposition of the metal gate). In practice, the required defect density can be estimated by the expression \( \rho_\mathrm{ins} \gtrsim - C_\mathrm{ins} \Delta V_\mathrm{H} 2 d_\mathrm{ins} / \Delta x^2 \), which is derived from Eq.~\ref{sec3:eq3} under the assumption that charges are uniformly distributed within the region \([d_\mathrm{ins} - \Delta x, d_\mathrm{ins}]\) near the gate.\\

A more likely mechanism resulting in CCW hysteresis is the drift of mobile charges within the insulator. We demonstrated that mobile charges in the insulator can lead to a characteristic kink in the \IDVG curve (see  \fig{fig:sec3:sec2:1}{c}), which can be used as a reliable indicator. Furthermore, the height of the hysteresis peaks is temperature independent and the temperature-induced shift of the hysteresis peaks follows an Arrhenius law if the transport obeys the conventional drift diffusion equation with a constant migration barrier. This behavior has been confirmed experimentally (see  \fig{fig:sec3:sec2:1}{e} and  \fig{fig:sec3:sec2:1}{f}) and could also be used as an indication for the presence of mobile charges in the insulator. Note that in the case of dispersive transport (e.g. distributed migration barriers) a different behavior is to be expected \cite{Grasser2006}. \\

Finally, ferroelectric materials in the gate stack can also lead to CCW hysteresis. Based on our theoretical investigation, we expect that the corresponding height of the hysteresis peak decreases when \(T_\mathrm{C}\) is approached. Moreover, we predict that the temperature-induced shift of the hysteresis peaks deviates from an Arrhenius law when \(T_\mathrm{C}\) is approached (see \fig{fig:sec3:sec3:1}{d}, \fig{fig:sec3:sec3:1}{e} and \fig{fig:sec3:sec3:1}{f}). Another indication of ferroelectricity is the ferroelectric phase transition itself, which could be detected experimentally by the characteristic discontinuity in the susceptibility, as described by the Curie-Weiss law \cite{KittelCharles2018Itss}. Therefore, the phase transition can be specifically explored by measuring the capacitance of the gate stack over an extended temperature range.

\section{Standardization and Normalization of Hysteresis Measurements}\label{sec4}

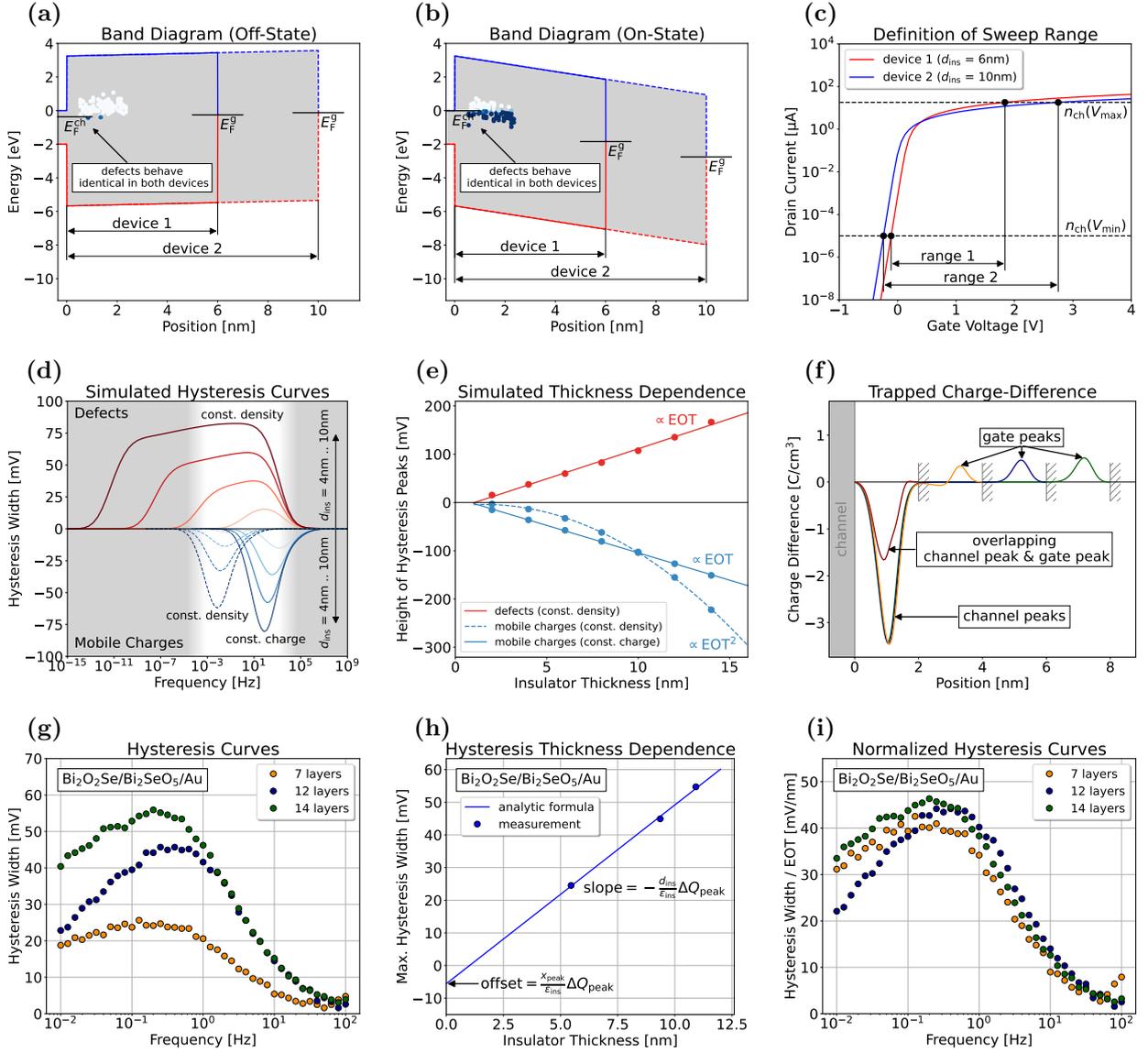
\begin{figure}[hbt!]
\begin{subfigure}[b]{.333\linewidth}
    \begin{tikzpicture}
    \node[inner sep=0pt] at (0,0) {
      \begin{minipage}{1.0\textwidth}
        \includegraphics[width=\textwidth]{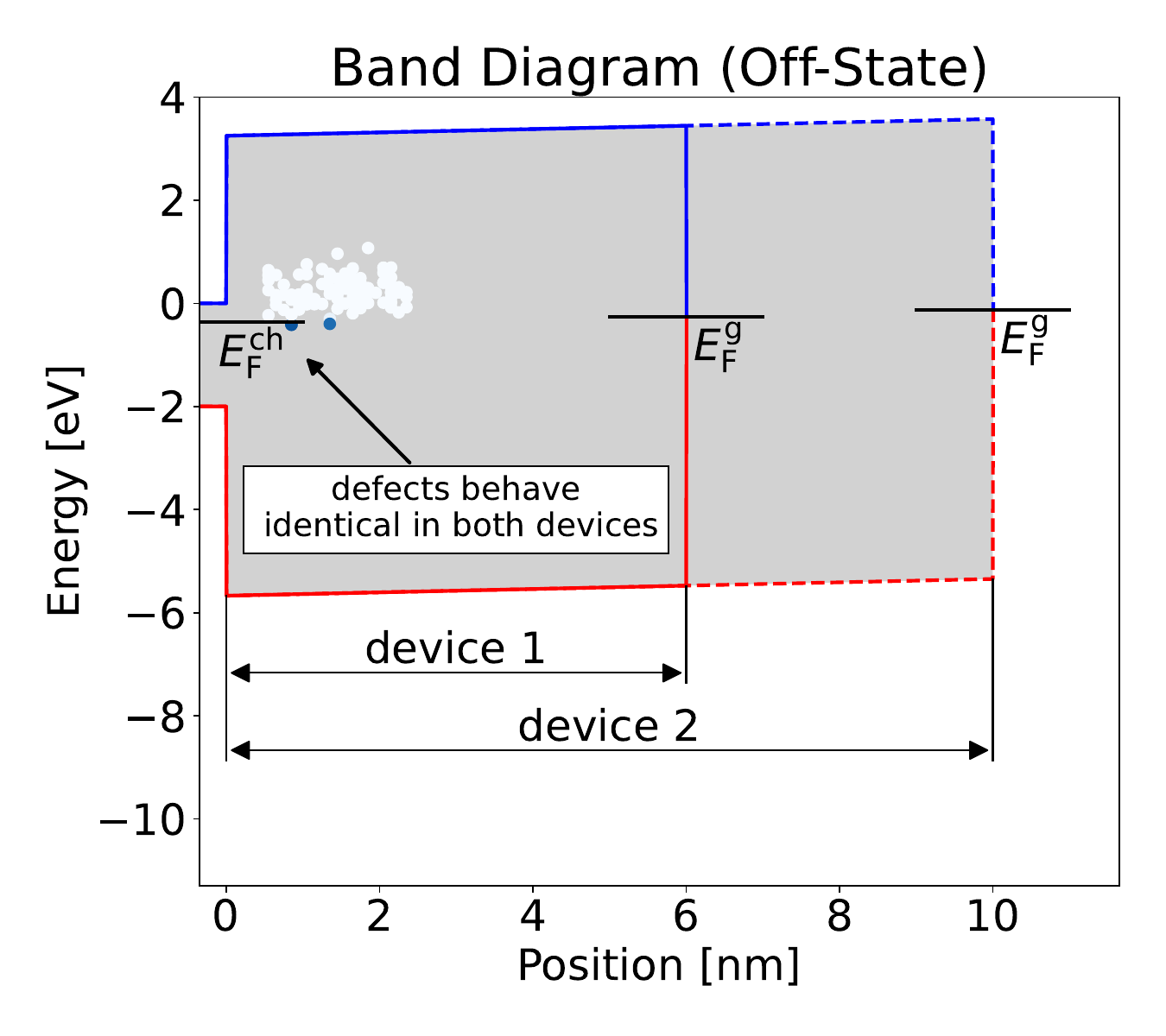} 
    \end{minipage}};
    \node[draw=none] at (-2.0,2.4) {\textbf{(a)}};
    \end{tikzpicture}
\end{subfigure}
\begin{subfigure}[b]{.333\linewidth}
    \begin{tikzpicture}
    \node[inner sep=0pt] at (0,0) {
      \begin{minipage}{1.0\textwidth}
        \includegraphics[width=\textwidth]{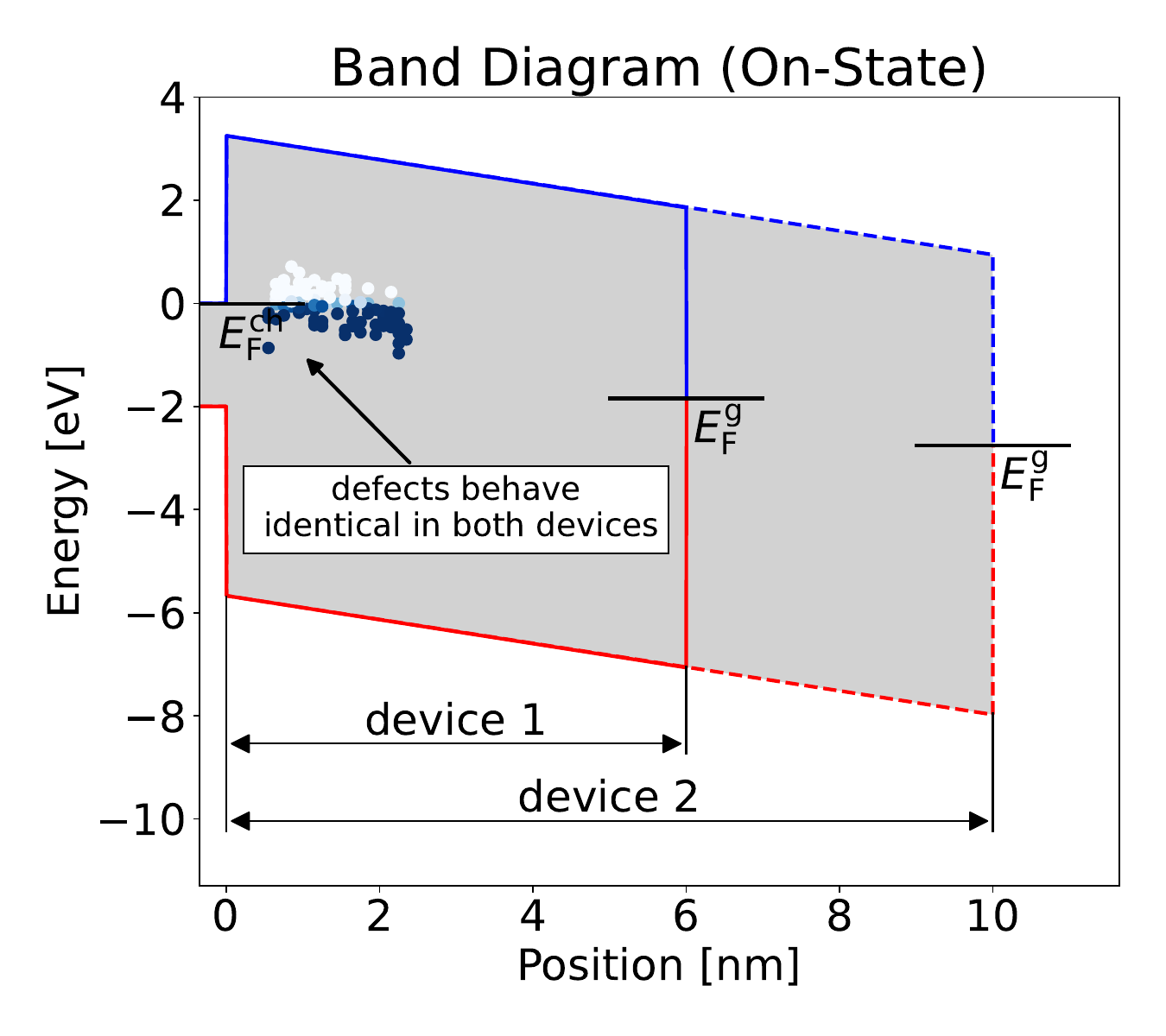}
    \end{minipage}};
    \node[draw=none] at (-2.0,2.4) {\textbf{(b)}};
    \end{tikzpicture}
\end{subfigure}
\begin{subfigure}[b]{.333\linewidth}
    \begin{tikzpicture}
    \node[inner sep=0pt] at (0,0) {
      \begin{minipage}{1.0\textwidth}
        \includegraphics[width=\textwidth]{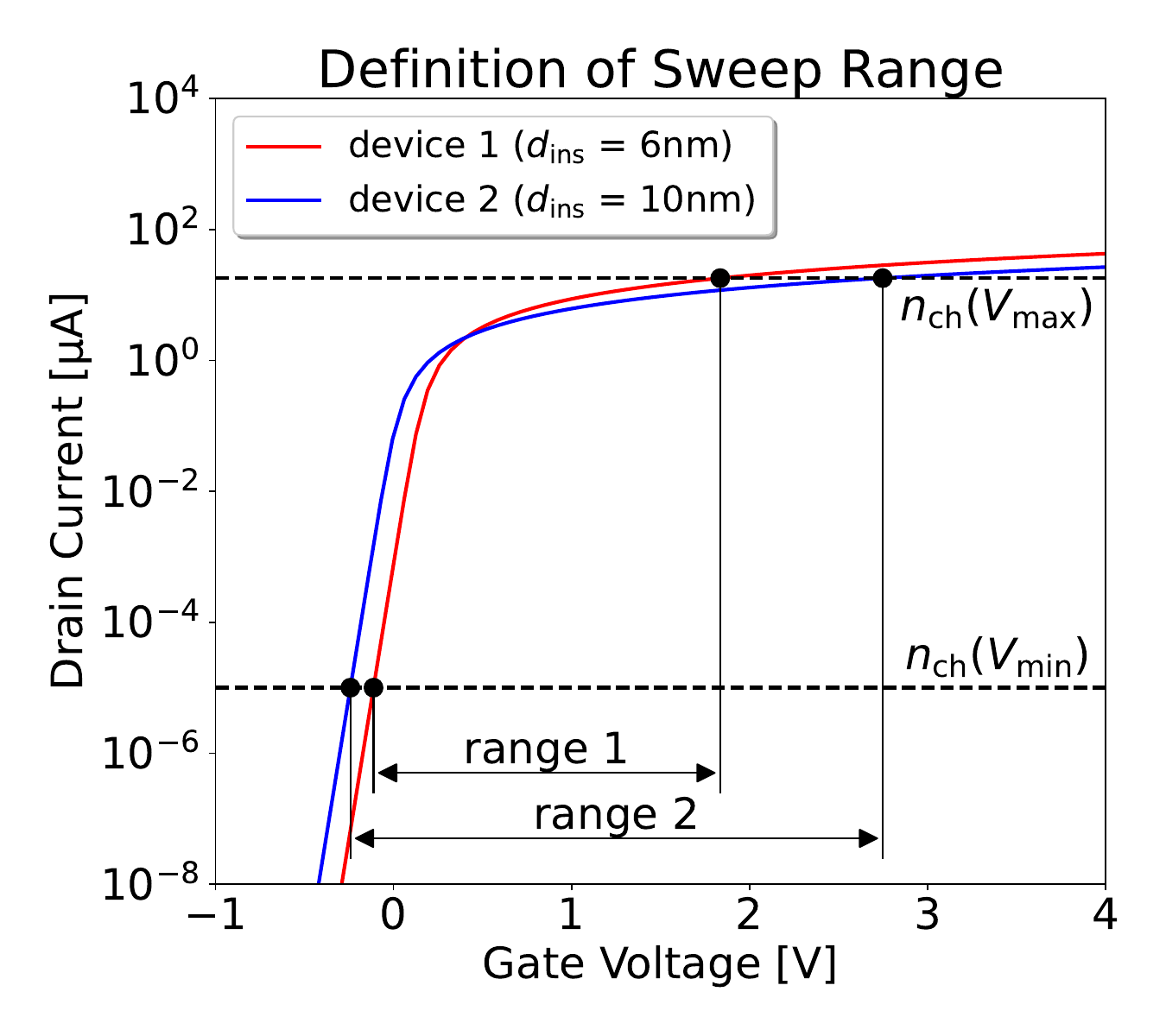}
    \end{minipage}};
    \node[draw=none] at (-2.0,2.4) {\textbf{(c)}};
    \end{tikzpicture}
\end{subfigure}
\begin{subfigure}[b]{.333\linewidth}
    \begin{tikzpicture}
    \node[inner sep=0pt] at (0,0) {
      \begin{minipage}{1.0\textwidth}
        \includegraphics[width=\textwidth]{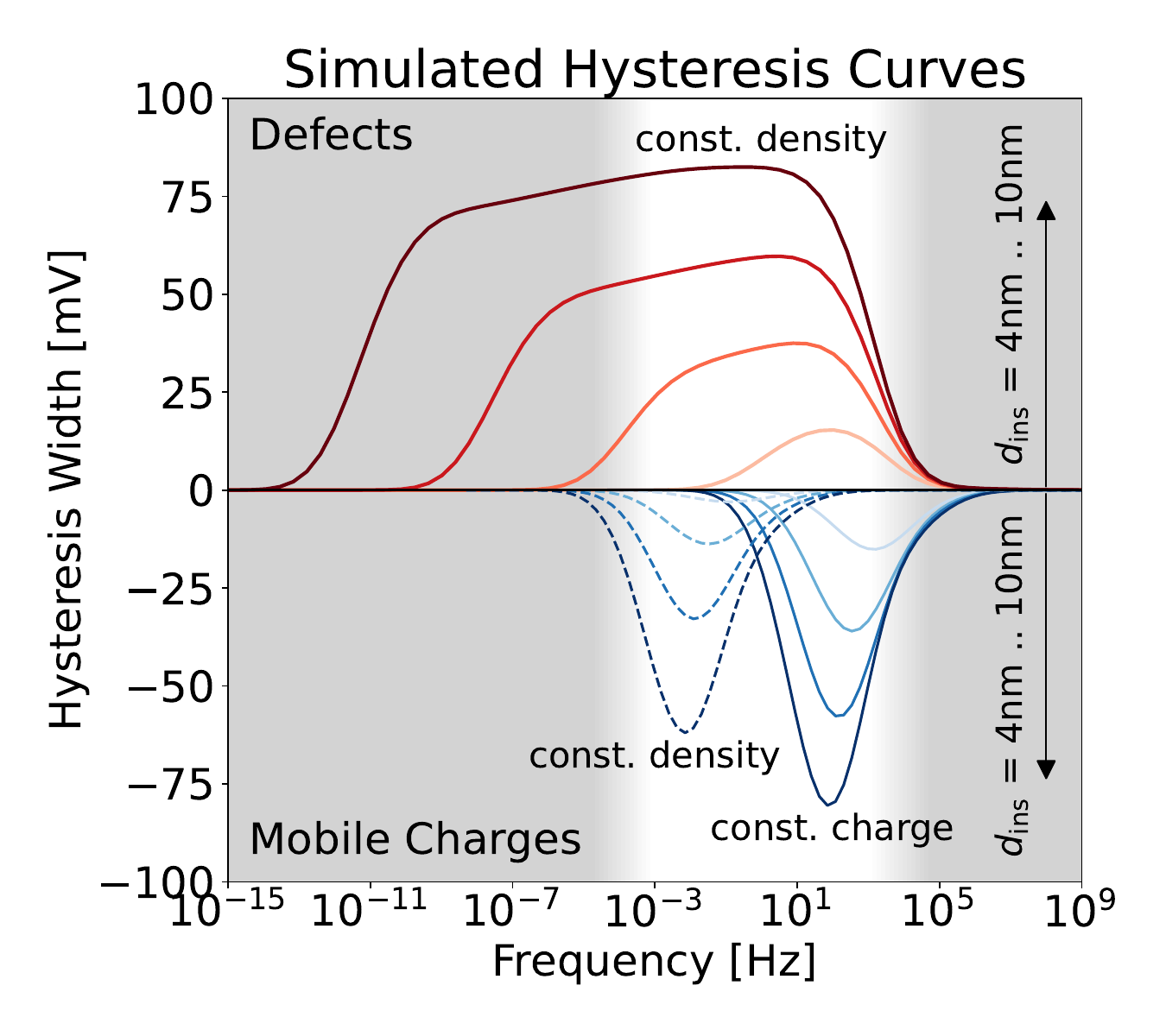}
    \end{minipage}};
    \node[draw=none] at (-2.0,2.4) {\textbf{(d)}};
    \end{tikzpicture}
\end{subfigure}
\begin{subfigure}[b]{.333\linewidth}
    \begin{tikzpicture}
    \node[inner sep=0pt] at (0,0) {
      \begin{minipage}{1.0\textwidth}
        \includegraphics[width=\textwidth]{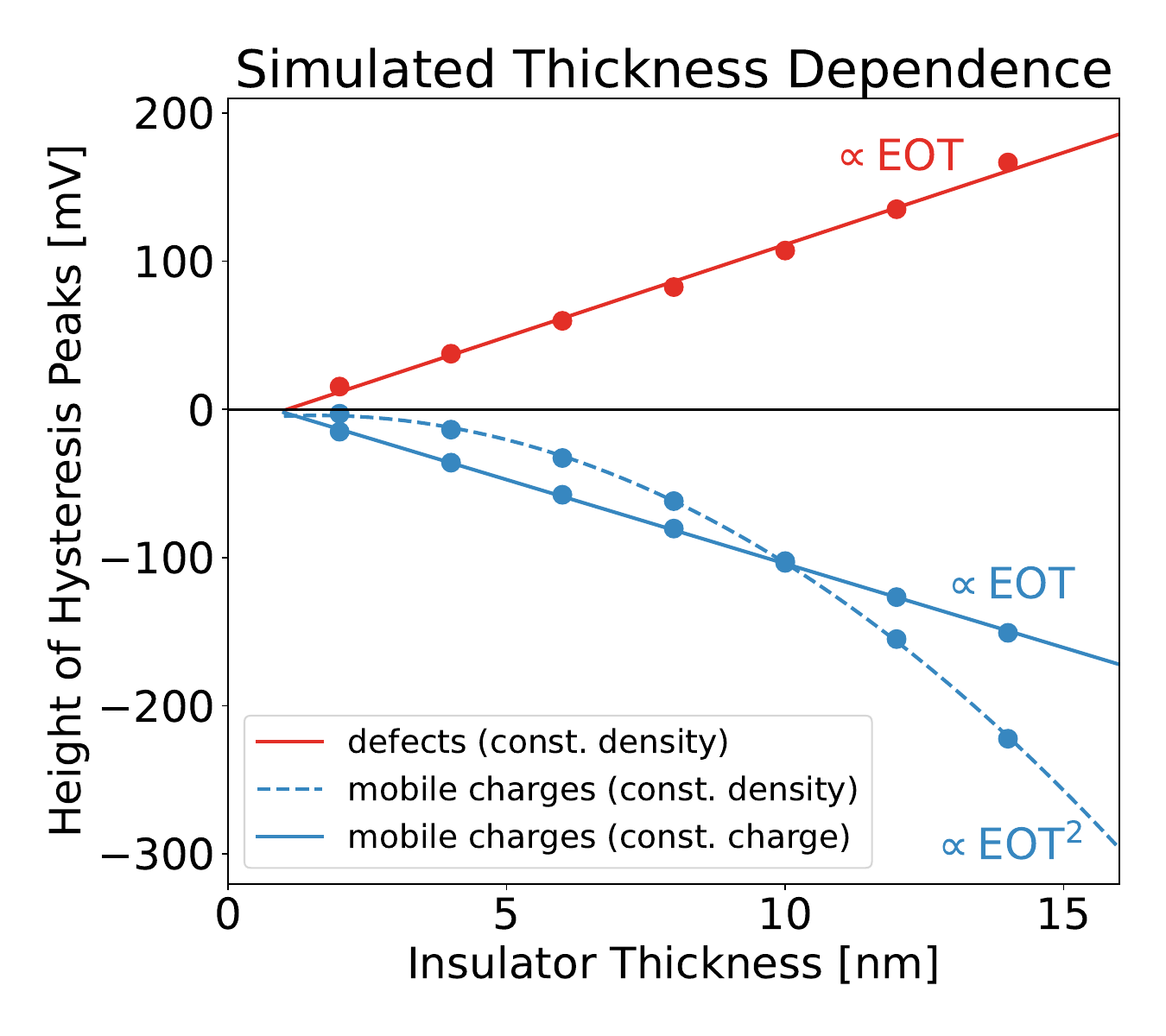}
    \end{minipage}};
    \node[draw=none] at (-2.0,2.4) {\textbf{(e)}};
    \end{tikzpicture}
\end{subfigure}
\begin{subfigure}[b]{.333\linewidth}
    \begin{tikzpicture}
    \node[inner sep=0pt] at (0,0) {
      \begin{minipage}{1.0\textwidth}
        \includegraphics[width=\textwidth]{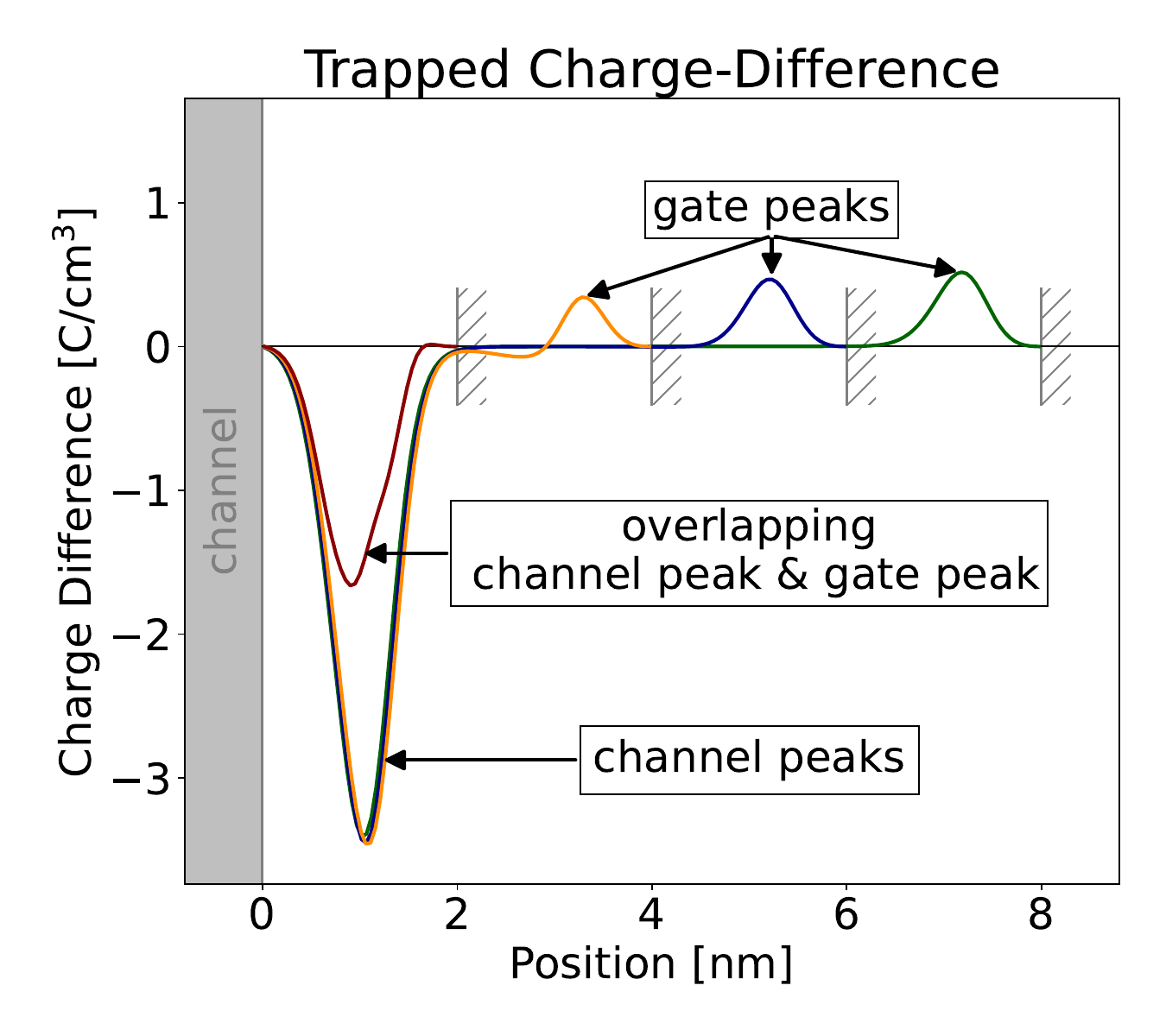}    \end{minipage}};
    \node[draw=none] at (-2.0,2.4) {\textbf{(f)}};
    \end{tikzpicture}
\end{subfigure}
\begin{subfigure}[b]{.333\linewidth}
    \begin{tikzpicture}
    \node[inner sep=0pt] at (0,0) {
      \begin{minipage}{1.0\textwidth}
        \includegraphics[width=\textwidth]{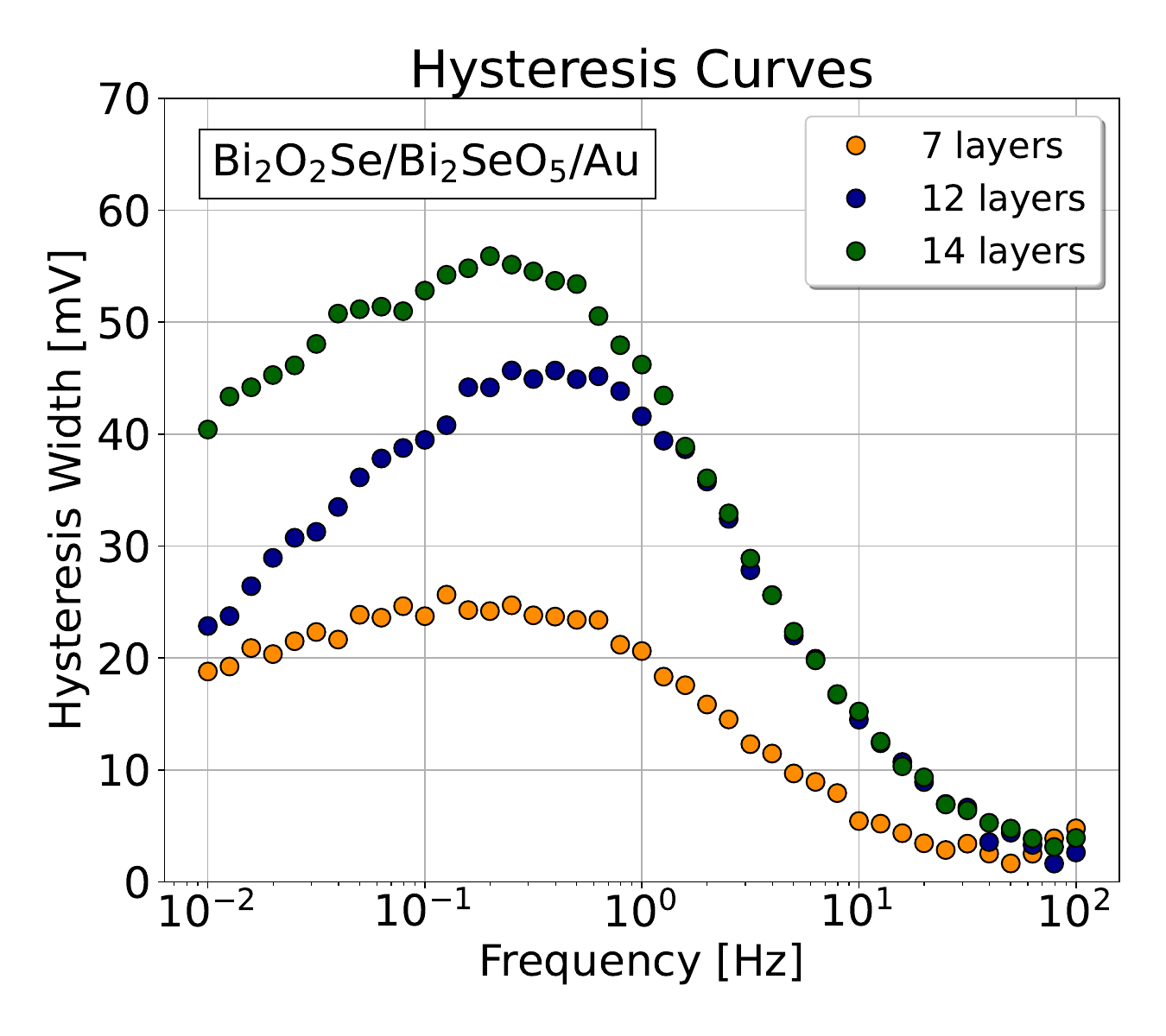}
    \end{minipage}};
    \node[draw=none] at (-2.0,2.4) {\textbf{(g)}};
    \end{tikzpicture}
\end{subfigure}
\begin{subfigure}[b]{.333\linewidth}
    \begin{tikzpicture}
    \node[inner sep=0pt] at (0,0) {
      \begin{minipage}{1.0\textwidth}
        \includegraphics[width=\textwidth]{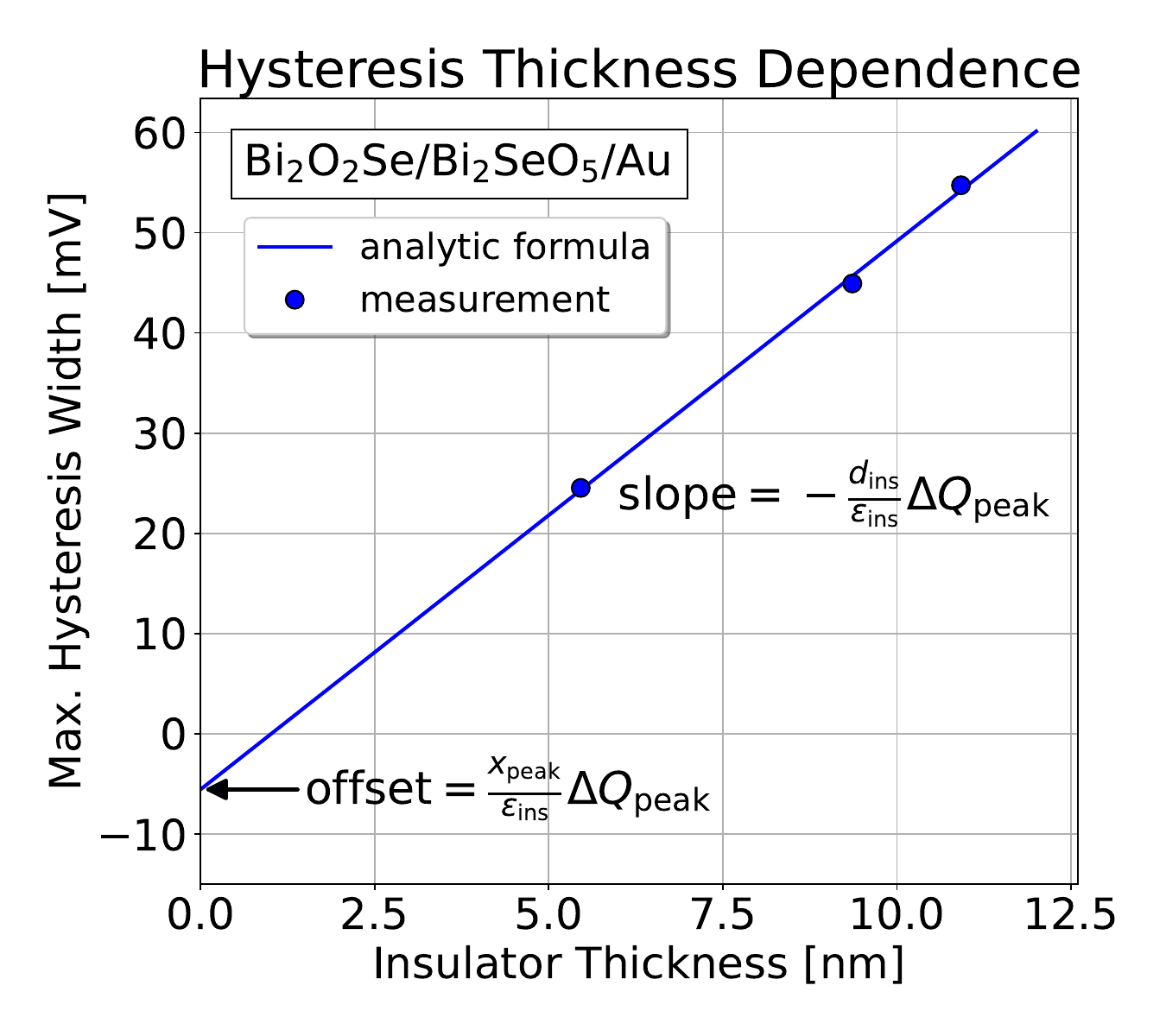}
    \end{minipage}};
    \node[draw=none] at (-2.0,2.4) {\textbf{(h)}};
    \end{tikzpicture}
\end{subfigure}
\begin{subfigure}[b]{.333\linewidth}
    \begin{tikzpicture}
    \node[inner sep=0pt] at (0,0) {
      \begin{minipage}{1.0\textwidth}
        \includegraphics[width=\textwidth]{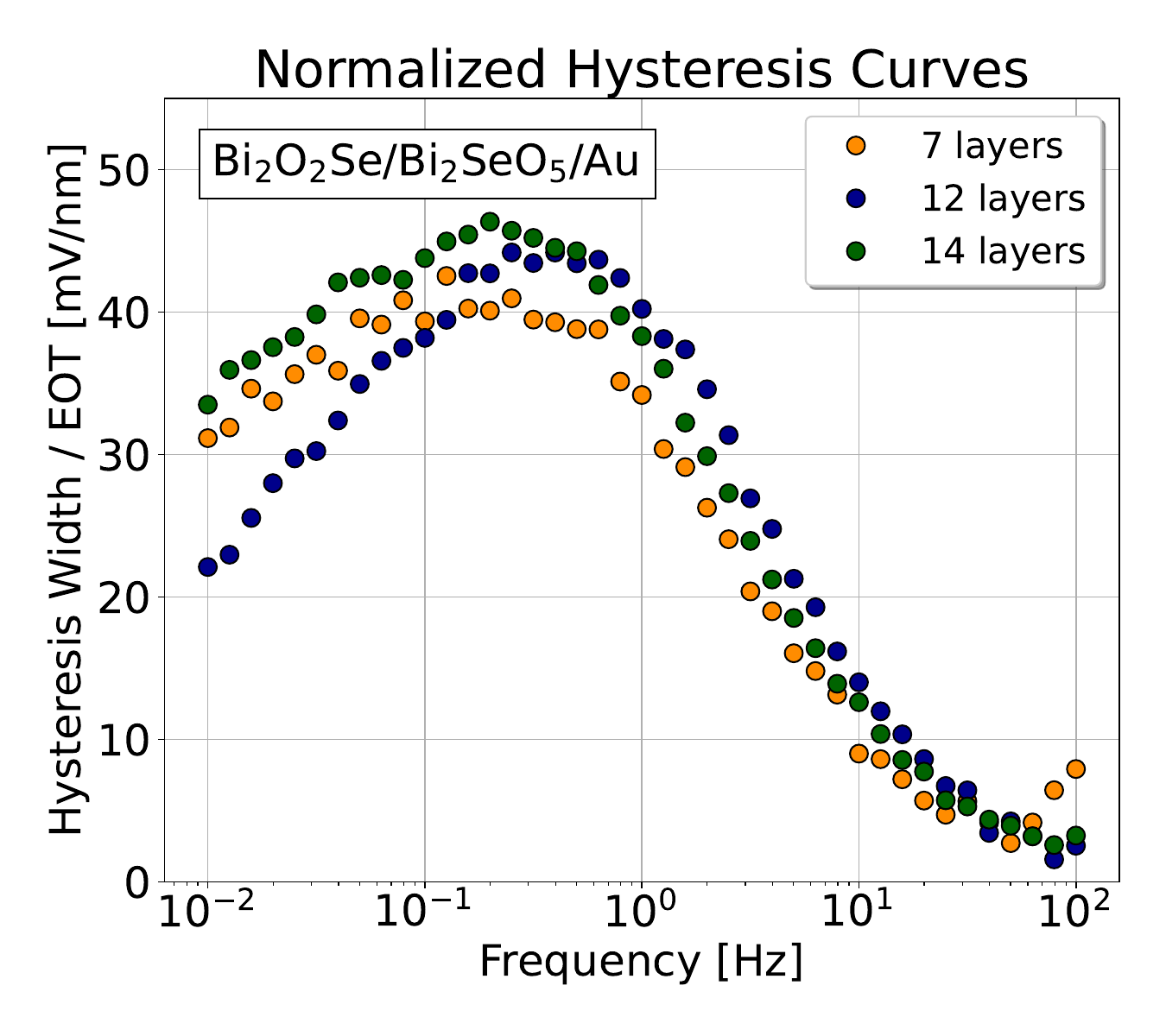}
    \end{minipage}};
    \node[draw=none] at (-2.0,2.4) {\textbf{(i)}};
    \end{tikzpicture}
\end{subfigure}
\caption{Superimposed off-state \textbf{(a)} and on-state \textbf{(b)} of two devices with different insulator thicknesses. \textbf{(c)} Standardized sweep range of hysteresis measurements, ensuring the alignment of the band diagrams of the devices in both the off-state and on-state. \textbf{(d)} Hysteresis curves simulated for varying insulator thicknesses \(d_\mathrm{ins}~\in~[\qty{2}{\nano\meter}, \qty{4}{\nano\meter}, \qty{6}{\nano\meter}, \qty{8}{\nano\meter}]\)). scenario~1 (red): hysteresis due to charge trapping by defects (\(E_\mathrm{T} = (-4.5 \pm 0.2) \qty{}{\electronvolt}\), \(E_\mathrm{R} = (2.5 \pm 0.2) \qty{}{\electronvolt}\), \(R = 1\)), scenario~2 (blue,~solid): hysteresis due to mobile charges with constant total charge (\(E_\mathrm{a} = \qty{0.55}{\electronvolt}\), \(D_\mathrm{0} = \qty{1.0E-10}{\meter^2\second^{-1}}\)), scenario~3 (blue,~dashed): hysteresis due to intrinsic mobile charges with constant average density (\(E_\mathrm{a} = \qty{0.30}{\electronvolt}\), \(D_\mathrm{0} = \qty{1.0E-10}{\meter^2\second}\)). \textbf{(e} Corresponding height of the hysteresis peaks as a function of thickness. \textbf{(f)} Charge difference due to  charge trapping in devices of varying insulator thicknesses during cyclo-stationary hysteresis sweeps with a frequency of \qty{1}{\hertz}. Provided that the sweep range is adapted to the EOT, the same defects are probed near the channel (or gate), regardless of the
insulator thickness. \textbf{(g)} Unscaled hysteresis curves measured on a device with a \ce{Bi2O2Se}/\ce{Bi2SeO5}/\ce{Au} gate stack \cite{Zhang2022} for 3 different insulator thicknesses (7 layers: \(d_\mathrm{ins} \approx \qty{5.6}{\nano\meter} \), 12 layers: \(d_\mathrm{ins} \approx \qty{9.6}{\nano\meter} \), 14 layers: \(d_\mathrm{ins} \approx \qty{11}{\nano\meter} \)). \textbf{(h)} Corresponding linear relationship between maximum hysteresis width and insulator thickness. \textbf{(h)} Corresponding normalized hysteresis curves.
} 
\label{fig:a4:1}
\end{figure}

Having discussed the main mechanisms contributing to hysteresis, we now focus on the standardization and normalization of hysteresis measurements to establish a suitable metric for device stability. We assume that in an optimized and stable technology the gate insulator consists of a high-quality dielectric, so that we can omit accidental ferroelectricity in the following discussion. The first observation is that the maximum hysteresis width \(\Delta V_\mathrm{H}^\mathrm{max}\) scales with EOT \cite{Park2017}. 
Since the experimental screening of a new semiconductor/insulator combination is typically conducted by various research groups using samples with different gate insulator thicknesses (and thus different EOT), \( \Delta V_\mathrm{H}^\mathrm{max} \) must be normalized for meaningful comparisons. \\

However, to properly normalize the hysteresis curves of devices with the same semiconductor/insulator combination but varying insulator thicknesses, it is essential that defects and mobile ions exhibit a similar temporal evolution in the devices under consideration. For this purpose, the sweep range \( [V_\mathrm{min}, V_\mathrm{max}] \) must be adjusted for each device to ensure that the band diagrams of the devices align in both the off- and on-state, as shown in \fig{fig:a4:1}{a} and \fig{fig:a4:1} {b}. This special condition is discussed in more detail in SI~Sec.~\ref{secA5} and can be approximately achieved by keeping the normalized voltage overdrive \(  (V_\mathrm{max} - V_\mathrm{th}') / \mathrm{EOT}\) as well as the on-off ratio \(I_\mathrm{d}(V_\mathrm{max})/ I_\mathrm{d}(V_\mathrm{min})  \) constant across all devices. These two conditions uniquely define the minimum and maximum voltage for each device and result in a relatively simple measurement rule:

\begin{tcolorbox}[colback=blue!5!white,colframe=black, ]

\begin{equation}\label{eq:sec4:1}
\begin{aligned}
\frac{V_\mathrm{max} - V_\mathrm{th}'}{\mathrm{EOT}} = \qty{3.5}{\mega\volt\centi\meter^{-1}} \qquad  \frac{I_\mathrm{d}(V_\mathrm{max})}{I_\mathrm{d}(V_\mathrm{min})} = \qty{E6}{}
\end{aligned}
\end{equation}
We recommend using a normalized voltage overdrive of \( \qty{3.5}{\mega\volt\centi\meter^{-1}}\) and an on-off ratio of \(\qty{E6}{} \) for standardized hysteresis measurements. This range reflects a realistic application scenario for 2D-MOSFETs, while being still applicable to most prototype devices.

\end{tcolorbox}
The proposed measurement procedure is illustrated in \fig{fig:a4:1}{c} and requires applying a significantly larger sweep range to a device with a high EOT compared to one with a low EOT. It is important to note that hysteresis measurements should be performed with a fixed current range on the SMU to ensure control over the sweep frequency. However, this method often leads to the minimum current, \(I_\mathrm{d}(V_\mathrm{min})\), falling below the measurement resolution of the selected current range. If \(I_\mathrm{d}(V_\mathrm{min})\) cannot be measured due to this limitation, we suggest determining \(V_\mathrm{min}\) through extrapolation of the subthreshold slope.

\fig{fig:a4:1}{d} displays simulated hysteresis curves for devices with varying insulator thicknesses, where the sweep range was adjusted as described above. The figure displays three distinct scenarios: In the first scenario (red), defects with a constant density were placed across the insulators of all devices. In the second scenario (blue, solid), mobile charges with a constant total charge \(Q_\mathrm{ins}\) were placed in the insulator of each device to simulate the external surface contamination by mobile ions. In the third scenario (blue, dashed), mobile charges with a constant average density \( \bar{\rho}_\mathrm{ins} = Q_\mathrm{ins} / d_\mathrm{ins}\) were placed in the insulators of each device, to simulate intrinsic mobile charges like charged oxygen vacancies. This figure again highlights the need for normalization to enable meaningful comparisons between devices with different insulator thicknesses, since the hysteresis scales with EOT. \fig{fig:a4:1}{e} shows the corresponding height of the hysteresis peaks as a function of the insulator thickness. The figure reveals that the hysteresis caused by defects scales linearly with EOT. Moreover, the hysteresis due to mobile charges scales linearly with EOT when the total charge is constant and quadratically with EOT when the average charge density is constant. These scaling laws are a direct consequence of the fact that mobile ions and defects behave similarly in all devices due to the adaptive sweep range and can be understood by the following considerations: \\

The hysteresis caused by mobile charges is determined by \( \Delta x_\mathrm{ins} =  x_\mathrm{ins}(t_\mathrm{down}) - x_\mathrm{ins}(t_\mathrm{up}) \), i.e. the shift of the charge center between up- and down-sweep (see Eq.~\ref{sec3:eq3}). Mobile charges strive for a mostly uniform distribution in the off-state and a distribution that is strongly localized at the channel or gate interface in the on-state. Consequently, the maximum displacement of the charge center is given in good approximation by \( ~d_\mathrm{ins} / 2 \) if the insulator is sufficiently thick. As a result, the maximum hysteresis exhibits the following scaling behavior:

\begin{equation}\label{eq:sec4:2}
    \Delta V_\mathrm{H, max} = \max \left|\frac{Q_\mathrm{ins}}{C_\mathrm{ins}} \frac{\Delta x_\mathrm{ins}}{d_\mathrm{ins}} \right| \approx
\begin{cases} 
        \, \mathrm{EOT} \, \frac{|Q_\mathrm{ins}|}{2 \varepsilon_\mathrm{\ce{SiO2}}},
    & \quad \text{if} \quad Q_\mathrm{ins} = \mathrm{const} \\ 
       \, \mathrm{EOT}^2 \, \frac{|\bar{\rho}_\mathrm{ins}| \varepsilon_\mathrm{ins}}{{2 \varepsilon_\mathrm{\ce{SiO2}}}^2} ,
    & \quad \text{if} \quad  \bar{\rho}_\mathrm{ins} = Q_\mathrm{ins} / d_\mathrm{ins} = \mathrm{const} \\
\end{cases}
\end{equation}
This aligns precisely with the behavior depicted in \fig{fig:a4:1}{e}. Conversely, plotting \( \Delta V_\mathrm{H, max} \)\ as a function of thickness can reveal whether \( \Delta V_\mathrm{H, max} \)\ arises from external contamination with mobile charges (\(\propto \mathrm{EOT}\)) or from intrinsic mobile charges (\(\propto \mathrm{EOT}^2\)). \\

Moreover, the hysteresis caused by charge trapping due to defects is determined by \( \Delta Q_\mathrm{ins} =  Q_\mathrm{ins}(t_\mathrm{down}) - Q_\mathrm{ins}(t_\mathrm{up}) \), i.e. the trapped charge difference between up- and down-sweep (see Eq.~\ref{sec3:eq3}). \fig{fig:a4:1}{f} illustrates the trapped charge difference within the devices of varying insulator thicknesses. For sufficiently thick insulators, defects primarily interact with either the channel or the gate, resulting in the formation of a separate channel-peak and gate-peak. In this case, the charge center \(x_\mathrm{ins}\) and the charge difference \(\Delta Q_\mathrm{ins}\) associated with the dominant channel peak become independent of the sample's EOT. Consequently, when the hysteresis is dominated by defects near the channel (\(x_\mathrm{ins} \ll d_\mathrm{ins}\)), the maximum hysteresis is given by (see Eq.~\ref{sec3:eq3}):
\begin{equation}\label{eq:sec4:3}
    \Delta V_\mathrm{H}^\mathrm{max} = \max \left| \frac{\Delta Q_\mathrm{ins}}{C_\mathrm{ins}}  \left(1 - \frac{x_\mathrm{ins}}{d_\mathrm{ins}}\right) \right| \approx \mathrm{EOT} \frac{|\Delta Q_\mathrm{ins}|}{\varepsilon_\mathrm{\ce{SiO2}}},
\end{equation}
which corresponds exactly to the behavior displayed in \fig{fig:a4:1}{e}. Experimental validation of this prediction is presented in \fig{fig:a4:1}{g} and \fig{fig:a4:1}{h}. Specifically, \fig{fig:a4:1}{g} illustrates the hysteresis curves for devices with a \ce{Bi2O2Se}/\ce{Bi2SeO5}/\ce{Au} gate stack, with  different insulator thicknesses. Meanwhile,  \fig{fig:a4:1}{h} shows the corresponding \( \Delta V_\mathrm{H}^\mathrm{max}\) plotted against insulator thickness, clearly illustrating the linear relationship between \(\Delta V_\mathrm{H}^\mathrm{max}\) and EOT. Finally, we point out that the height of the hysteresis peaks remains nearly constant when the hysteresis curves are normalized by EOT as shown in \fig{fig:a4:1}{i}. \\

Based on these considerations, we propose to normalize the hysteresis curves by \(\mathrm{EOT} \) when the hysteresis is dominated by defects near the channel or by externally introduced mobile charges (e.g., surface contamination) whose total charge does not scale with insulator thickness. Conversely, if hysteresis is dominated by intrinsic mobile charges whose total charge scales linearly with insulator thickness, we suggest to normalize the hysteresis curves by \(\mathrm{EOT}^2 \). This normalization ensures that the height of the hysteresis curves becomes independent of insulator thickness, enabling meaningful comparisons of measurement data across devices with varying insulator thicknesses. Furthermore, the corresponding metrics \( \Delta V_\mathrm{EOT} = \Delta V_\mathrm{H}^\mathrm{max} / \mathrm{EOT} \) and \( \Delta V_\mathrm{EOT^2} = \Delta V_\mathrm{H}^\mathrm{max} / \mathrm{EOT^2} \) are to a good approximation independent of the insulator thickness (Eq.~\ref{eq:sec4:2} and Eq.~\ref{eq:sec4:3}) and thus serve as a reliable measure of device stability. Finally, the proposed measurement scheme can be used to estimate the absolute hysteresis for devices with scaled insulators, based on large EOT prototype devices fabricated during early development stages. For instance, if the normalized hysteresis width is \( \Delta V_\mathrm{EOT} = \qty{10}{\milli\volt\per\nano\meter} \), a device with an EOT of \qty{2}{\nano\meter} will show a maximum hysteresis of \(\Delta V_\mathrm{H}^\mathrm{max} \approx \qty{20}{\milli\volt}\), while a device with an EOT of \qty{0.5}{\nano\meter} will exhibit a maximum hysteresis of \(\Delta V_\mathrm{H}^\mathrm{max} \approx \qty{5}{\milli\volt}\) within the standard measurement window.

\section{Conclusions}\label{sec7}

We investigated the primary mechanisms contributing to hysteresis in 2D-MOSFETs, focusing on charge trapping, mobile charge drift, and ferroelectricity. We demonstrated that these hysteresis mechanisms can be distinguished based on their physical signatures, facilitating the targeted development of effective countermeasures. Our study demonstrates that, in n-channel devices, charge trapping near the channel leads to clockwise hysteresis, whereas defects near the gate, mobile charges, and ferroelectric gate insulators produce counterclockwise hysteresis (in p-channel devices, the hysteresis orientation is reversed due to the polarity of the applied voltages). \\

We emphasized the importance of standardized hysteresis measurements, noting that currently published data is often collected under arbitrary conditions, which makes cross-device comparisons nearly impossible. To solve this issue, we proposed a standardized measurement scheme that adjusts the sweep range according to the insulator thickness and normalizes the obtained data using the equivalent oxide thickness (EOT). This approach ensures comparability across devices with varying insulator thicknesses and, most importantly, facilitates the extrapolation of hysteresis data from large EOT prototype devices fabricated during early development stages to scaled devices with small EOT. \\

These advancements establish hysteresis as a reliable diagnostic tool for evaluating device quality and stability. These insights are expected to facilitate the development of more stable and reliable devices that is urgently needed for advancing 2D-material based devices towards stacked 2D-MOSFETs at ultra-scaled technology nodes.

\section{Methods}\label{sec8}

The simulations presented in this work, including transfer characteristics, hysteresis curves caused by charge trapping, mobile charge drift, and ferroelectric gate insulators, were conducted using the Python-based framework Comphy (Compact Physics) \cite{RZEPA201849, WALDHOER2023115004}, jointly developed by \href{https://www.iue.tuwien.ac.at/}{TU Wien} and \href{https://www.imec-int.com/en}{imec}. The source code of the extended version (Comphy V4.0) will be made publicly accessible at \href{https://comphy.eu/}{https://comphy.eu/} after publication of this manuscript. \\
Comphy is a one-dimensional simulation tool that treats the electric field in the channel direction as a small perturbation compared to the field in the direction perpendicular to the channel. This approximation is valid for most experimental prototype devices, where the dimensions \( W \) and \( L \) are typically on the scale of micrometers, while the gate dielectric thickness measures only a few nanometers. As a result, the electrostatics of the device can be effectively modeled along a one-dimensional cross-section in the direction perpendicular to the channel. The framework allows the simulation of various dynamic processes along this direction, such as charge trapping at insulator defects, mobile charge drift, and the switching behavior of ferroelectric gate insulators. Additional details on these individual models are provided in the supplementary information.

\backmatter

\bmhead{Acknowledgements}
We would like to express our gratitude to Huawei Technologies R\&D Belgium for their generous financial support, which was instrumental in advancing this research. Furthermore this work was supported by the European Research Council under Grant agreement no. 101055379 F2GO. Moreover, we acknowledge support from the Singapore Ministry of Education (MOE) Academic Research Fund Tier 3 grant (MOE-MOET32023-0003) "Quantum Geometric Advantage". Furthermore, invaluable discussions with Ben Kaczer and Jacopo Franco (imec) are gratefully acknowledged. 

\clearpage

\setcounter{section}{0}
\setcounter{equation}{0}
\setcounter{figure}{0}
\section{Supplementary Information}

\subsection{Drain Current Model}\label{secA1}


\begin{figure}[hbt!]
\begin{subfigure}[b]{.333\linewidth}
    \begin{tikzpicture}
    \node[inner sep=0pt] at (0,0) {
      \begin{minipage}{1.0\textwidth}
        \includegraphics[width=\textwidth]{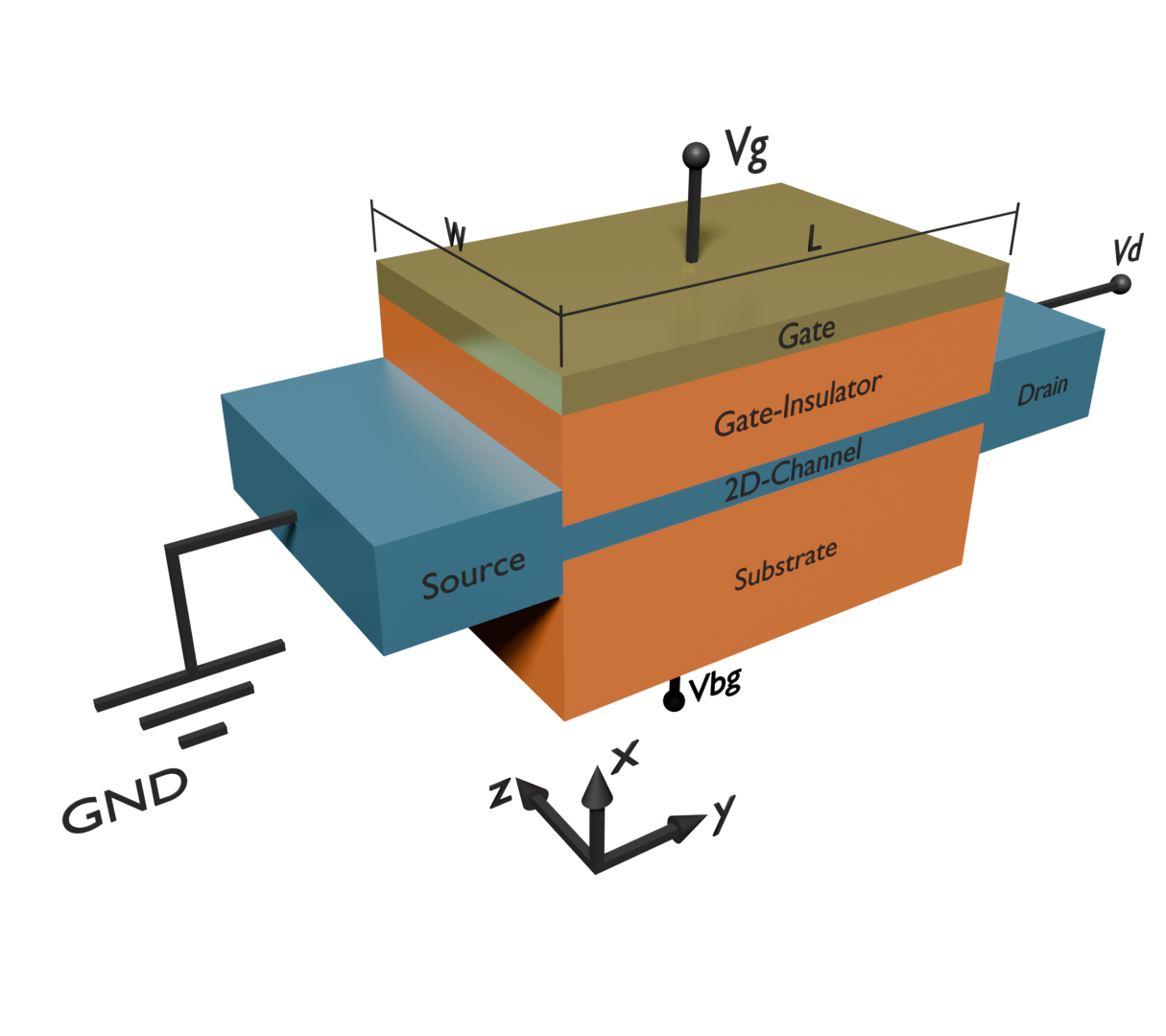}
    \end{minipage}};
    \node[draw=none] at (-2.0,2.5) {\textbf{(b)}};
    \end{tikzpicture}
\end{subfigure}
\begin{subfigure}[b]{.333\linewidth}
    \begin{tikzpicture}
    \node[inner sep=0pt] at (0,0) {
      \begin{minipage}{1.0\textwidth}
        \includegraphics[width=\textwidth]{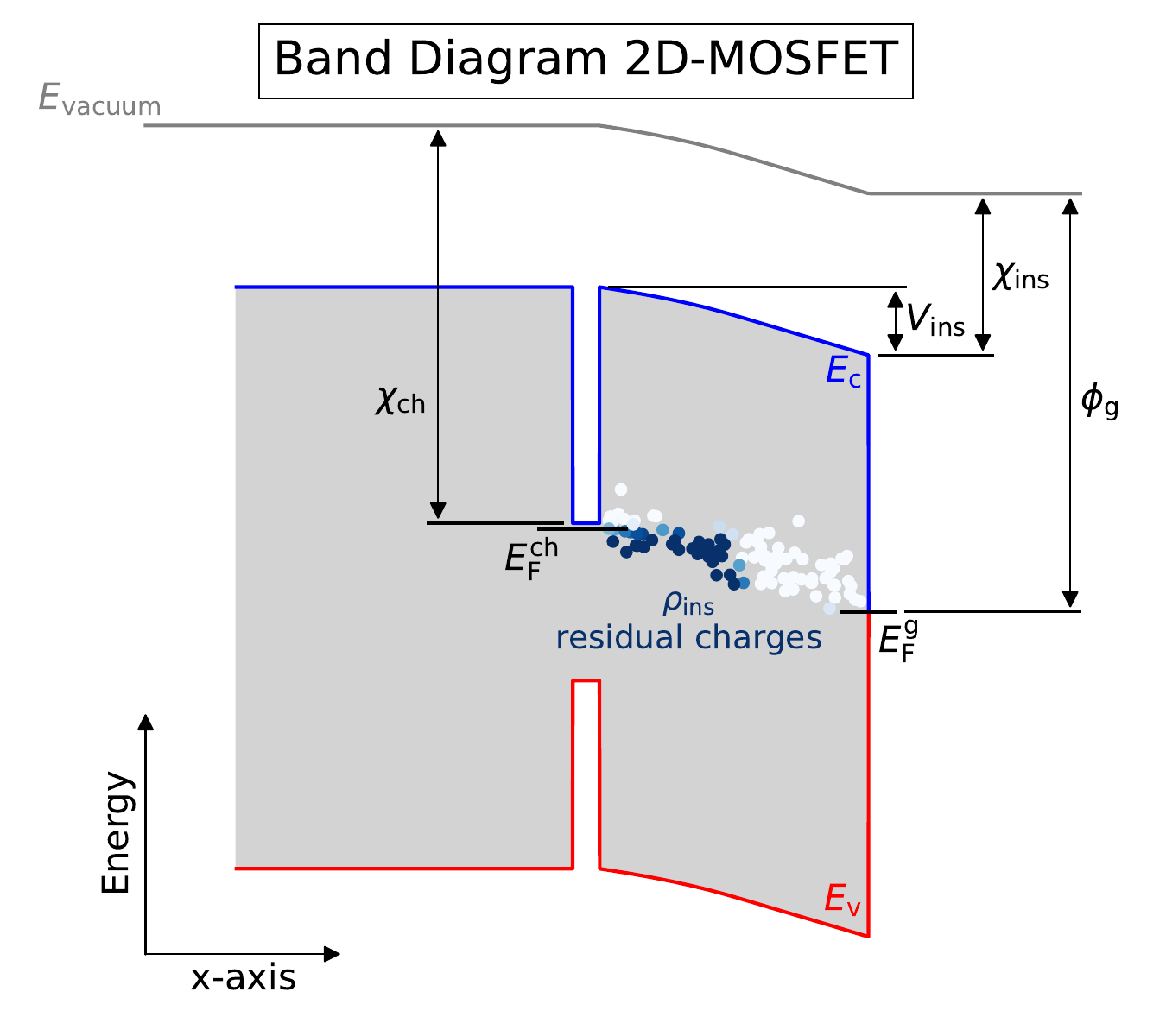}
    \end{minipage}};
    \node[draw=none] at (-2.0,2.5) {\textbf{(b)}};
    \end{tikzpicture}
\end{subfigure}
\begin{subfigure}[b]{.333\linewidth}
    \begin{tikzpicture}
    \node[inner sep=0pt] at (0,0) {
      \begin{minipage}{1.0\textwidth}
        \includegraphics[width=\textwidth]{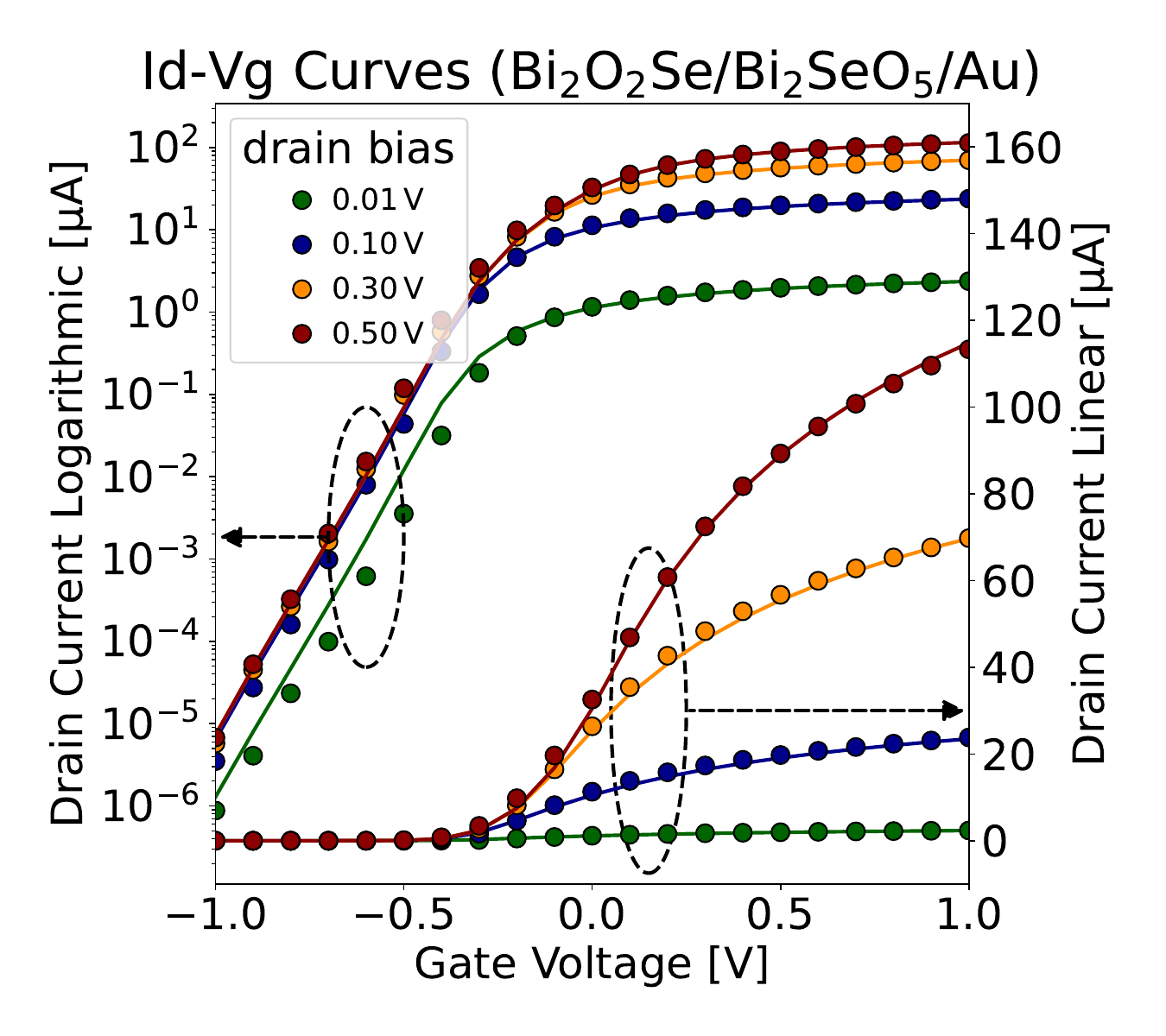}
    \end{minipage}};
    \node[draw=none] at (-2.0,2.5) {\textbf{(c)}};
    \end{tikzpicture}
\end{subfigure}
\caption{\textbf{(a)} Typical device structure of a top-gated MOSFET with a 2D-channel. \textbf{(b)} Corresponding band diagram, highlighting the band bending caused by residual charges in the gate insulator. \textbf{(c)} Measured (circles) and simulated \IDVG curves (lines) of a device with a \ce{Bi2O2Se}/\ce{Bi2SeO5}/\ce{Au} gate stack \cite{Zhang2022}, demonstrating excellent agreement between the drain current model and the measurement data. The simulation considers the impact of fast interface defects with a density of \(\qty{1E13}{\centi\meter^{-2}}\), fixed insulator charges with a density of \(\qty{5E19}{\centi\meter^{-3}}\) and  ohmic contact resistances of \qty{9}{\kilo\ohm\micro\meter}. }
\label{fig:a1:1}
\end{figure}

In the following section we derive a drain current model for the single-gated 2D-MOSFET shown in \fig{fig:a1:1}{a}, where we assume that the gate dielectric contains a certain density \( \rho_\mathrm{ins} \) of residual charges, causing a band bending  in the gate insulator. The width of the device is denoted by \(W\), the length of the channel by \(L\) and the thickness of the gate insulator by \(d_\mathrm{ins}\). During operation, the voltage \( V_\mathrm{g} \) is applied to the gate  and  \( V_\mathrm{d} \) to the drain contact, while the source contact serves as common ground. We assume that the electric field in the \(y\)-direction is negligibly small compared to the electric field in the \(x\)-direction and can be treated as a small perturbation. For experimental prototype devices which typically have dimensions \( W \) and \( L \) on the scale of a few micrometers, while the gate dielectric thickness is only a few nanometers, this approximation generally holds. This approximation allows us to describe the electrostatics in the device on a one-dimensional cut along the \(x\)-direction. \\

\fig{fig:a1:1}{b} displays the band diagram along the described cut ranging from the substrate to the gate with the charges in the gate insulator highlighted in blue. \(E_\mathrm{C}^\mathrm{ch} \) represents the conduction band minimum of the channel, \(E_\mathrm{V}^\mathrm{ch} \) the valence band maximum of the channel, \(\phi_\mathrm{g}\) the work function of the gate, \( \chi_\mathrm{ch} \) the electron affinity of the channel, and \( V_\mathrm{ins} \) the voltage drop across the gate insulator. Moreover, \( E_\mathrm{F}^\mathrm{ch}\) denotes the position dependent electron quasi Fermi level in the 2D-channel, \( E_\mathrm{F}^\mathrm{g} \) the Fermi level in the gate, \(E_\mathrm{F}^\mathrm{d} \) the Fermi level in the drain contact and \( E_\mathrm{F}^\mathrm{s}\) the Fermi level in the source contact. In a first step we express the gate voltage \( V_\mathrm{g} \) and the drain voltage \( V_\mathrm{d} \) in terms of the respective Fermi levels:
\begin{equation}\label{eq:secA1:1}
 V_\mathrm{g} = - \frac{E_\mathrm{F}^\mathrm{g} -E_\mathrm{F}^\mathrm{s}}{q}, \qquad V_\mathrm{d} = -\frac{E_\mathrm{F}^\mathrm{d} - E_\mathrm{F}^\mathrm{s}}{q}, 
\end{equation}
Furthermore we introduce abbreviations for the ideal threshold voltage  \( V_\mathrm{th} \), the Fermi potential \( V_\mathrm{C}(y) \) relative to the channel's conduction band and the Fermi potential  \( V_\mathrm{F}(y) \) relative to the source contact:
\begin{equation}\label{eq:secA1:2}
    V_\mathrm{th} = \frac{\phi_\mathrm{g} -\chi_\mathrm{ch}}{q}, \qquad   V_\mathrm{C}(y) = - \frac{E_\mathrm{F}^\mathrm{ch}(y) - E_\mathrm{C}^\mathrm{ch}(y)}{q}, \qquad V_\mathrm{F}(y) = -\frac{E_\mathrm{F}^\mathrm{ch}(y) - E_\mathrm{F}^\mathrm{s}}{q}.
\end{equation}
Based on the band diagram shown in \fig{fig:a1:1}{b}, the potential drop across the gate insulator can be expressed as follows:
\begin{equation}\label{eq:secA1:3}
    V_\mathrm{ins}(y)=(V_\mathrm{g}-V_\mathrm{th} + V_\mathrm{C}(y) - V_\mathrm{F}(y)).
\end{equation}
Moreover, we assume that the charge distribution is uniformly distributed in the \(y\)-\(z\) plane and hence only depends on the \(x\)-position (\(\rho_\mathrm{ins}(x,y,z) \approx \rho_\mathrm{ins}(x) \)). This approximation (also known as charge sheet approximation) allows the Poisson equation in the gate insulator to be solved on the one-dimensional cut, leading to an explicit expression for \(V_\mathrm{ins}\):
\begin{equation}\label{eq:secA1:4}
	 V_\mathrm{ins} = \frac{Q_\mathrm{g} + Q_\mathrm{ins}}{ C_\mathrm{ins}} + \Delta V_\mathrm{ins},
\end{equation}
\begin{equation}\label{eq:secA1:5}
     \Delta V_\mathrm{ins} = - \frac{Q_\mathrm{ins}}{C_\mathrm{ins}}  \left(1 - \frac{x_\mathrm{ins}}{d_\mathrm{ins}}\right), \qquad 
      Q_\mathrm{ins} =  \int_{0}^{d_\mathrm{ins}} \rho_\mathrm{ins}(x)  \,\mathrm{d}x, \qquad      x_\mathrm{ins} = \frac{1}{Q_\mathrm{ins}} \int_{0}^{d_\mathrm{ins}} \rho_\mathrm{ins}(x) x  \,\mathrm{d}x, 
\end{equation}
Here \( Q_\mathrm{g} \) denotes the charge density at the gate, \( Q_\mathrm{ins} \) the charge density contained in the gate insulator and  \( C_\mathrm{ins} = \varepsilon_\mathrm{ins} / d_\mathrm{ins} \) the areal capacitance of the gate insulator. Furthermore \( \Delta V_\mathrm{ins} \) represents the potential shift caused by the charges in the gate insulator. By applying Gauss's law to an enclosure around the 2D-channel we find that the charge density in the channel is given by \(Q_\mathrm{ch} = - Q_\mathrm{g} + Q_\mathrm{ins}\), which allows us to combine Eq.~\ref{eq:secA1:1}-\ref{eq:secA1:5} to a single equation:
\begin{equation}\label{eq:secA1:6}
    Q_\mathrm{ch}(V_\mathrm{C}) = - C_\mathrm{ins} (V_\mathrm{g} - V_\mathrm{th} + V_\mathrm{C}(y) - V_\mathrm{F}(y) - \Delta V_\mathrm{ins})
\end{equation}
In general the total charge density \( Q_\mathrm{ch} \) in an n-type 2D-material consists of multiple contributions:  the electron density \(n_\mathrm{ch} \), the density of ionized donor-like defects \(N_\mathrm{d}^+ \) and the density of ionized acceptor-like defects \(N_\mathrm{a}^- \). In order to make use of Eq.~\ref{eq:secA1:5}, we need a physical model for the electron density in the channel as a function of the Fermi potential \(V_\mathrm{C} \). Here, we follow the approach of Marin et al. \cite{8288825} and Pasadas et al. \cite{pasadas_large-signal_2019} and describe the channel as 2D electron gas with the quantum capacitance:
\begin{equation}\label{eq:secA1:7}
\begin{aligned}
    C_\mathrm{q} =  g_\mathrm{s} g_\mathrm{v} \frac{m_\mathrm{eff} q^2}{2 \pi \hbar^2},
\end{aligned}
\end{equation}
where \(g_\mathrm{v}\) represents the valley degeneracy factor,  \(g_\mathrm{s}\) the spin degeneracy factor and \( m_\mathrm{eff}\) the effective mass associated with the parabolic energy dispersion of the electron gas. Based on this approach the total charge in the channel is given by:
\begin{equation}\label{eq:secA1:8}
\begin{aligned}
    Q_\mathrm{ch}(V_\mathrm{C})  = q \left( N_\mathrm{d}^+ - N_\mathrm{a}^-  - n_\mathrm{ch}(V_\mathrm{C}) \right), \qquad q n_\mathrm{ch}(V_\mathrm{C}) = V_\mathrm{T} C_\mathrm{q} \ln{ \left( 1 + \exp{\left(-\frac{V_\mathrm{C}}{ V_\mathrm{T}}\right)}\right)},
\end{aligned}
\end{equation}
where \(V_\mathrm{T} = k_\mathrm{B}T/q \) denotes the thermal voltage and \(q \) the elementary charge.  Finally, Eq.~\ref{eq:secA1:6} and Eq.~\ref{eq:secA1:8} can be combined into a single equation, to obtain an implicit relationship between the effective threshold voltage \(V_\mathrm{th}' \), the electron density \(n_\mathrm{ch}(y) \) and the Fermi potential \(V_\mathrm{F}(y)\) in the channel:
\begin{equation}\label{eq:secA1:9}
   \exp{\left( \frac{q n_\mathrm{ch}(y)}{V_\mathrm{T}C_\mathrm{ins}} \right)}  \left(    \exp{\left( \frac{ q n_\mathrm{ch} (y) }{ V_\mathrm{T} C_\mathrm{q}} \right)} -1  \right) = \exp{\left( \frac{V_\mathrm{g} - V_\mathrm{th}' - V_\mathrm{F}(y)}{ V_\mathrm{T}}\right) },
\end{equation}
\begin{equation}\label{eq:secA1:10}
V_\mathrm{th}' = V_\mathrm{th}  -  q(N_\mathrm{d}^+ - N_\mathrm{a}^-)/C_\mathrm{ins} + \Delta V_\mathrm{ins}
\end{equation}
In the off state, the left side of Eq.~\ref{eq:secA1:9} can be expanded in terms of \(n_\mathrm{ch}\) while retaining only the leading order term. Conversely, in the on state the ``1'' in the bracket of Eq.~\ref{eq:secA1:9} can be ignored. This approach leads to the following local approximations of \(n_\mathrm{ch}\):
\begin{equation}\label{eq:secA1:11}
    q n_\mathrm{ch} \quad = \quad
\begin{cases}
    \quad  C_\mathrm{q} V_\mathrm{T} \exp{\left( \frac{V_\mathrm{g} - V_\mathrm{th}' - V_\mathrm{F}}{ V_\mathrm{T}} \right)},  
    & \text{if} \quad V_\mathrm{g} \ll V_\mathrm{th}' \\
    \quad \left(\frac{1}{C_\mathrm{ins}} + \frac{1}{C_\mathrm{q}}\right)^{-1} \left( V_\mathrm{g} - V_\mathrm{th}' - V_\mathrm{F} \right), 
    & \text{if} \quad  V_\mathrm{g} \gg V_\mathrm{th}' \\
\end{cases}
\end{equation}
 Finally, the current density, based on the drift-diffusion model, is given by the expression \(j_\mathrm{d} = n_\mathrm{ch} \mu \, \mathrm{d}E_\mathrm{F}^\mathrm{ch} / \mathrm{d}y\), where \(\mu\) denotes the electron mobility of the 2D-material \cite{Sze2006}. To calculate the total current, we multiply the current density by the width of the device and integrate over the whole channel length, which allows the current to be expressed in terms of the known boundary values \(V_\mathrm{F}(0) = 0\) and \(V_\mathrm{F}(L) = V_\mathrm{d}\):
\begin{equation}\label{eq:secA1:12}
    I_\mathrm{d}(V_\mathrm{g}) = q \mu \frac{W}{L} \int_{0}^{L}  n_\mathrm{ch} \frac{1}{q}\frac{\mathrm{d}E_\mathrm{F}^\mathrm{ch}}{\mathrm{d}y} \mathrm{d}y = - q \mu \frac{W}{L} \int_{V_\mathrm{F}(0) = 0}^{V_\mathrm{F}(L) = V_\mathrm{d}}  n_\mathrm{ch}\mathrm{d}V_\mathrm{F}
\end{equation}
 
Unfortunately, this integral does not have an analytical solution due to the implicit nature of the relationship between \( n_\mathrm{ch}(y) \) and \( V_\mathrm{F}(y) \) given by Eq.~\ref{eq:secA1:9}. However, the integral can be easily computed numerically. Nevertheless, to enhance the understanding of the device behavior, we present approximate solutions for both the off-state and the on-state. For this purpose, we employ the local approximations for \(n_\mathrm{ch}\) given by Eq.~\ref{eq:secA1:11}, leading to the following local current expressions:

\begin{equation}\label{eq:secA1:13}
    I_\mathrm{d}(V_\mathrm{g}) \quad = \quad
\begin{cases}
    \quad - \mu \frac{W}{L} C_q V_\mathrm{T}^2 \exp{\left( \frac{V_\mathrm{g} - V_\mathrm{th}'}{ V_\mathrm{T}} \right)} \left( \exp{\left( - \frac{ V_\mathrm{d}}{ V_\mathrm{T}}  \right)} - 1 \right),  
    & \text{if} \quad V_\mathrm{g} \ll V_\mathrm{th}' \\
    \quad - \mu  \frac{W}{L} \left( \frac{1}{C_\mathrm{ins}} + \frac{1}{C_q}   \right) ^{-1} \left( V_\mathrm{g} - V_\mathrm{th}' - \frac{V_\mathrm{d}}{2} \right)  V_\mathrm{d}, 
    & \text{if} \quad  V_\mathrm{g} \gg V_\mathrm{th}' \\
\end{cases}
\end{equation}
Similar to a conventional 3D MOSFET, the drain current in the subthreshold region depends exponentially on the voltage overdrive, while in the on state, this dependence becomes linear.  \\

Finally, we extend the compact model to account for the impact of source and drain contact resistances. When the source and drain contacts have a non-negligible contact resistance \(R_\mathrm{S/D}\), the effective voltage drop \(V_\mathrm{d,eff}\) across the channel is reduced by the voltage drop across these contacts relative to the externally applied drain voltage \(V_\mathrm{d}\). This effect is captured by the equation 
\begin{equation}\label{eq:secA1:13}
 V_\mathrm{d}^\mathrm{eff} = V_\mathrm{d} -  I_\mathrm{d}(V_\mathrm{d,eff}) R_\mathrm{S} - I_\mathrm{d}(V_\mathrm{d,eff}) R_\mathrm{D},
\end{equation}
which serves as an implicit equation for \(I_\mathrm{d}\) as a function of \(V_\mathrm{d}\). In the off state, \(I_\mathrm{d}\) is small, leading to a negligible voltage drop across the contacts, and the device behaves like an ideal device (\(V_\mathrm{d} \approx V_\mathrm{d}^\mathrm{eff}\)). However, in the on state, \(I_\mathrm{d}\) becomes significant, causing a substantial voltage drop across the contact resistances, leading to a deviation from the ideal device behavior (\(V_\mathrm{d} > V_\mathrm{d}^\mathrm{eff}\)).

The accuracy of the proposed compact model is demonstrated in \fig{fig:a1:1}{c}, which displays a comparison between simulated and measured \IDVG curves of a recently fabricated device with a \ce{Bi2O2Se}/\ce{Bi2SeO5}/Au stack \cite{Zhang2022}. The consistency across different drain biases underscores the model's ability to effectively capture the device behavior under varying operational conditions. 

\subsection{Simulation of Charge Transfer Reactions}\label{secA2}

The dynamics of the charge trapping process can be described by the non-radiative multiphonon (NMP) theory. The capture coefficient between a single delocalized state and a localized defect state can be calculated from first principles within the first order of electron-phonon coupling as demonstrated by Alkauskas et. al. \cite{PhysRevB.90.075202} or Turiansky et. al. \cite{TURIANSKY2021108056},  leading to the following expression:
\begin{equation}\label{eq:secA2:1}
\begin{aligned}
C_{if} = \frac{2 \pi }{\hbar} W_{if}^2 \sum_\alpha  w_\alpha(T)  \sum_\beta  \left\lvert \left\langle  \chi_{i, \alpha} |  \hat{Q} - Q_\mathrm{0}  | \chi_{f, \beta} \right\rangle \right\rvert^2\delta (E_{i, \alpha} - E_{f,\beta}),
\end{aligned}
\end{equation}
for the capture coefficient $C_{if}$. It is described in terms of the normalized electron-phonon matrix element \( W_{if} \) and the sum over all possible transitions from an initial state $i,\alpha$ to a final state $f,\beta$, with \( w_\alpha(T) \) accounting for the thermal occupation  of the initial states. The vibrational states \(\chi_{i, \alpha} \) and \(\chi_{f, \beta} \) are approximated by the wave functions of two displaced harmonic potentials, as previously illustrated in \fig{fig:sec3:sec1:1}{c}. These potentials are typically parametrized by the energy difference \( \Delta E \), the configuration coordinate shift \( \Delta Q \), the relaxation energy \( E_\mathrm{R} \), and the curvature ratio \( R \).\\

Calculating the capture coefficient according to Eq.~\ref{eq:secA2:1} for a simulation with thousands of time steps and defects would be too time-consuming, so suitable approximations must be used:

\begin{enumerate}

    \item The greatest computational effort is caused by the calculation and summation of the phonon matrix elements. However, it has been shown that for defects with sufficiently large \(\Delta Q \) (\(\gtrsim\) \qty{2}{\atomicmassunit^{1/2}\angstrom}) at sufficiently high temperatures (\(\gtrsim\) \qty{300}{\kelvin}), the double sum in Eq.~\ref{eq:secA2:1} can be replaced by the classical lineshape function:
        \begin{equation}\label{eq:secA2:3}
        \xi_\mathrm{classic} = \sqrt{\frac{\beta} {4 \pi}} \frac{R q_{21}^2}{\sqrt{E_\mathrm{R} + \Delta E (R^2 - 1) }}  \exp\left(-\beta \varepsilon_\mathrm{21}\right),
    \end{equation}
    which drastically reduces the computational effort \cite{GOES2018286}. In this expression \((q_{21}, \varepsilon_{21}) \) denotes the intersection point of both parabolic potentials (see \fig{fig:sec3:sec1:1}{c}), which itself can be expressed in terms of \( \Delta E \), \( \Delta Q\),  \( E_\mathrm{R} \) and  \( R \). \\

    \item The matrix element \(W_{if} \) depends on the overlap of the initial and final electronic wave function and typically decreases exponentially with the distance of the defect from the considered charge reservoir. Therefore, it is common practice to approximate the matrix element as:
    \begin{equation}\label{eq:secA2:4}
        W_{if}^2(x_\mathrm{T}) \approx W_{0}^2 T_\mathrm{WKB}(x_\mathrm{T}),
    \end{equation}
    where \(W_{0}\) is a constant and \(T_\mathrm{WKB}\) represents the tunnelling probability, which accounts for the exponential position dependence. In this work the tunneling probability is calculated according to the Wentzel–Kramers–Brillouin (WKB) approximation, indicated by the subscript WKB \cite{GRASSER201239}. The main advantage of this approximation is that the matrix element \(W_{if}\) can be calculated for defects at different positions once the constant \(W_{0}\) has been determined using first principles methods. \\

\end{enumerate}
These approximations lead to the following analytic expression for the capture coefficient:
\begin{equation}\label{eq:secA2:5}
\begin{aligned}
C_{if} = \frac{2 \pi }{\hbar} W_\mathrm{0}^2 T_\mathrm{WKB} \sqrt{\frac{\beta} {4 \pi}} \frac{R q_{21}^2}{\sqrt{E_\mathrm{R} + \Delta E (R^2 - 1) }}  \exp({-\beta \varepsilon_\mathrm{21}})
\end{aligned}
\end{equation}
Finally, the total capture rate between a continuum of delocalized band states and the localized defect state is given by the integral 
\begin{equation}\label{eq:secA2:6}
\begin{aligned}
\int C_{if}(E) \, \text{DOS}(E) \, f_\text{FD}(E) \, \mathrm{d}E,
\end{aligned}
\end{equation}
which takes into account the density of states (DOS) of the band and its occupation probability. This derivation focuses on interactions with the conduction band, as the procedure for calculating the rates for the valence band and the gate is identical. To evaluate the integral over the conduction band, we use the band edge approximation, where \( C_{if}(E) \) is replaced by \( C_{if}(E_\mathrm{C}^\mathrm{ch}) \), representing the capture coefficient for an electron located exactly at the conduction band edge. Under this approximation, the electron capture rate \( k^\text{CB}_{21} \) and the electron emission rate \( k^\text{CB}_{12} \) for interactions with the conduction band are given by:

\begin{equation}\label{eq:secA2:7}
\begin{aligned}
k_{12}^\mathrm{CB} &= n_\mathrm{ch} \frac{2 \pi }{\hbar} W_\mathrm{0}^2 T_\mathrm{WKB} \sqrt{\frac{\beta} {4 \pi}} \frac{R q_{21}^2}{\sqrt{E_\mathrm{R} + \Delta E (R^2 - 1) }}  \exp\left(-\beta (\varepsilon_\mathrm{21} +  E_\mathrm{F}^\mathrm{ch} - E_\mathrm{T}^\mathrm{eff} )\right) \\
k_{21}^\mathrm{CB} &= n_\mathrm{ch} \frac{2 \pi }{\hbar} W_\mathrm{0}^2 T_\mathrm{WKB} \sqrt{\frac{\beta} {4 \pi}} \frac{R q_{21}^2}{\sqrt{E_\mathrm{R} + \Delta E (R^2 - 1) }}  \exp\left(-\beta \varepsilon_\mathrm{21} \right)
\end{aligned}
\end{equation}
The rates can be expressed in terms of the device quantities \( n_\mathrm{ch} \), \( E_\mathrm{C}^\mathrm{ch} \) and \( E_\mathrm{F}^\mathrm{ch} \), the electronic matrix element \( W_\mathrm{0} \), the tunneling probability \( T_\mathrm{WKB} \), as well as the NMP parameters \( \Delta E \), \( \Delta Q\),  \( E_\mathrm{R} \) and  \( R \).  The energy difference \( \Delta E = E_\mathrm{C}^\mathrm{ch} - E_\mathrm{T}^\mathrm{eff} \) between the two parabolas is determined by the effective charge transition level \( E_\mathrm{T}^\mathrm{eff} \) of the defect (see \fig{fig:sec3:sec1:1}{c}). At flatband conditions, \( E_\mathrm{T}^\mathrm{eff} \)  corresponds to the energy \( E_\mathrm{T} \) at which the formation energies of the two charge states are equal \cite{GOES2018286}. However, when an electric field \(E_\mathrm{ins}\) is present in the insulator, the band diagram tilts, causing the defect to be energetically shifted by the energy difference \( ~ q E_\mathrm{ins} x_\mathrm{T} \) with respect to the channel, where \(x_\mathrm{T}\) denotes defect's distance from the channel. As a result, the effective charge transition level that enters the capture and emission rates is given by:
\begin{equation}\label{eq:secA2:2}
    E_\mathrm{T}^\mathrm{eff} = E_\mathrm{T} - q E_\mathrm{ins} x_\mathrm{T}.
\end{equation}
Consequently, the parabolas depicted in \fig{fig:sec3:sec1:1}{c} shift vertically relative to each other depending on the applied field in the insulator, resulting in the strong gate bias dependence of the capture and emission rates observed experimentally. A more detailed description of the NMP model parameters and their influence on the rates can be found in the literature \cite{GOES2018286} and \cite{GRASSER201239}. \\

In general, every defect interacts not only with the conduction band but also with the valence band and gate. Thus the total rates for capture and emission are given by \(k_{21} = k_{21}^\mathrm{CB} + k_{21}^\mathrm{VB} + k_{21}^\mathrm{G} \) and \(k_{12} = k_{12}^\mathrm{CB} + k_{12}^\mathrm{VB} + k_{12}^\mathrm{G} \) respectively, yielding the characteristic capture time \(\tau_\mathrm{c} = 1 / k_{21} \) and emission time \(\tau_\mathrm{e} = 1 / k_{12} \) of the defect \cite{RZEPA201849}.
Finally, the dynamics of the defect is described by the Pauli master equation using the total capture and emission rates:
\begin{equation}\label{eq:secA2:8}\begin{aligned}\frac{\mathrm{d}f_2(t)}{\mathrm{d}t} = + k_{12}(t) f_1(t) - k_{21}(t) f_2(t),
\end{aligned}
\end{equation} 
\begin{equation}\label{eq:secA2:9}
\begin{aligned}
\frac{\mathrm{d}f_1(t)}{\mathrm{d}t} = - k_{12}(t) f_1(t) + k_{21}(t) f_2(t),
\end{aligned}
\end{equation} 
where \(f_2\) denotes the probability that the defect is empty and \(f_1\) that it is occupied. The temporal evolution of the defects is generally very complex and is solved in practice numerically by an implicit Euler method.\\

Finally, we address the impact of defects in the gate insulator on the transfer characteristic. As described by Eq.~\ref{eq:secA1:9}, defects can induce a shift in the effective threshold voltage, potentially leading to hysteresis in the transfer characteristic. However, insulator defects with time constants well above \(1/f_\mathrm{min} = \qty{E3}{s}\) do not change their charge state during the hysteresis measurement. These defects therefore do not contribute to the hysteresis in the transfer characteristic, since they affect both the up- and down-sweep equally. In contrast, defects with time constants well below \(1/f_\mathrm{max} = \qty{E-3}{s}\) adiabatically follow their equilibrium occupation. Although these defects can alter the shape of the \IDVG curve, particularly the subthreshold slope,  they do not contribute to hysteresis since they also affect both the up-sweep and down-sweep equally. These fast-responding defects are typically found at the semiconductor interface or in the semiconductor itself due to the high tunneling probability and the generally low relaxation energy \cite{PhysRevB.90.075202, TURIANSKY2021108056}. As a result, hysteresis measurements primarily detect defects located in the insulator's border region, whose time constants are well within the measurement window.

\subsection{Simulation of Drift \& Diffusion of Mobile Charges}\label{secA3}
From a microscopic view point the diffusion of mobile charges occurs as the charges hop between stable lattice sites, whereby they have to overcome a certain energy barrier \(E_\mathrm{a}\), also known as migration barrier (see \fig{fig:sec3:sec2:1}{a}). Macroscopically, the movement of mobile charges in the direction orthogonal to the channel can be described by a 1D drift-diffusion (DD) equation \cite{heitjans_diffusion_2005}:

\begin{equation}\label{eq:secA3:1}
    \frac{1}{q} \frac{\partial \rho_\mathrm{ins}(x,t)}{\partial t} = - \frac{\partial J(x,t)}{\partial x}, \qquad  J(x,t) = - D \frac{1}{q} \frac{\partial \rho_\mathrm{ins}(x,t)}{\partial x} \pm  \mu E_\mathrm{ins}(x, t) \frac{1}{q} \rho_\mathrm{ins}(x,t)
\end{equation}
The diffusion coefficient \(D\) of most mobile charges in solids follows an exponential law \( D = D_\mathrm{0} \exp(-E_\mathrm{a} / k_\mathrm{B} T)\)~\cite{heitjans_diffusion_2005} \cite{Hillen1979} \cite{Greeuw1984}. Furthermore, the mobility \(\mu\) can be expressed via the diffusion coefficient using the Einstein relation \(D/\mu = k_\mathrm{B}T/q\) where \(q\) represents the charge of the particles \cite{Sze2006}. In general, the temporal evolution of the charge distribution \(\rho_\mathrm{ins}(x, t)\) can be quite complex, because the electric field \( E_\mathrm{ins}(x, t) \) depends on the instantaneous charge distribution of all particles. In this work, the DD equation is therefore solved numerically using an implicit backward Euler method for the temporal update. 

\begin{figure}[hbt!]
\begin{subfigure}[b]{.333\linewidth}
    \begin{tikzpicture}
    \node[inner sep=0pt] at (0,0) {
      \begin{minipage}{1.0\textwidth}
        \includegraphics[width=\textwidth]{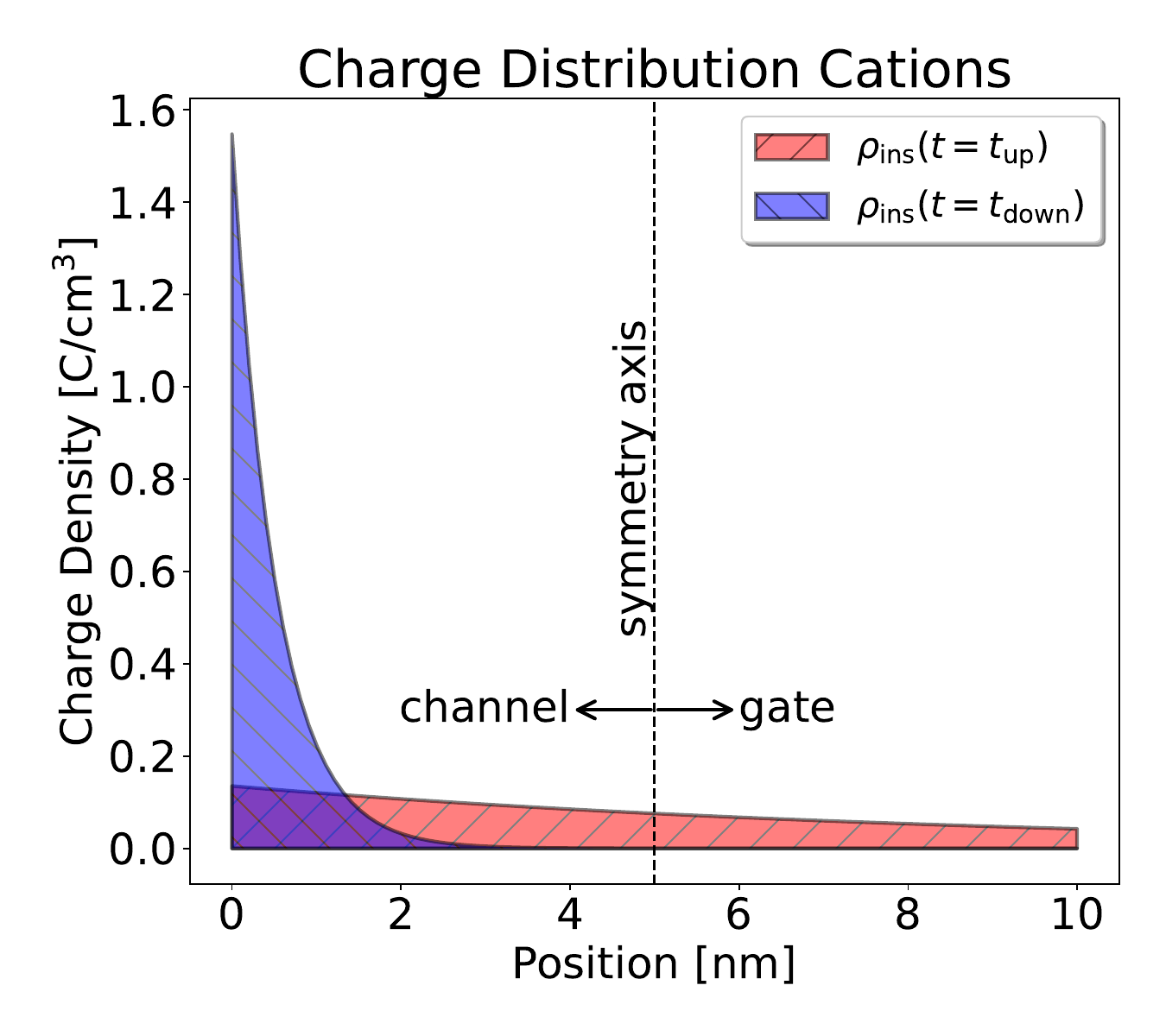} 
    \end{minipage}};
    \node[draw=none] at (-2.0,2.4) {\textbf{(a)}};
    \end{tikzpicture}
\end{subfigure}
\begin{subfigure}[b]{.333\linewidth}
    \begin{tikzpicture}
    \node[inner sep=0pt] at (0,0) {
      \begin{minipage}{1.0\textwidth}
        \includegraphics[width=\textwidth]{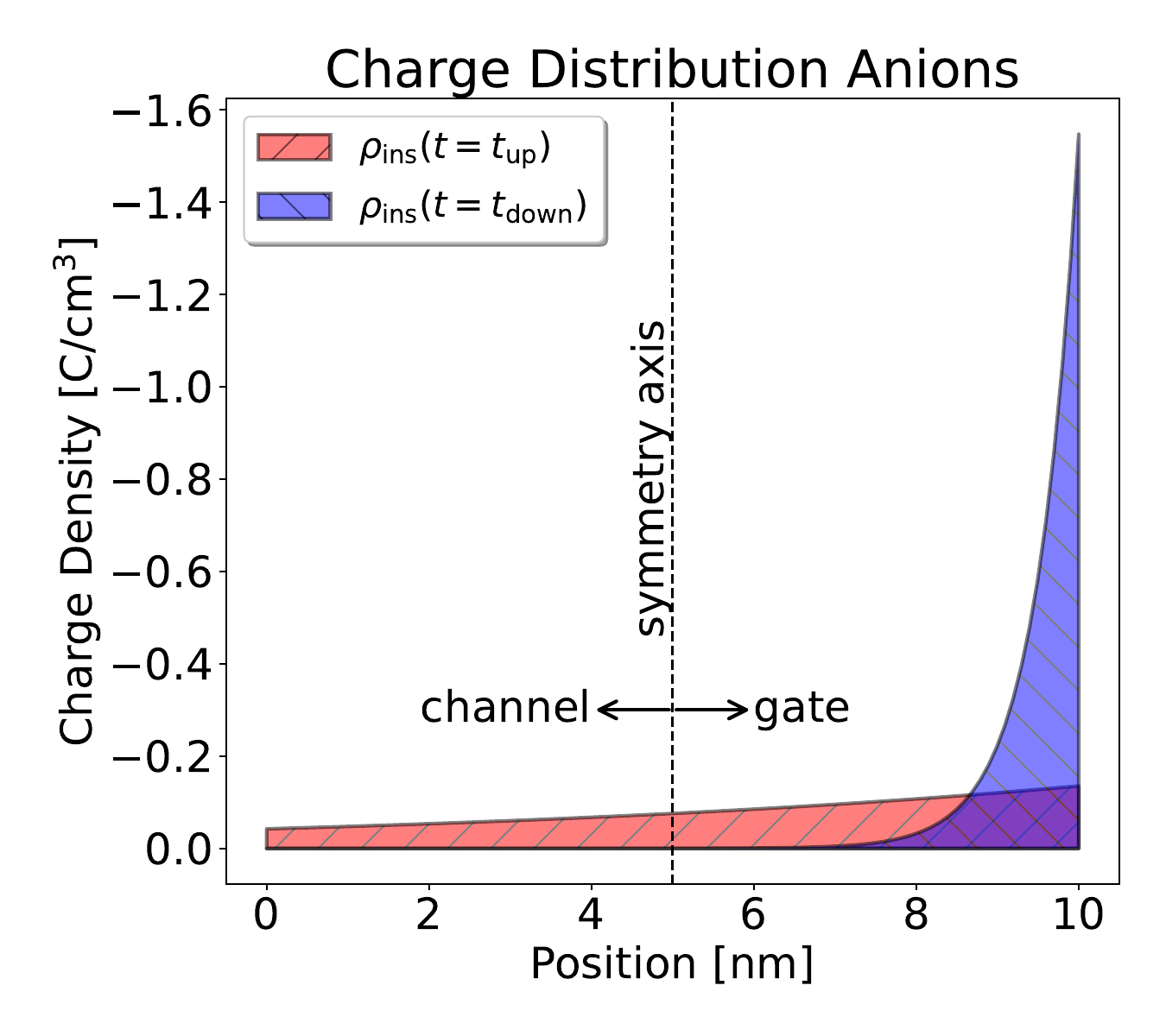} 
    \end{minipage}};
    \node[draw=none] at (-2.0,2.4) {\textbf{(b)}};
    \end{tikzpicture}
\end{subfigure}
\begin{subfigure}[b]{.333\linewidth}
    \begin{tikzpicture}
    \node[inner sep=0pt] at (0,0) {
      \begin{minipage}{1.0\textwidth}
        \includegraphics[width=\textwidth]{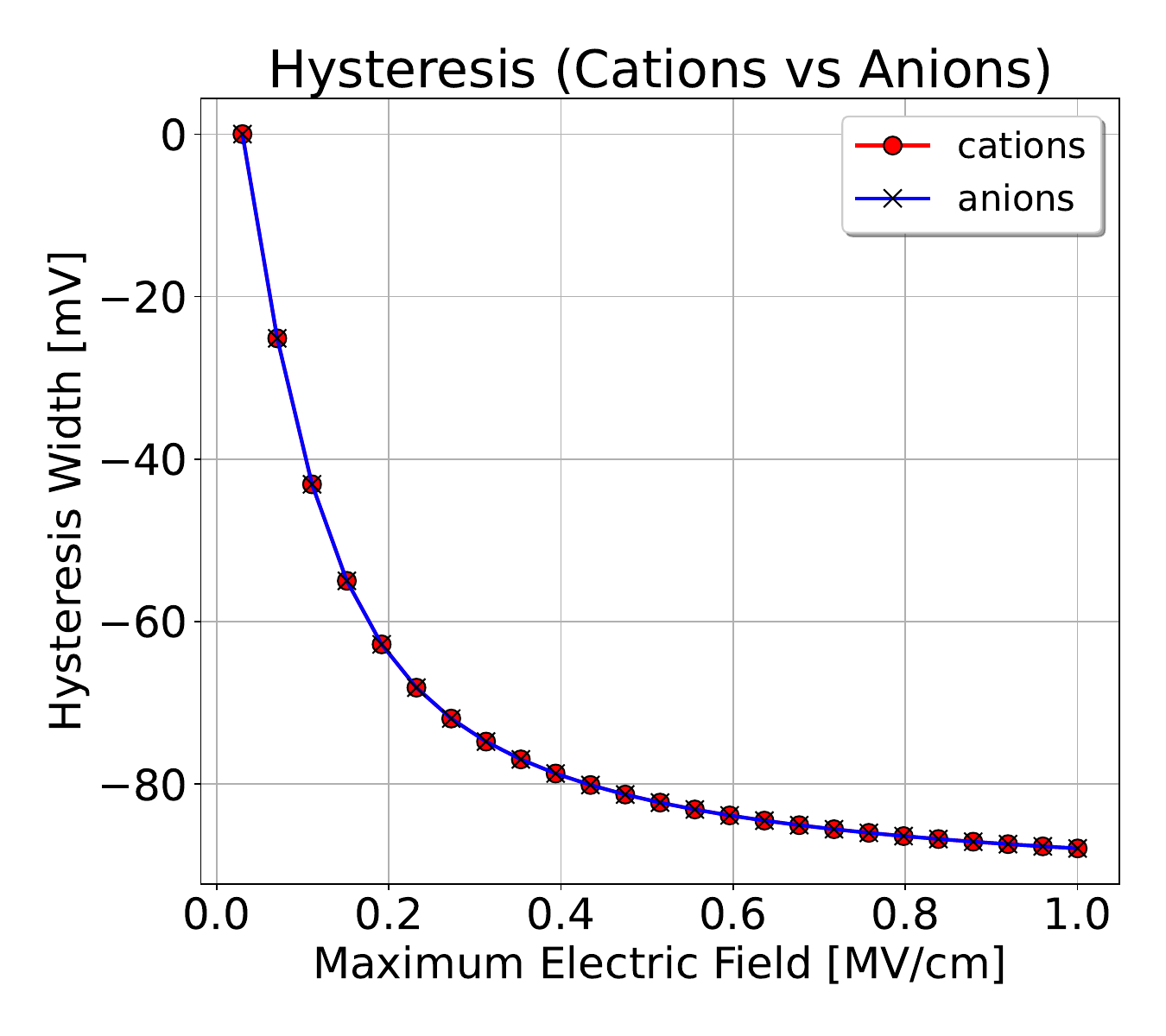} 
    \end{minipage}};
    \node[draw=none] at (-2.0,2.4) {\textbf{(c)}};
    \end{tikzpicture}
\end{subfigure}
\caption{\textbf{(a)} Distribution of cations during an up- and down-sweep. \textbf{(b)} Distribution of anions during up- and down-sweep. \textbf{(c)} Hysteresis due to anions and cations.}
\label{fig:a3:1}
\end{figure}

While a general analytical solution is unavailable, it is important to note that both positive and negative charges contribute to CCW hysteresis. To illustrate this, we consider two devices—one with cations and the other with anions in the gate insulator—each containing the same total number of ions. We assume that the time evolution of the electric field is similar in both devices. This condition holds if the total number of charges is small enough that the electric field generated by the ions is only a minor perturbation of the field of the ion-free device. When the devices are switched off, both cations and anions adopt a roughly uniform distribution within the insulator, because the field is small. However, when the devices are switched on, the cations are driven toward the channel (see \fig{fig:a3:1}{a}), while the anions are pushed in the opposite direction, towards the gate (see \fig{fig:a3:1}{b}). In both cases, the sweep results in an effective positive charge difference near the channel, resulting in CCW hysteresis, as demonstrated by the following considerations:  Under the conditions mentioned, the cyclo-stationary distributions of cations and anions determined according to the DD equation obeys the following symmetry:
\begin{equation}\label{eq:secA3:2}
\rho_\mathrm{ins}(x, t)\big|_\mathrm{cations} \approx -\rho_\mathrm{ins}( d_\mathrm{ins} - x, t)\big|_\mathrm{anions}, 
\end{equation}
which is also evident in \fig{fig:a3:1}{a} and \fig{fig:a3:1}{b}. Due to this symmetry, the total charge \(Q_\mathrm{ins}\) and the displacement of the charge center \(x_\mathrm{ins}\) reverse their sign when all cations are swapped for anions:
\begin{equation}\label{eq:secA3:3}
\begin{split}
Q_\mathrm{ins}\big|_\mathrm{cations} &= -Q_\mathrm{ins}\big|_\mathrm{anions}, \\ \left(
x_\mathrm{ins}(t_\mathrm{down}) - x_\mathrm{ins}(t_\mathrm{up}) \right)\bigg|_\mathrm{cations} &= -\left( x_\mathrm{ins}(t_\mathrm{down}) - x_\mathrm{ins}(t_\mathrm{up}) \right)\big|_\mathrm{anions}.
\end{split}
\end{equation}
As a result of this symmetry, the hysteresis induced by cations and anions is the exactly the same for low ions concentration and to a good approximation otherwise:
\begin{equation}\label{eq:secA3:4}
     \Delta V_\mathrm{H} =  \frac{Q_\mathrm{ins}}{C_\mathrm{ins}}  \left( \frac{x_\mathrm{ins}(t_\mathrm{down})}{d_\mathrm{ins}} - \frac{x_\mathrm{ins}(t_\mathrm{up})}{d_\mathrm{ins}}\right) \qquad    \Rightarrow \qquad     \Delta V_\mathrm{H} \bigg|_\mathrm{cations} = \Delta V_\mathrm{H} \bigg|_\mathrm{anions} < 0
\end{equation}

In summary, mobile charges in the insulator induce a CCW hysteresis, regardless of their charge polarity. This result is highlighted in \fig{fig:a3:1}{c} that displays the hysteresis caused by cations and anions as a function of the maximum electric field in the insulator. \\

\subsection{Simulation of Ferroelectric Switching Dynamics }\label{secA4}

Ferroelectricity generally occurs below a critical temperature \(T_\mathrm{C}\), also known as the Curie temperature (in analogy to ferromagnetism). When the Curie temperature is exceeded the material loses its ferroelectric properties and typically transitions to a paraelectric phase. This transition can be described phenomenologically by the Landau-Devonshire theory, where the free energy is expanded as a series in terms of the polarization \(P\) and the applied electric field \(E\) in the layer. When neglecting effects such as the depolarization field, mechanical strain and crystal defects, the free energy can be expressed as \cite{Chandra2007} 
\begin{equation}\label{eq:secA4:1}
 F(P, E, T) = \left(\frac{1}{2} a P^2 + \frac{1}{4} b P^4 + ... - P E\right) V,
\end{equation}
where \(a\), \(b\), and \(c\) are material-specific coefficients that may vary with temperature, and \(V\) represents the volume under consideration. Below the Curie temperature,  \(a\) is assumed to be negative, and the free energy has the shape of a double well (see  \fig{fig:sec3:sec3:1}{a}), whose local minima \(P_1\) and \(P_2\) correspond to the possible equilibrium polarization states of the ferroelectric phase. Without electric field, the double well is symmetric, where \(P_\mathrm{S} = \left| P_\mathrm{1} \right| =  \left| P_\mathrm{2} \right| \) denotes the spontaneous polarization and \(W_\mathrm{B} = W_\mathrm{21} / V =  W_\mathrm{12} / V \) the normalized energy barrier between both polarization states. Another key assumption of the Landau-Devonshire model is that \( a(T) = a_0 (T - T_\mathrm{C})\) holds, while the remaining coefficients are assumed to be temperature independent. This leads to the following temperature dependence of the spontaneous polarization and the energy barrier:
\begin{equation}\label{eq:secA4:2}
 P_\mathrm{S} \approx \sqrt{ \frac{a_0}{b}(T_\mathrm{C} - T)} , \qquad  W_\mathrm{B} \approx  \frac{1}{4}  \frac{a_0^2 (T_\mathrm{C} - T)^2}{b},
\end{equation}
Both spontaneous polarization and the energy barrier approach zero as the Curie temperature is reached, indicating that the material loses its ferroelectric property to maintain a spontaneous polarization. \\

When an electric field is applied, the symmetry of the free energy landscape is broken (see  \fig{fig:sec3:sec3:1}{a}) and the energy barriers are modified to first order as \(W_\mathrm{21} \approx (W_\mathrm{B} + P_\mathrm{S} E) V \) and \(W_\mathrm{12} \approx (W_\mathrm{B} - P_\mathrm{S} E) V\). In this work we follow the approach of Vopsaroiu et. al. \cite{PhysRevB.82.024109}, which has proven successful in reproducing experimental data of thin films \cite{CHEN2020100919, Zhu2016, Nomura2015}, and model the transition between both polarization states as a thermally activated process with following rates:

\begin{equation}\label{eq:secA4:3}
 k_{12} = \nu_0 \exp\left( \frac{W_\mathrm{12}}{k_\mathrm{B}T} \right) \approx \nu_0 \exp\left( \frac{(W_\mathrm{B} - P_\mathrm{S} E) V}{k_\mathrm{B}T} \right),
\end{equation}
\begin{equation}\label{eq:secA4:4}
 k_{21} = \nu_0 \exp\left( \frac{W_\mathrm{21}}{k_\mathrm{B}T} \right) \approx \nu_0 \exp\left( \frac{(W_\mathrm{B} + P_\mathrm{S} E) V}{k_\mathrm{B}T} \right),
\end{equation}
where \(\nu_0\) represents the attempt frequency of the transition. The asymmetry of the transition rates as a function of the electric field enables the field to induce a transition from one polarization state to the another. For instance, if the system begins in polarization state 1 and the applied electric field is strong enough to alter the barriers such that \(W_\mathrm{12} \ll W_\mathrm{21}\), the system will switch to polarization state 2.

Finally, the dynamics of the polarization is described by the Pauli master equation:
\begin{equation}\label{eq:secA4:5}\begin{aligned}\frac{\mathrm{d}f_2(t)}{\mathrm{d}t} = + k_{12}(t) f_1(t) - k_{21}(t) f_2(t),
\end{aligned}
\end{equation} 
\begin{equation}\label{eq:secA4:6}
\begin{aligned}
\frac{\mathrm{d}f_1(t)}{\mathrm{d}t} = - k_{12}(t) f_1(t) + k_{21}(t) f_2(t),
\end{aligned}
\end{equation} 
Here, \(f_2\) represents the probability of the layer being polarized in the positive direction, while \(f_1\) represents the probability of polarization in the negative direction. The temporal evolution of polarization is solved numerically in practice using an implicit Euler method.

\subsection{Standardized Hysteresis Measurement Scheme}\label{secA5}

In this section we discuss the theory and limitations of the standardized hysteresis measurement scheme, employed in the main part of the work. Please note that the drain voltage is kept small (\( q V_\mathrm{d} \lesssim k_\mathrm{B} T \)) during hysteresis measurements, to keep the electric field along the channel direction small. While not strictly necessary, this restriction enables us to approximate quantities along the channel—such as the quasi Fermi level \(E_\mathrm{F}^\mathrm{ch}\)—as position-independent, thereby simplifying the following analysis. \\

The goal of a standardized hysteresis measurement scheme is to ensure that defects and mobile charges exhibit a similar temporal evolution in devices that differ only in their insulator thickness. While the dynamics of mobile charges depend primarily on the electric field \(E_\mathrm{ins}\) in the insulator (see SI~Sec.~\ref{secA3}), the dynamics of defects depend on both the electric field \(E_\mathrm{ins}\) and the Fermi level \(E_\mathrm{F}^\mathrm{ch}\) (see SI~Sec.~\ref{secA2}). Strictly speaking, one must therefore keep both the range of the electric field \([E_\mathrm{ins}(V_\mathrm{min}), E_\mathrm{ins}(V_\mathrm{max})] \) 
 and the range of the Fermi level \([E_\mathrm{F}^\mathrm{ch}(V_\mathrm{min}), E_\mathrm{F}^\mathrm{ch}(V_\mathrm{max})] \) consistent across the devices under consideration. However, Eq.~\ref{eq:secA1:8} can be rewritten in a slightly different form: 
 \begin{equation}\label{eq:secA5:1}
    E_\mathrm{ins}(n_\mathrm{ch})  = \frac{q}{\varepsilon_\mathrm{ins}} \left( N_\mathrm{d}^+ - N_\mathrm{a}^-  - n_\mathrm{ch}\right), 
\end{equation}
 \begin{equation}\label{eq:secA5:2}
     E_\mathrm{F}^\mathrm{ch}(n_\mathrm{ch})  = E_\mathrm{C}^\mathrm{ch} + q V_\mathrm{T} \ln{ \left(\exp{\left(\frac{q n_\mathrm{ch}}{  V_\mathrm{T} C_\mathrm{q}}\right)} - 1\right)}, 
\end{equation}
In other words, the electric field in the insulator and the Fermi level in the channel can each be uniquely controlled by the carrier concentration in the channel. Let us first examine the ideal case where the devices differ only in their insulator thickness. In this scenario, the ranges \([E_\mathrm{ins}(V_\mathrm{min}), E_\mathrm{ins}(V_\mathrm{max})]\) and \([E_\mathrm{F}^\mathrm{ch}(V_\mathrm{min}), E_\mathrm{F}^\mathrm{ch}(V_\mathrm{max})]\) naturally remain consistent across the devices, provided that the minimum carrier concentration \(n_\mathrm{ch}(V_\mathrm{min})\) and the maximum carrier concentration \(n_\mathrm{ch}(V_\mathrm{max})\) are maintained constant. \\

Consequently, based on Eq.~\ref{eq:secA1:11}, we obtain the implicit condition \(q n_\mathrm{ch}(V_\mathrm{max}) = \left(1/C_\mathrm{ins} + 1/C_\mathrm{q}\right)^{-1} \left( V_\mathrm{max} - V_\mathrm{th}' \right) = \mathrm{const} \) for the maximum voltage. However, in most cases the quantum capacitance \( C_\mathrm{q}\) is considerably larger than the gate insulator capacitance \( C_\mathrm{ins}\) and can therefore be neglected, simplifying the condition for the maximum voltage to: 
\begin{equation}\label{eq:secA5:3}
     \frac{V_\mathrm{max} - V_\mathrm{th}'}{\mathrm{EOT}} = \mathrm{const} 
\end{equation}

\begin{figure}[hbt!]
\begin{center}
\begin{subfigure}[b]{.333\linewidth}
    \begin{tikzpicture}
    \node[inner sep=0pt] at (0,0) {
      \begin{minipage}{1.0\textwidth}
        \includegraphics[width=\textwidth]{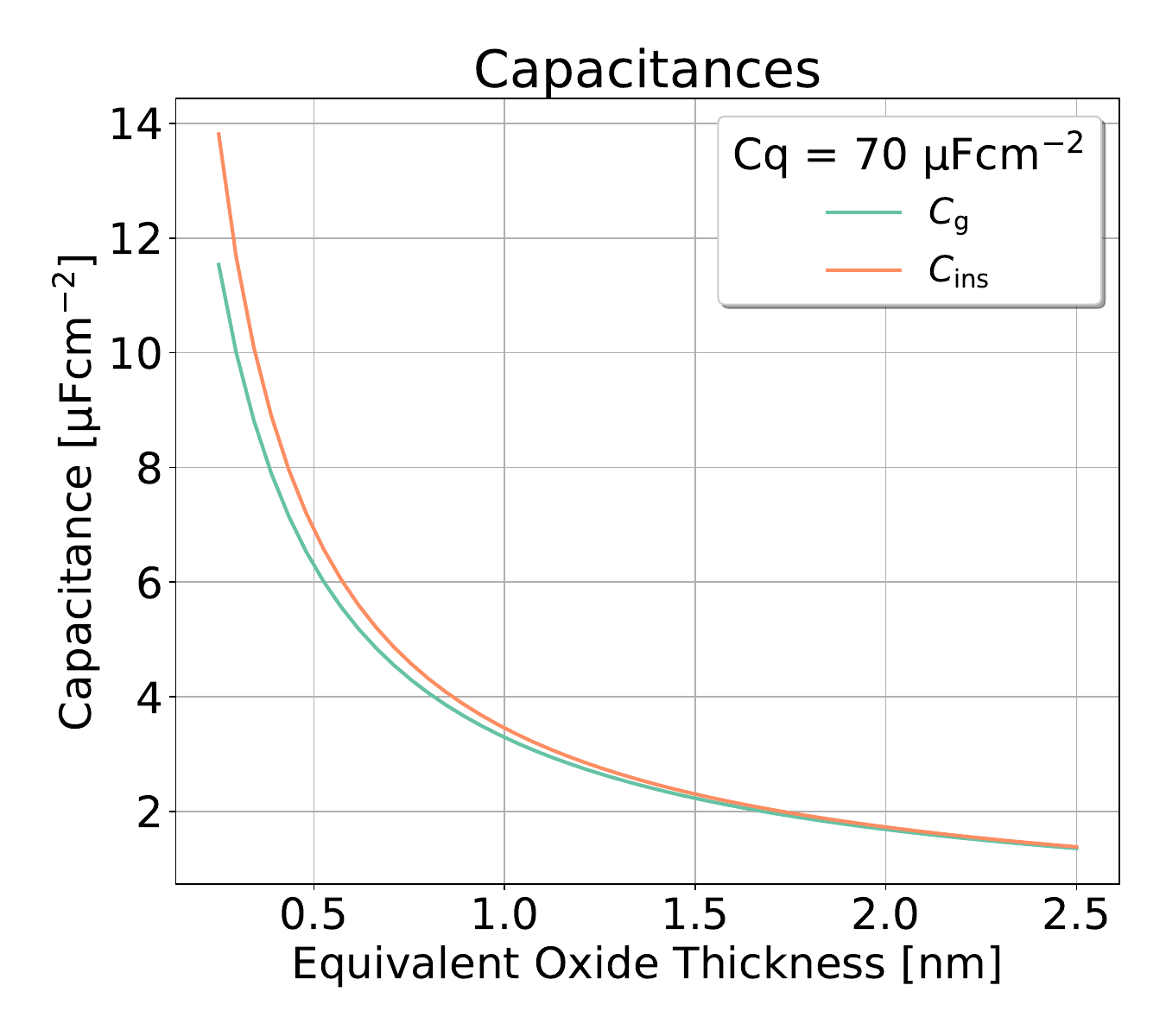} 
    \end{minipage}};
    \node[draw=none] at (-2.0,2.4) {\textbf{(a)}};
    \end{tikzpicture}
\end{subfigure}
\begin{subfigure}[b]{.333\linewidth}
    \begin{tikzpicture}
    \node[inner sep=0pt] at (0,0) {
      \begin{minipage}{1.0\textwidth}
        \includegraphics[width=\textwidth]{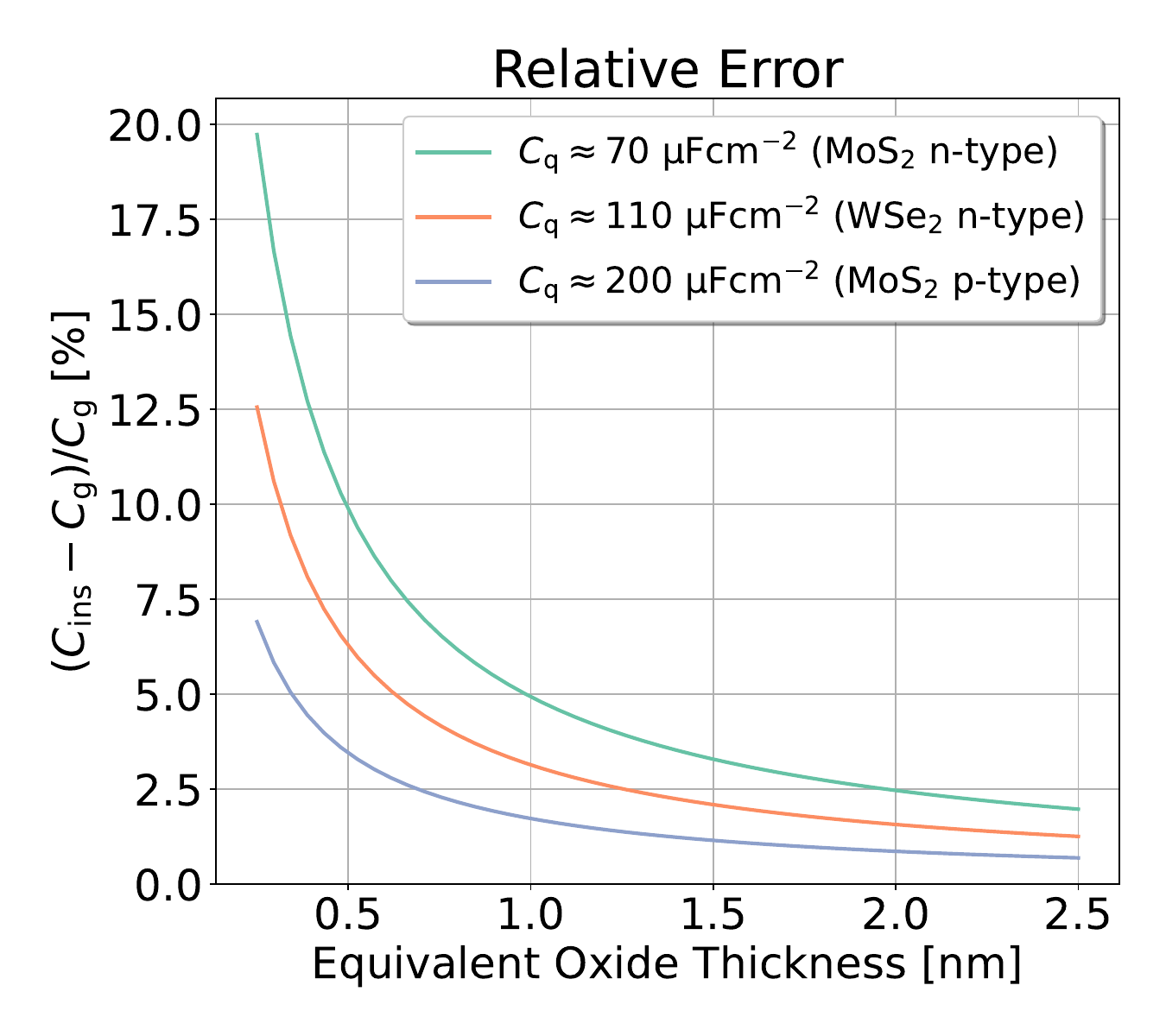} 
    \end{minipage}};
    \node[draw=none] at (-2.0,2.4) {\textbf{(b)}};
    \end{tikzpicture}
\end{subfigure}
\end{center}
\caption{\textbf{(a)} The total capacitance given by \(C_\mathrm{g} = (1/C_\mathrm{q} + 1/C_\mathrm{ins})^{-1}\), and the gate insulator capacitance \(C_\mathrm{ins}\) plotted as functions of the equivalent oxide thickness (EOT), assuming a quantum capacitance of \(C_\mathrm{q} = \qty{70}{\micro\farad\centi\meter^{-2}}.\) \textbf{(b)} Relative error in approximating the total capacitance \(C_\mathrm{g} = (1/C_\mathrm{q} + 1/C_\mathrm{ins})^{-1}\) with the gate insulator capacitance \(C_\mathrm{ins}\) plotted for various channel materials.\cite{Bera2019}\cite{Bennett2023}.}
\label{fig:a5:1}
\end{figure}

This simplification is particularly valid for prototype devices with thick insulators. For example the quantum capacitance for n-type monolayer \ce{MoS2} has been estimated to be \(C_\mathrm{q} = \qty{70}{\micro\farad\centi\meter^{-2}} \) \cite{Bennett2023}. \fig{fig:a5:1}{a} displays the corresponding total capacitance \(C_\mathrm{g} = (1/C_\mathrm{q} + 1/C_\mathrm{ins})^{-1}\) and the gate insulator capacitance \(C_\mathrm{ins}\) plotted as functions of the EOT. The figure shows that for EOT values greater than \qty{1}{\nano\meter}, the total capacitance can be accurately approximated by the insulator capacitance. Moreover, \fig{fig:a5:1}{b} presents the relative error introduced by substituting the total capacitance with the insulator capacitance for various 2D-materials.  The figure indicates that for typical EOTs in prototype devices (\( \qty{2}{\nano\meter} - \qty{10}{\nano\meter}\)), the approximation error is generally less than a few percent. If this approximation is not valid, the quantum capacitance can still be calculated from the effective mass of the 2D-material according to Eq.~\ref{eq:secA1:7} or measured directly using a three-electrode method as demonstrated by Xia et al. \cite{Xia2009}. \\

Moreover, for small drain voltages the relationship \(  n_\mathrm{ch} (V_\mathrm{max}) /  n_\mathrm{ch} (V_\mathrm{min}) \approx I_\mathrm{d} (V_\mathrm{max}) /  I_\mathrm{d} (V_\mathrm{min})\) can be obtained from Eq.~\ref{eq:secA1:11} and Eq.~\ref{eq:secA1:13}. This means that the minimum carrier concentration can be kept constant in practice by maintaining a fixed on/off ratio, which leads to an implicit condition for the minimum voltage:

\begin{equation}\label{eq:secA5:4}
      n_\mathrm{ch}(V_\mathrm{min}) \approx n_\mathrm{ch}(V_\mathrm{max})  \frac{I_\mathrm{d}(V_\mathrm{min}) }{I_\mathrm{d}(V_\mathrm{max}) } = \mathrm{const},
\end{equation}

Eq.~\ref{eq:secA5:3} and Eq.~\ref{eq:secA5:4} establish the foundation of the proposed measurement scheme and specify the sweep range for hysteresis measurements in terms of normalized voltage overdrive \(\left( V_\mathrm{max} - V_\mathrm{th}' \right) / \mathrm{EOT} \) and the on-off ratio \( I_\mathrm{d}(V_\mathrm{max})  / I_\mathrm{d}(V_\mathrm{min}) \). We recommend using a normalized voltage overdrive of \(\qty{2.0}{\mega\volt\centi\meter^{-1}}\) and an on-off ratio of \(\qty{E6}{} \) for standardized hysteresis measurements. This range reflects a realistic application scenario for 2D-MOSFETs, while being still applicable to most prototype devices. The determination of the sweep range based on Eq.~\ref{eq:secA5:3} and Eq.~\ref{eq:secA5:3} only requires the effective threshold voltage as well as the EOT of the sample, making this measurement specification relatively straightforward to implement. However, since the effective threshold voltage is itself time-dependent and prone to slow drifts, it is advisable to determine the threshold voltage and sweep range shortly before performing the hysteresis measurement. Additionally, these parameters should be updated for any subsequent hysteresis measurements.\\

\begin{figure}[hbt!]
\begin{subfigure}[b]{.333\linewidth}
    \begin{tikzpicture}
    \node[inner sep=0pt] at (0,0) {
      \begin{minipage}{1.0\textwidth}
        \includegraphics[width=\textwidth]{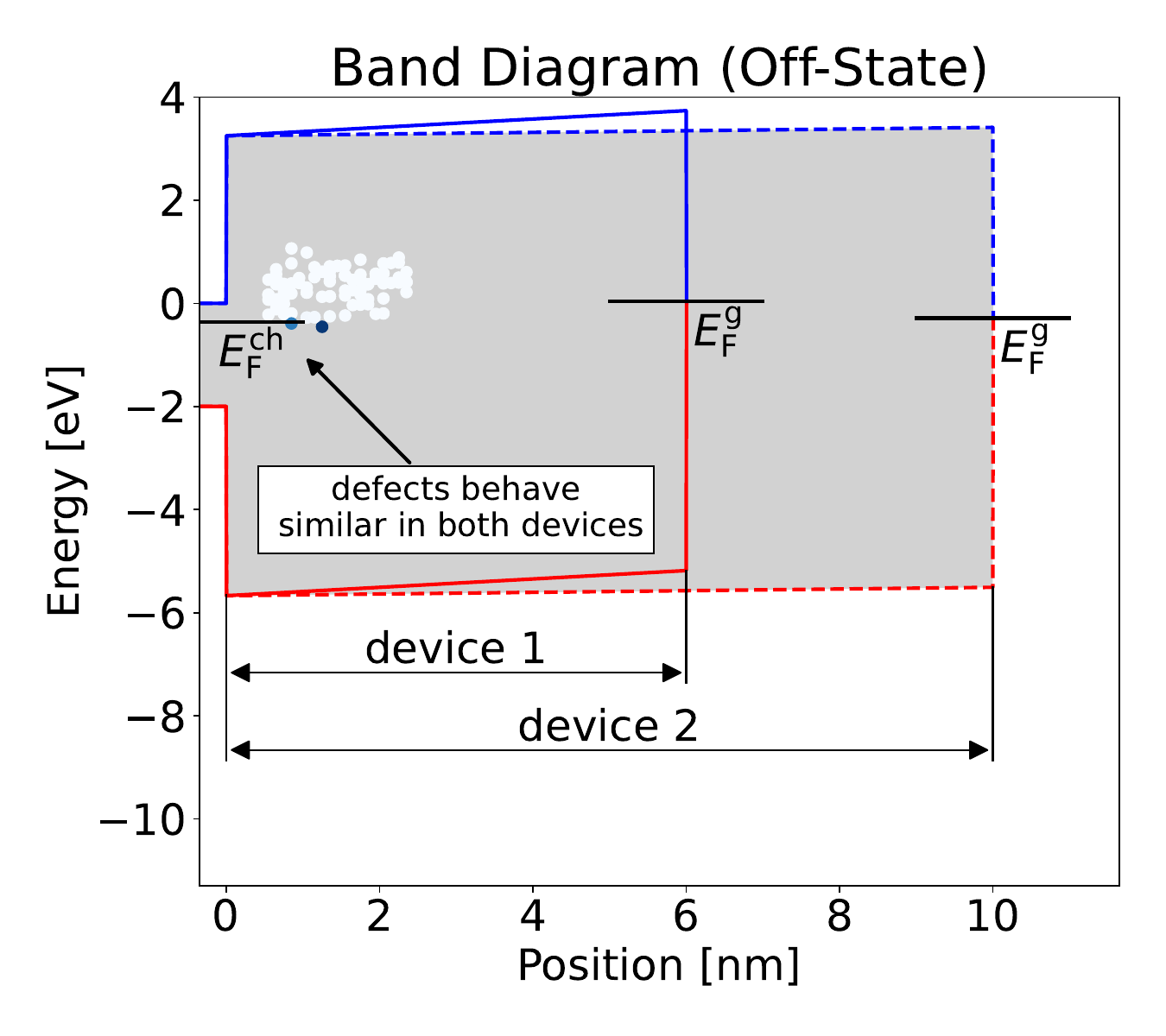} 
    \end{minipage}};
    \node[draw=none] at (-2.0,2.4) {\textbf{(a)}};
    \end{tikzpicture}
\end{subfigure}
\begin{subfigure}[b]{.333\linewidth}
    \begin{tikzpicture}
    \node[inner sep=0pt] at (0,0) {
      \begin{minipage}{1.0\textwidth}
        \includegraphics[width=\textwidth]{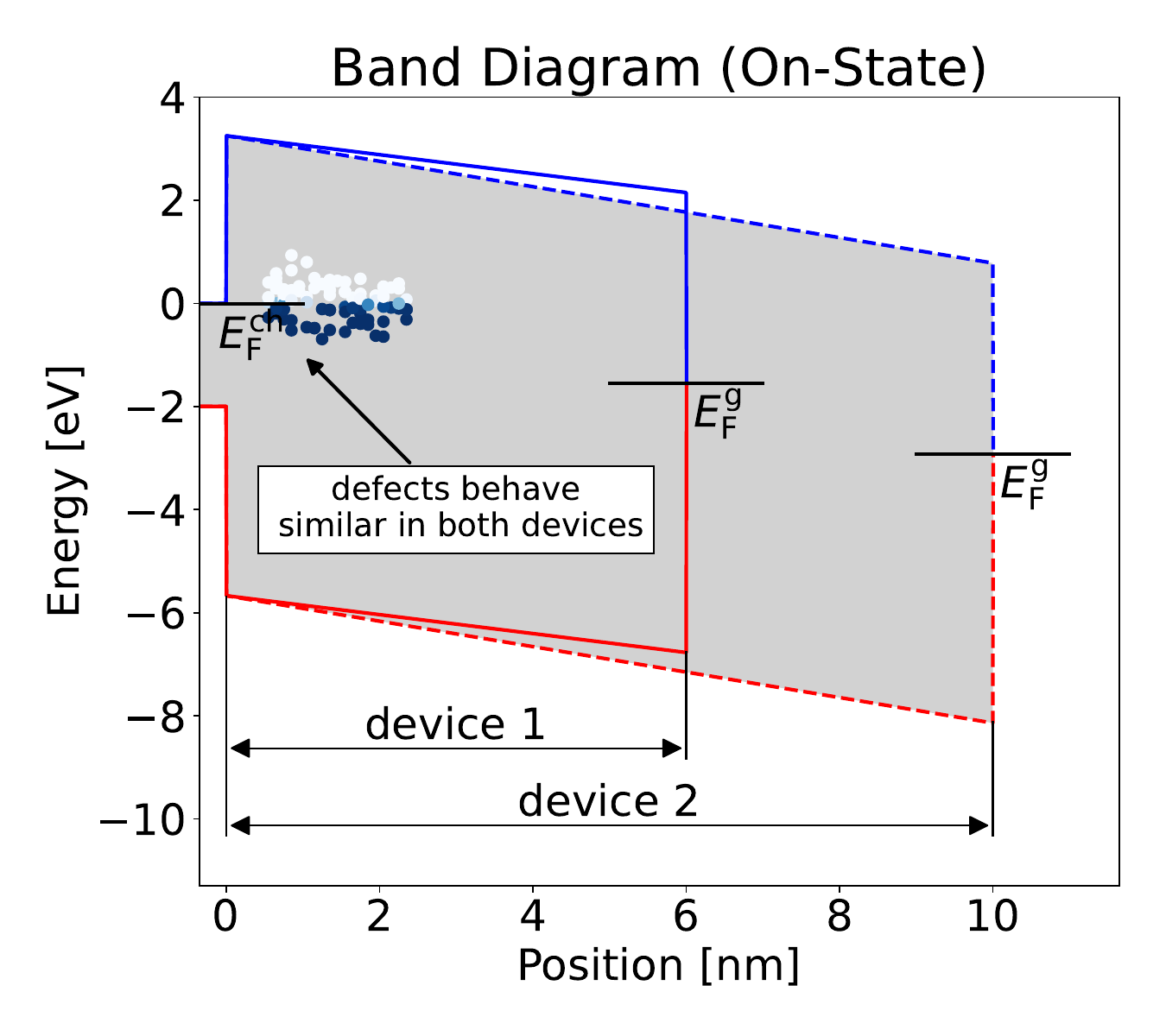} 
    \end{minipage}};
    \node[draw=none] at (-2.0,2.4) {\textbf{(b)}};
    \end{tikzpicture}
\end{subfigure}
\begin{subfigure}[b]{.333\linewidth}
    \begin{tikzpicture}
    \node[inner sep=0pt] at (0,0) {
      \begin{minipage}{1.0\textwidth}
        \includegraphics[width=\textwidth]{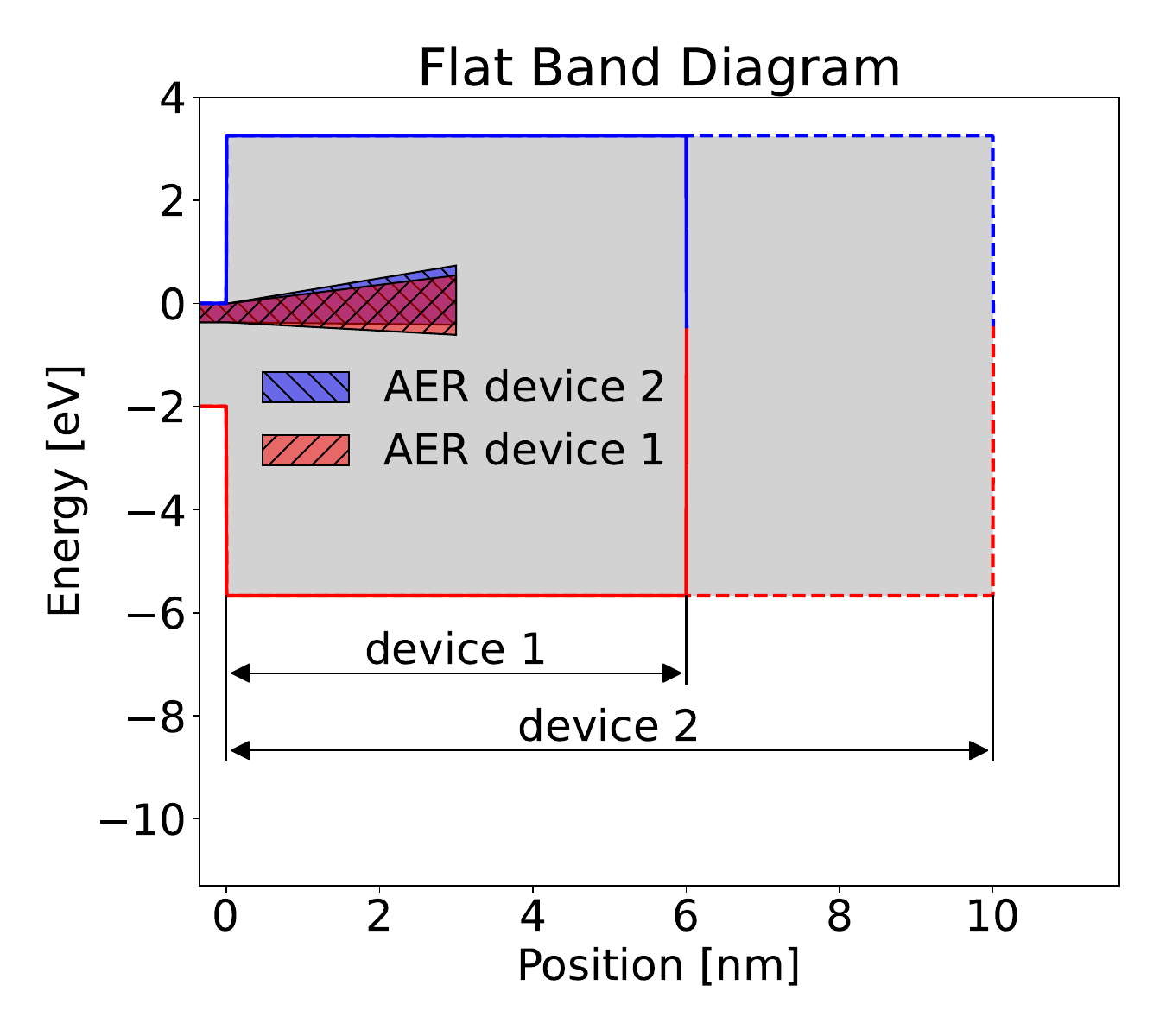} 
    \end{minipage}};
    \node[draw=none] at (-2.0,2.4) {\textbf{(c)}};
    \end{tikzpicture}
\end{subfigure}
\caption{Impact of varying dopant (impurity) concentrations on the measurement scheme, demonstrated by the comparison of two devices with distinct dopant concentrations (device~1: \( d_\mathrm{ins} = \qty{6}{\nano\meter}\) and \( N_\mathrm{d} = \qty{1.75E12}{\centi\meter^{-2}}\), device~2: \( d_\mathrm{ins} = \qty{10}{\nano\meter}\) and \(  N_\mathrm{d} = \qty{3.5E11}{\centi\meter^{-2}}\)). \textbf{(a)} Superimposed band diagrams of both devices in their off state. \textbf{(b)} Superimposed band diagrams of both devices in their on state. \textbf{(c)} Comparison of the AERs of both devices. The comparison demonstrates that the AERs of both devices are slightly shifted with respect to each other.}
\label{fig:a5:2}
\end{figure}

So far, we have considered the ideal scenario where the devices only differ in their thickness. If the devices also vary in other properties, it may be impossible to keep the range \([E_\mathrm{ins}(V_\mathrm{min}), E_\mathrm{ins}(V_\mathrm{max})] \) and \([E_\mathrm{F}^\mathrm{ch}(V_\mathrm{min}), E_\mathrm{F}^\mathrm{ch}(V_\mathrm{max})] \) simultaneously consistent across the devices. To illustrate this, \fig{fig:a5:2}{} compares two devices with distinct dopant (impurity) concentrations: device~1 (\( d_\mathrm{ins} = \qty{6}{\nano\meter},\ N_\mathrm{d} = \qty{1.75E12}{\centi\meter^{-2}} \)) and device~2 (\( d_\mathrm{ins} = \qty{10}{\nano\meter},\ N_\mathrm{d} = \qty{3.5E11}{\centi\meter^{-2}} \)). Although the standardized measurement scheme still ensures that both devices reach the same minimum and maximum Fermi levels, their band diagrams are misaligned in both the off-state and on-state (see \fig{fig:a5:2}{a} and \fig{fig:a5:2}{b}). This misalignment occurs because the same Fermi level is now mapped to different electric fields according to Eq.~\ref{eq:secA5:1}. Consequently, the active energy regions of the two devices are slightly shifted relative to each other (see \fig{fig:a5:2}{c}), which implies that the defects in both devices will no longer behave completely identically. \\

This analysis demonstrates that variations in the dopant (impurity) concentrations in the channel introduce slight variations in the AER between nominally identical devices. Although the changes in the AER remain small, in practice hysteresis measurements should always be carried out on several nominally identical devices. The mean or median value of the metric \(\Delta V_\mathrm{EOT} = \Delta V_\mathrm{H}^\mathrm{max} \)~/~EOT (i.e., the maximum hysteresis within the measurement window normalized by the EOT), can then be determined for these devices, which is much more robust against strong outliers and thus effectively reflects the average trend of the device type.



\clearpage

\bibliography{refs}

\end{document}